\documentclass[11pt]{article}        % base layout
\usepackage[a4paper,margin=1in]{geometry} % similar margins
\usepackage{authblk}                 % author/affiliation footnotes
\usepackage[english]{babel}
\usepackage[utf8]{inputenc}  % Encoding
\usepackage[T1]{fontenc}     % Font encoding
\usepackage{amssymb,amsthm,amsmath,amsfonts, bm} % Load AMS packages first
\usepackage{newtxtext}     % OK to keep before
\usepackage{newtxmath}     % Load after amssymb to avoid \Bbbk conflict
\usepackage{graphicx}        % Figures
\usepackage{algorithm,algpseudocode}  % Algorithms
\usepackage{subcaption}
\usepackage{booktabs,gensymb}
\usepackage{float}
\usepackage{csquotes}
\usepackage[
  backend=biber,
  style=numeric-comp,    
  sorting=nyt,           
  giveninits=false,
  maxbibnames=5,         
  doi=true,
  url=false,
  eprint=false,
  isbn=false
]{biblatex}
\DeclareFieldFormat{volume}{\textbf{#1}}
\renewbibmacro*{journal+issuetitle}{%
  \printfield{journaltitle}% italic by default
  \setunit*{\addspace}%
  \printfield{volume}% now bold
  \setunit{\addspace}%
  \printfield{pages}}

\renewbibmacro{in:}{%
  \ifentrytype{article}{}{\printtext{\bibstring{in}\intitlepunct}}}
\usepackage{hyperref}           
\hypersetup{colorlinks=true,linkcolor=blue,
            citecolor=blue,urlcolor=blue}
\usepackage{microtype}               
\newtheorem{theorem}{Theorem}[section]
\newtheorem{proposition}[theorem]{Proposition}  
\title{Incorporating Correlated Nugget Effects in Multivariate Spatial Models: An Application to Argo Ocean Data}

\author[1]{\textsc{Damilya Saduakhas}%
  \thanks{Corresponding author.\;E-mail: \texttt{damilya.saduakhas@kaust.edu.sa}}}
\author[1]{\textsc{David Bolin}}
\author[1]{\textsc{Xiaotian Jin}}
\author[1]{\textsc{Alexandre B.~Simas}}
\author[2]{\textsc{Jonas Wallin}}

\affil[1]{Statistics Program, CEMSE Division,  
          King Abdullah University of Science and Technology (KAUST),  
          Thuwal 23955-6900, Kingdom of Saudi Arabia}
\affil[2]{Department of Statistics, Lund University,  
          SE-221 00 Lund, Sweden}

\date{}  

\begin{document}
\maketitle
\vspace{-3em} 
\begin{abstract} 
Accurate analysis of global oceanographic data, such as temperature and salinity profiles from the Argo program, requires geostatistical models capable of capturing complex spatial dependencies. 
This study introduces Gaussian and non-Gaussian hierarchical multivariate Matérn-SPDE models with correlated nugget effects to account for small-scale variability and measurement error correlations. Using simulations and Argo data, we demonstrate that incorporating correlated nugget effects significantly improves the accuracy of parameter estimation and spatial prediction in both Gaussian and non-Gaussian multivariate spatial processes.
When applied to global ocean temperature and salinity data, our model yields lower correlation estimates between fields compared to models that assume independent noise. This suggests that traditional models may overestimate the underlying field correlation. By separating these effects, our approach captures fine-scale oceanic patterns more effectively.
These findings show the importance of relaxing the assumption of independent measurement errors in multivariate hierarchical models. 
\end{abstract}

\vspace{1em}
\noindent \textbf{Keywords:} non-Gaussian random fields; SPDE approach; Argo project; multivariate random fields; nugget effect.

%%%%%%%%%%%%%%%%%%%%%%%%%%%%%%%%%%%%%%%%%%%%%%
%%%% Main text entry area:
\section{Introduction}
Monitoring and modeling oceanic processes are critical to understanding climate variability, predicting environmental changes, and informing policies to mitigate climate impacts. 
Given that these processes are inherently multivariate, it is essential to model variables such as temperature and salinity simultaneously rather than treating them independently. In fact, joint modeling of temperature and salinity is crucial for understanding key oceanographic phenomena such as density stratification and heat transport \parencite{Talley2011, Siedler2013}. This enables the direct estimation of derived quantities, such as potential density, and avoids biases from treating fields independently \parencite{Yarger2022}. In this work, we propose a new class of Gaussian and non-Gaussian spatial multivariate models with a correlated nugget effect. We demonstrate the practical benefits of our method and apply it to Argo data, a comprehensive dataset of in-situ ocean measurements, which serves as an excellent example of the challenges posed by irregular spatio-temporal sampling.

The Argo Program, a global network of almost 4,000 active, fully autonomous profiling floats (sensors), first deployed in 1999, has revolutionized oceanography by providing high-resolution temperature and salinity measurements throughout the upper 2,000 meters of the ocean \parencite{ArgoProgram2017}. These data are indispensable for tracking climate-related oceanic changes, such as rising sea levels, ocean heat content, and circulation patterns, and for integration into Earth system models to enhance climate projections under diverse scenarios (e.g., \textcite{Durack2012}; \textcite{Chang2013}).

\begin{figure}
    \centering
    \includegraphics[width=\linewidth,trim=0cm 1cm 0cm 1cm,clip]{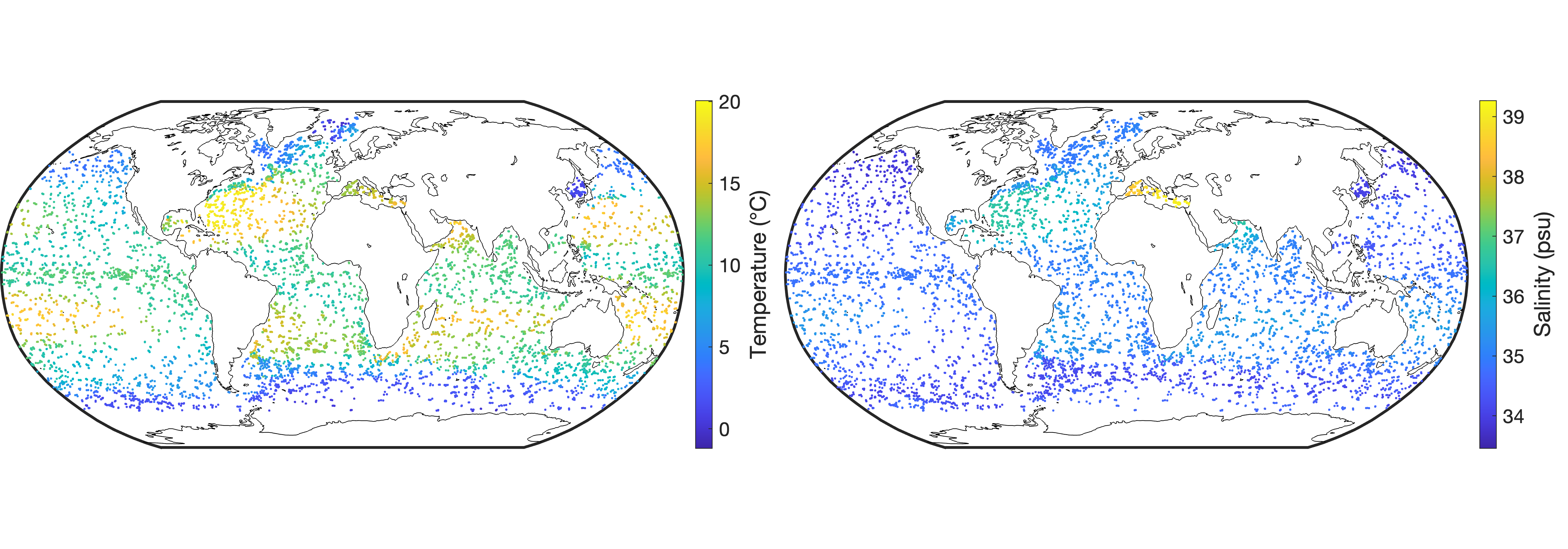}
    \caption{Argo temperature and salinity data at 300 dbar for January 2018. %The map illustrates the irregular spatial sampling characteristic of float trajectories. Robust interpolation methods are critical to derive continuous fields for climate studies.
    }
    \label{fig:map_data_example}
\end{figure}

The Argo floats drift with ocean currents, cyclically descending and ascending to collect vertical profiles of temperature and salinity, recording pressure in decibars (dbar) as a proxy for depth, where approximately 1 dbar corresponds to 1 meter. These vertical profiles, along with the coordinates and time stamps of the float, are transmitted by satellite to data processing centers \parencite{Wong2020}. This near-global sampling has transformed our understanding of marine physical dynamics \parencite{ArgoReview2022}. However, the analysis of Argo data presents significant challenges. Floats follow non-uniform trajectories dictated by currents, resulting in irregular spatio-temporal sampling. Furthermore, instrument-specific noise and systematic biases introduce variability that complicates reliable prediction, especially for salinity measurements \parencite{SaltyDrift2024, Wong2023}. Thus, robust statistical frameworks are essential to interpolate these sparse observations into dense, accurate gridded products for global ocean analyses (see Figure~\ref{fig:map_data_example}).

Various interpolation methods have been proposed for Argo data, ranging from oceanographic techniques to geostatistical approaches. Among these, kriging \parencite{Cressie2011}, also known as optimal interpolation or objective mapping in geosciences and oceanography \parencite{Barth2008,Bretherton1976}, has been widely adopted for its ability to specify mean and covariance structures and minimize the variance of prediction errors. A natural extension of this method to a multivariate context is cokriging \parencite{matheron1979recherche}. By explicitly incorporating the cross-covariance between correlated variables, such as temperature and salinity, cokriging leverages extra information from related processes to potentially improve prediction accuracy. However, implementing cokriging is more challenging than standard kriging, as it requires the joint modeling of the correlated variables.

Several widely used gridded products demonstrate the application of kriging-based methods for Argo data. For example, the widely used Roemmich--Gilson (RG) climatology, one of the first monthly products derived from Argo data, estimates the mean field using weighted local regression before applying kriging to map anomalies \parencite{Roemmich2009}. Other products, such as the In Situ Analysis System (ISAS) and the Grid Point Value of the Monthly Objective Analysis using Argo data (MOAA GPV), both apply optimal interpolation techniques. ISAS directly interpolates from Argo temperature and salinity profiles \parencite{ISAS}, whereas MOAA GPV \parencite{Hosoda2008} integrates additional oceanographic data sources, including Argo floats, to generate monthly gridded datasets of temperature and salinity fields. Despite their utility, these methods often treat temperature and salinity as independent fields, limiting their ability to capture important covariance structures.

Another limitation is the widespread assumption of independent and identically distributed (i.i.d.) measurement errors. Although Argo sensors are generally accurate, occasional problematic float measurements, sensor drift, and calibration offsets can introduce temporally and spatially correlated errors. For example, an analysis of 482 Atlantic Argo floats in real-mode found that 15\% of the data showed sensor drift or offset \parencite{Gaillard2009}. Furthermore, pressure sensor errors that misalign vertical profiles induce errors in both temperature and salinity, which then show a similar spatial distribution \parencite{Barker2011}. Notably, even delayed-mode salinity data, which undergo expert quality control,  can retain unadjusted biases, as shown by \textcite{Wong2023}. Another assessment of gridded products showed that these biases in the observations then resulted in unrealistic increases in global mean salinity from 2015 to 2019 \parencite{SaltyDrift2024}. These errors, if unaccounted for, propagate into gridded products, which lead to misestimated parameters, underestimated uncertainties, and compromised reliability in climate analyses. 

To overcome these challenges, one effective strategy is to model temperature and salinity jointly and integrate both processes with correlated observational uncertainties. Specifically, we propose a multivariate statistical framework that extends the Matérn Stochastic Partial Differential Equation (SPDE) model to jointly interpolate temperature and salinity and explicitly model correlated measurement noise.  This approach simultaneously interpolates the temperature and salinity fields while explicitly modeling their interdependence, which enables more accurate predictions. Unlike standard methods, we relax the assumption of independence in the noise structure to provide a more realistic representation of Argo data. Additionally, we address the non-Gaussian variability frequently observed in environmental datasets by incorporating non-Gaussian driving noise in the SPDE, specifically normal-inverse Gaussian (NIG) noise. This extension allows our model to better account for extreme values and provides robust uncertainty quantification.

Building on localized modeling approaches by \textcite{Kuusela2018} and \textcite{Park2021}, our framework adopts a moving-window strategy to estimate spatial covariance structures within localized neighborhoods. This approach enables the model to adapt to the nonstationary nature of Argo data to capture regional variations in ocean dynamics. Furthermore, we leverage sparse matrix operations and parallelized algorithms to ensure computational efficiency that makes the framework scalable for large-scale environmental datasets. To facilitate its use by practitioners, we implement our methodology in an open-source \textit{ngme2} R package, that allows seamless application to a variety of spatial processes, including but not limited to Argo data \parencite{ngme2}.
Through simulations and global analysis of Argo data (2007--2020), we demonstrate that correlated measurement noise significantly impacts parameter estimates and prediction accuracy. To our knowledge, this is the first study to provide a global bivariate analysis of Argo temperature and salinity fields using a non-Gaussian SPDE framework.

The rest of the paper is organized as follows. Section~\ref{sec:theory} introduces the proposed model and the theoretical foundation of the bivariate Matérn SPDEs. Section~\ref{sec:models} discusses parameter estimation methods and model evaluation. Section~\ref{sec:simulation} presents a simulation study to evaluate the impact of correlated measurement noise. Section~\ref{sec:application} applies the proposed model to Argo data, focusing on temperature and salinity measurements. Finally, Section~\ref{sec:conclusion} discusses the results, implications, and future research directions. Our code is publicly available online at \url{https://github.com/d-saduakhas/Argo-SPDE} 
to support reproducibility and facilitate reuse of the proposed framework.

\section{Multivariate Matérn Fields and Systems of SPDEs}\label{sec:theory}
In this section, we develop a hierarchical model for bivariate spatial data such as the Argo dataset and define the theoretical foundations. Section~\ref{sec:General_model} introduces the hierarchical bivariate model that links the Argo observations to a latent random field. In Section~\ref{sec:Multivariate_models} we then present a Gaussian bivariate Matérn–SPDE framework for that field, while Section~\ref{sec:BivariateModel} generalizes the construction to non-Gaussian noise. Finally, Section~\ref{sec:discretization} details the finite-element discretization used for numerical implementation.

\subsection{Model description}\label{sec:General_model}
Let $\mathbf{X}(\mathbf{s})  = (X_1(\mathbf{s}), X_2(\mathbf{s}))^{\top}$ be a bivariate spatial process on a spatial domain $\mathcal{D}$, specified as
\begin{equation}\label{eq:Xmodel}
\mathbf{X}(\mathbf{s}) = \mathbf{m}(\mathbf{s}) + \mathbf{u}(\mathbf{s}) \,, s\in \mathcal{D}, 
\end{equation}
where $\mathbf{u}(\mathbf{s}) = (u_1(\mathbf{s}),u_2(\mathbf{s}))^\top$
is a centered spatial random field, 
and $\mathbf{m}(\mathbf{s}) = (m_1(\mathbf{s}), m_2(\mathbf{s}))^\top$
is the mean value of $\mathbf{X}$ modeled as regressions on some explanatory variables,  $z_1(\mathbf{s}),\ldots, z_m(\mathbf{s})$, where $m\in \mathbb{N}$. Specifically,
\[
m_1(\mathbf{s}) = \sum_{k=1}^m \beta_{1,k} z_k(\mathbf{s}) \quad \text{and} \quad
m_2(\mathbf{s}) = \sum_{k=1}^m \beta_{2,k} z_k(\mathbf{s}).
\]

We then model observations of $\mathbf{X}$ from different locations as Gaussian and conditionally independent given the latent field. That is, if we have an observation $Y_1$ of $X_1$ at some location $\mathbf{s}_1$, this is modeled as $Y_{1} | \mathbf{X} \sim N(X_1(\mathbf{s}_1),\sigma_1^2)$, where $\sigma_1$ represent the measurement noise and small-scale variability, the so-called nugget effect for $X_1$ \parencite{Cressie2011}. If we have an observation $Y_2$ of $X_2$ at some other location $\mathbf{s}_2$, this is modeled as $Y_2 | \mathbf{X} \sim N(X_2(\mathbf{s}_2),\sigma_2^2)$ for a possibly different measurement noise variance $\sigma_2^2$. Finally, if we have an observation $\mathbf{Y}_3 = (Y_{3,1}, Y_{3,2})^{\top}$ of both $X_1$ and $X_2$ at the same location $\mathbf{s}_3$, we assume that $\mathbf{Y}_3 | \mathbf{X} \sim N(\mathbf{X}(\mathbf{s}_3), \mathbf{\Sigma})$ for 
$$
\mathbf{\Sigma} = \begin{bmatrix}
    \sigma_{\varepsilon,1}^2 & \sigma_{\varepsilon,1}\sigma_{\varepsilon,2}\rho_{\epsilon} \\
    \sigma_{\varepsilon,1}\sigma_{\varepsilon,2}\rho_{\epsilon} & \sigma_{\varepsilon,2}^2
\end{bmatrix}.
$$
The standard geostatistical models often assume independent and identically distributed noise, meaning that $\rho_\varepsilon = 0$ in the matrix above. Including this extra correlation parameter to allow for correlation between measurements at the same location is, as we will see later, crucial for Argo data. 

We choose a bivariate SPDE-based formulation for the random field $\mathbf{u}(\mathbf{s})$, because of its flexibility and good computational properties, and give more details about it in the next section.

\subsection{Bivariate Gaussian Matérn fields}\label{sec:Multivariate_models}

The Matérn covariance family is a class of isotropic covariance functions commonly used in geostatistical applications due to its flexibility and ability to effectively capture real-world spatial correlations. The Matérn covariance function is defined as follows \parencite{Matern1960}:
\begin{equation}\label{eq:Matern_cov}
c\left(\mathbf{s}, \mathbf{t}\right) = \sigma^{2} M\left(\left\|\mathbf{s} - \mathbf{t}\right\| \,\big|\, \kappa, \nu\right) = \frac{\sigma^{2}}{2^{\nu - 1} \Gamma(\nu)} \left( \kappa \left\|\mathbf{s} - \mathbf{t}\right\| \right)^{\nu} K_{\nu} \left( \kappa \left\|\mathbf{s} - \mathbf{t}\right\| \right), \quad \mathbf{s}, \mathbf{t}\in \mathbb{R}^d,
\end{equation}
where \( \Gamma(\cdot) \) is the Gamma function, \( K_{\nu}(\cdot) \) is the modified Bessel function of the second kind, \( \nu > 0 \) controls the smoothness of the function, \( \kappa > 0 \) is a spatial scale or decay parameter that affects the correlation range, and \( \sigma^2 \) is the variance.

\textcite{Gneiting2010} expanded equation~\eqref{eq:Matern_cov} to a multivariate setting, introducing two models with valid multivariate covariance functions: a parsimonious model and a general bivariate model. The parsimonious model constrains the parameters \( \kappa_{ij} = \kappa \) and \( \nu_{ij} = (\nu_{ii} + \nu_{jj})/2 \) in the general cross-correlation function \( \rho_{ij} M\left(\left\|\mathbf{s} - \mathbf{t}\right\| \,\big|\, \kappa_{ij}, \nu_{ij}\right) \) to ensure valid theoretical properties.

\textcite{Hu2016} and \textcite{Bolin2020} instead proposed defining multivariate Gaussian fields through systems of SPDEs. The motivation for this is that a centered Gaussian  field with Matérn covariance function can be viewed as a solution to the SPDE 
\begin{equation}\label{eq:spde}
\left( \kappa^{2} - \Delta \right)^{\alpha/2} (\tau u) = \mathcal{W} \quad \text{on} \quad \mathbb{R}^{d},
\end{equation}
where \( \tau^2 = \Gamma(\nu)/(\sigma^2\Gamma(\alpha)(4\pi)^{d/2}\kappa^{2\nu})\), \( \Delta \) is the Laplace operator, \( \alpha = \nu + d/2 \), and \( \mathcal{W} \) is Gaussian white noise \parencite{Whittle1963}. 
\textcite{Bolin2020} introduced a bivariate Matérn-SPDE field \( \mathbf{u}(\mathbf{s}) = [ u_{1}(\mathbf{s}),\ u_{2}(\mathbf{s}) ]^\top \) as a solution to the system of SPDEs
\begin{equation}\label{eq:multiSPDE}
\mathbf{D}( \theta,\rho) \begin{bmatrix}
c_1 \left( \kappa_{1}^{2} - \Delta \right)^{\alpha_1/2} & 0 \\
0 & c_2 \left( \kappa_{2}^{2} - \Delta \right)^{\alpha_2/2}
\end{bmatrix} \begin{bmatrix}
u_{1} \\
u_{2}
\end{bmatrix} = \begin{bmatrix}
\mathcal{W}_1 \\
\mathcal{W}_2
\end{bmatrix}  \quad \text{on} \quad \mathbb{R}^{d},
\end{equation}
where \( c_{i} = \sqrt{ \sigma_{i}^{-2} (4 \pi)^{-d/2} \kappa_{i}^{-2 \nu_{i}} \Gamma\left( \nu_{i} \right) / \Gamma\left( \alpha_{i} \right) } \) for \( i = 1, 2 \),  \( \mathcal{W}_1 \) and \( \mathcal{W}_2 \) are independent Gaussian noises and 
\begin{equation}\label{eq:dependenceMatrix}
\mathbf{D}( \theta, \rho ) = \begin{bmatrix}
\cos( \theta ) + \rho \sin( \theta ) & -\sin( \theta ) \sqrt{ 1 + \rho^{2} } \\
\sin( \theta ) - \rho \cos( \theta ) & \cos( \theta ) \sqrt{ 1 + \rho^{2} }
\end{bmatrix},
\end{equation}
is a dependence matrix where \( \theta \in [0, 2\pi] \) and \( \rho \in \mathbb{R} \) controls the dependence between \( u_1(\mathbf{s}) \) and \( u_2(\mathbf{s}) \).
Given that driving noise satisfies the isometry
$E\,\!\bigl[\mathcal{W}_i(h)\,\mathcal{W}_i(g)\bigr] 
= \int_{\mathbb{R}^d} h(\mathbf{s})\,g(\mathbf{s})\,\mathrm{d}\mathbf{s},
\, i = 1,2$, it can be shown that the cross-covariance functions of $u_1$ and $u_2$ are
\begin{equation}\label{eq:covarMulti}
\operatorname{Cov}\left( u_{i}(\mathbf{s}),\ u_{j}(\mathbf{t}) \right) = 
\begin{cases}
   \sigma_i^2 M\left(\|\mathbf{s}-\mathbf{t}\|\mid\kappa_i,\nu_i\right)& i=j,\\
    \frac{\rho}{c_1\,c_2\,\sqrt{1+\rho^2}\,(2\pi)^d}\,
\mathcal{F}^{-1}(S)\left(\|\mathbf{s}-\mathbf{t}\|\right) & i\neq j,
\end{cases}
\end{equation}
where \( \mathcal{F}^{-1} \) is the inverse Fourier transform, and 
$S(\mathbf{k}) = (\kappa_1^2 + \|\mathbf{k}\|^2)^{-\alpha_1/2}(\kappa_2^2 + \|\mathbf{k}\|^2)^{-\alpha_2/2}$. Based on this expression, we note that $u_1$ and $u_2$ have marginally Matérn covariance functions, and if \( \kappa = \kappa_1 = \kappa_2 \), the model coincides with the parsimonious Matérn model by \textcite{Gneiting2010}.

It should be noted that \( \rho \) is not the standard Pearson correlation coefficient, which ranges between \(-1\) and \(1\); therefore, values greater than one are possible and \( \rho \) rather describes a more general dependence between fields. Henceforth, we refer to $\rho$ as the correlation parameter, whereas the designation Pearson correlation is reserved exclusively for the classical coefficient $\rho_{u_1,u_2}$. This coefficient is computed for the special case of $d=2$ and $\alpha=2$ in the following proposition, and the proof is available in the Appendix~\ref{sec:supp_theory}.

\begin{proposition}\label{prop:PearsonCorrelation}
For spatial dimension $d = 2$ and smoothness parameter $\alpha = 2$,  
the Pearson correlation coefficient between the fields $u_1(s)$ and $u_2(s)$ is
\begin{equation}\label{eq:PearsonCorrelation}
\rho_{u_1, u_2}(h)=
\begin{cases}
\displaystyle
\frac{2\rho\,\kappa_1\kappa_2}{\sqrt{1+\rho^2}}\,
\frac{\ln\!\bigl(\kappa_1/\kappa_2\bigr)}{\kappa_1^2-\kappa_2^2},
& \text{if } \kappa_1 \neq \kappa_2,\\[8pt]
\displaystyle
\frac{\rho}{\sqrt{1+\rho^2}},
& \text{if } \kappa_1 = \kappa_2.
\end{cases}
\end{equation}
\end{proposition}

Finally, note that the parameter $\theta$ in the dependence matrix is not identifiable for Gaussian models, as one obtains the same covariance functions for any value of $\theta$. Thus, for Gaussian models, one can simply set $\theta = 0$, whereas this parameter will have an effect for the non-Gaussian models introduced in the next subsection.

\subsection{Non-Gaussian Matérn-SPDE fields}\label{sec:BivariateModel}

Most applications of multivariate models have used multivariate Gaussian processes (GP) to model spatial data. Although they are generally flexible enough to fit well in most applications \parencite{Gelfand2016}, some additional flexibility is required to address non-Gaussian dependence or exponential tail behavior that often arises in real data. 

\textcite{Bolin2014} suggested extending univariate Matérn fields to a class of non-Gaussian Mat\'ern fields by replacing the Gaussian noise \( \mathcal{W} \) with non-Gaussian noise \( \dot{\mathcal{M}} \), specifically Laplace noise, in equation~\eqref{eq:spde}. \textcite{Wallin2015} considered geostatistical models based on these fields and also proposed using NIG noise as an alternative. The advantage of using NIG or generalized asymmetric Laplace (GAL) noise is that we can allow for asymmetry and heavier tails in marginal distributions. We focus on these two noises because they are the only subclasses of the generalized hyperbolic family that remain closed under convolution, a property required for SPDE models \parencite{Jonas2016}.

Based on this idea, \textcite{Bolin2020} suggested to introduce multivariate non-Gaussian models by replacing the Gaussian noise in \eqref{eq:multiSPDE} with independent NIG or GAL noise terms $\mathcal{\dot{M}}_1$ and $\mathcal{\dot{M}}_2$, and in this work we focus on NIG noise, for which the probability density function is always differentiable. 
In the non-Gaussian case, the field has the same cross-covariance functions as in the Gaussian case, and the parameter $\theta$ is then identifiable and determines the shape of the bivariate marginal distributions of $\mathbf{u}(s)$.

The NIG distribution can be represented as a normal variance–mean mixture. Specifically, let 
$z \sim N(0,1)$ and $v \sim \operatorname{IG}(\eta), \eta>0$,
where the mixing variable \(v\) is distributed according to an inverse Gaussian distribution with density
$$
\operatorname{IG}\left(x ; \eta\right)=\sqrt{\frac{ \eta}{2 \pi x^3}} \exp \left\{-\frac{\eta}{2} x-\frac{\eta}{2 x}+ \eta\right\}, \quad \quad x>0.
$$
Then, $E(v) = 1$ and a random variable $\gamma + v \,\mu + \sqrt{v}\, z$ with  $\gamma,\mu\in\mathbb{R}$, follows a NIG distribution with probability density function \parencite{BarndorffNielsen1997}
$$
\operatorname{NIG}\left(x ; \gamma, \mu, \eta\right)=\frac{\exp \left\{\eta+\mu(x-\gamma)\right\} \sqrt{ \eta \mu^2+\eta^2}}{\pi \sqrt{ \eta+(x-\gamma)^2}} K_1\left(\sqrt{\left\{\eta+(x-\gamma)^2\right\}\left(\mu^2+\eta\right)}\right),
$$
where \(K_1(\cdot)\) is the modified Bessel function of the second kind and $x>0$. For \( k = 1,2 \), we assume that \( \dot{\mathcal{M}}_k \) is NIG noise with parameters $\mu_k, \eta_k, \gamma_k$. We set \(\gamma_k = -\mu_k\) to ensure that the process \(u_k(s)\) has zero mean. In this formulation, the parameter $\mu_k$ serves as a shift parameter that quantifies the degree and direction of skewness in the mixing distribution, and the shape parameter $\eta_k$ controls its dispersion and tail behavior for each field $u_k$. As $\eta_k \rightarrow \infty$, $k = 1,2$, the NIG density converges to the Gaussian density. Thus, in this sense, the Gaussian model can be seen as a limiting case as $\eta_k \rightarrow \infty$, $k = 1,2$.

\subsection{Discretization}\label{sec:discretization}
%Using the connection between the SPDE above~\eqref{eq:spde} and Gaussian processes with Matérn covariance functions, \textcite{Lindgren2011} proposed computationally efficient Gaussian Markov random field approximations of \( u(\mathbf{s}) \) on a bounded domain \( \mathcal{D} \subsetneq \mathbb{R}^{d} \). The random field \( u(\mathbf{s}) \) is approximated as 
%\( u(\mathbf{s}) = \sum_{i=1}^{n} w_{i} \psi_i(\mathbf{s}) \)
%where \( \mathbf{w} \) is a vector of stochastic weights and \( \psi_{i}(\mathbf{s}) \) are piecewise linear and continuous basis functions obtained by a triangulation of the domain. See Appendix~\ref{sec:Appendix} for more detail of these representations. 

Throughout this work, we fix \( \alpha_1 = \alpha_2 = 2 \), which is the standard choice for the smoothness parameter for SPDE models. Although the multivariate field induced by \eqref{eq:multiSPDE} is defined on the entire \( \mathbb{R}^{d} \), to apply the model to real-world data, we restrict ourselves to a bounded domain \( \mathcal{D} \subset \mathbb{R}^{d} \) when implementing the fields numerically. To do that, we supplement the differential operators in \eqref{eq:multiSPDE} with Neumann boundary conditions and approximate solutions using finite element discretization. Specifically, both $u_1$ and $u_2$ are represented through weighted sums of basis functions, \( u_1(\mathbf{s}) = \sum_{i=1}^{n} w_{1,i} \psi_i(\mathbf{s}) \) and \( u_2(\mathbf{s}) = \sum_{i=1}^{n} w_{2,i} \psi_i(\mathbf{s}) \), where $\{\psi_i\}$ are piecewise linear and continuous basis functions obtained from a triangulation of the domain \( \mathcal{D} \) and $n$ is the dimension of the finite-element space. The distribution of the weights is then computed using a Galerkin finite element method 
as originally proposed by \textcite{Lindgren2011} for univariate SPDE models and for these bivariate SPDEs by \textcite{Bolin2020}. For details, we refer the reader to the Appendix~\ref{sec:supp_theory}.

When the driving noise is Gaussian white noise, the stochastic weights 
\( \mathbf{w} = (\mathbf{w}_1^\top,\,\mathbf{w}_2^\top)^\top \) with 
\(\mathbf{w}_k = (w_{k,1},\dots,w_{k,n})^\top\), \(k=1,2\)
are distributed as
\begin{equation}\label{eq:NormalWeights}
\mathbf{w} \sim N\left( \mathbf{0},\ \mathbf{K}^{-1} \operatorname{diag}\left(( \mathbf{h},\ \mathbf{h} )\right)\ \mathbf{K}^{- \top } \right).
\end{equation}
Here, \(\mathbf{h}\) is a vector defined as $\mathbf{h} = (h_1, \ldots, h_n)^\top,$
with each element \(h_i = |\mathcal{D}_i|\) is the area of the region $\mathcal{D}_i = \{ \mathbf{s}: \psi_i(\mathbf{s}) \geq \psi_j(\mathbf{s}) \,\forall \, i \neq j\},\,i=1,\ldots,n.$ Throughout the remainder of the paper, we write $\operatorname{diag}(\cdot)$ for the (block-)diagonal operator that places its vector or matrix arguments on the main diagonal and sets all off-diagonal entries to zero. The discretized operator matrix is $$
\mathbf{K}=\left(\mathbf{D} \otimes \mathbf{I}_n\right) \operatorname{diag}\left(\mathbf{L}_1\left(\sigma_1, \kappa_1\right), \mathbf{L}_2\left(\sigma_2, \kappa_2\right)\right),
$$
where $\mathbf{I}_n$ is the $n \times n$ identity matrix and $\mathbf{D}$ is defined in \eqref{eq:dependenceMatrix}.
For \(k = 1,2\), the discretized operator for the \(k\)th field is $\mathbf{L}_k\left(\sigma_k, \kappa_k\right)=c_k\left(\mathbf{G}+\kappa_k^2 \mathbf{C}\right)$, where $c_k$ is the normalizing constant defined in \eqref{eq:multiSPDE}. For the case $\alpha_k=2$, $d=2$ this reduces to $c_k=(2\sqrt{\pi}\,\sigma_k\kappa_k)^{-1}$.
The elements of the matrices \(\mathbf{C}\) and \(\mathbf{G}\) are defined by
\begin{align*}
 \quad \mathbf{C}_{ij} = \int_{\mathcal{D}} \psi_i( \mathbf{s} )\, \psi_j( \mathbf{s} )\, d\mathbf{s}, \quad \mathbf{G}_{ij} = \int_{\mathcal{D}} \nabla \psi_i( \mathbf{s} ) \cdot \nabla \psi_j( \mathbf{s} )\, d\mathbf{s}, \quad i,j,=1, \ldots, m,
\end{align*}
where \( \mathbf{C} \) and \( \mathbf{G} \) are commonly referred to as mass and stiffness matrices, respectively, in finite element method theory, and $\nabla$ denotes the gradient operator.

In the NIG case, we instead have:
\begin{align}
\mathbf{w} \mid \mathbf{v}_1,\ \mathbf{v}_2 &\sim N\left( \mathbf{K}^{-1} \begin{bmatrix}
\mu_1 (\mathbf{v}_1 - \mathbf{h}) \\
\mu_2 (\mathbf{v}_2 - \mathbf{h}) \\
\end{bmatrix},\ \mathbf{K}^{-1} \operatorname{diag}( (\mathbf{v}_1,\ \mathbf{v}_2 ))\ \mathbf{K}^{- \top } \right), \label{eq:NIGWeights}
\end{align}
where $\mathbf{v}_1 = (v_{1,1},\ldots, v_{1,n})^{\top}$ is a vector with independent variables $v_{1,i}\sim IG( \nu_1,\ \nu_1 h_i^2)$, where \( IG \) denotes the inverse Gaussian distribution. Similarly, $\mathbf{v}_2 = (v_{2,1},\ldots, v_{2,n})^{\top}$ is a vector with independent variables $v_{2,i}\sim IG( \nu_2,\ \nu_2 h_i^2)$ which are also independent of $\mathbf{v}_1$.  
%The parameters $\mu_1$ and $\mu_2$ respectively control the shift (skewness) of the inverse Gaussian mixing distributions for each field. 
% \textcolor{blue}{
% $\mathbf{Q}_x = \mathbf{K}^\top \mathbf{C}^{-1} \mathbf{K}$
% By using the Galerkin finite element method, the density of the weights is given by
% \begin{align*}
%     &\quad \mathbf{w} \sim N\left( \mathbf{0},\ \mathbf{K}^{-1} \mathbf{C} \mathbf{K}^{-1} \right),\quad \text{where} \\
% \end{align*}
% }

\section{Geostatistical estimation and prediction}\label{sec:models}

In this section, we introduce the methods used for likelihood-based inference and prediction.

\subsection{Model description}
To use the models from the previous section for the Argo data,% we let $\mathbf{X}(\mathbf{s})  = (X_1(\mathbf{s}), X_2(\mathbf{s}))^{\top}$, where $X_1$ denote temperature and $X_2$ salinity. We model $\mathbf{X}(\mathbf{s})$ as 
suppose now that we have $N_1$ observations $\mathbf{Y}_1 = (Y_{1,1}\ldots, Y_{1,N_1})$ of $X_1$ at locations $\mathbf{s}_{1,1},\ldots, \mathbf{s}_{1,N_1}$ and $N_2$ observations $\mathbf{Y}_2 = (Y_{2,1}\ldots, Y_{2,N_2})
$ of $X_2$ at locations $\mathbf{s}_{2,1},\ldots, \mathbf{s}_{2,N_2}$, where some observation locations may be shared for the two fields. 
Collecting all $N = N_1 + N_2$ observations in a vector $\mathbf{Y}_N = (\mathbf{Y}_1, \mathbf{Y}_2)^{\top}$,  letting 
$$
\mathbf{X}_N = (X_1(\mathbf{s}_{1,1}), \ldots, X_1(\mathbf{s}_{1,N_1}), X_2(\mathbf{s}_{2,1}), \ldots, X_2(\mathbf{s}_{2,N_2}))^{\top}, 
$$
and  discretizing the SPDE model using the finite element approach, we can write the complete model from Section~\ref{sec:General_model} in vector form as 
\begin{equation}\label{eq:field_def}
\begin{split}
\mathbf{X}_N &= \boldsymbol{m}_N + \mathbf{A} \mathbf{w},  \\
\mathbf{Y}_N \mid \mathbf{X}_N &\sim N\big( \mathbf{X}_N,\ \boldsymbol{\Sigma}_{\varepsilon} \big), 
\end{split}
\end{equation}
where the stochastic weights \( \mathbf{w} \) follow equation~\eqref{eq:NormalWeights} for Gaussian noise or equation~\eqref{eq:NIGWeights} for NIG noise and  
\( \boldsymbol{m}_N \) is the mean value evaluated at the measurement locations. In addition, \( \mathbf{A} = \operatorname{diag}( \mathbf{A}_1,\ \mathbf{A}_2 ) \) is a block-diagonal projector matrix with elements \( (\mathbf{A}_1)_{ij} = \psi_{i}(s_{1,j}) \) and \( (\mathbf{A}_2)_{ij} = \psi_{i}(s_{2,j}) \) respectively. Further, $\mathbf{\Sigma}_{\varepsilon}$ is the sparse covariance matrix for the measurement noise, which also notably has a sparse inverse  $\mathbf{Q}_{\varepsilon} = \mathbf{\Sigma}_{\varepsilon}^{-1}$.
Also, if we would have $N_1 = N_2$ observations of both fields at the same locations, 
$$\mathbf{\Sigma}_{\varepsilon} = \mathbf{\Sigma}\otimes \mathbf{I}_{N_1}
$$
where $\mathbf{I}_{N_1}$ is an $N_1 \times N_1$ identity matrix, and $\mathbf{Q}_{\varepsilon} = \mathbf{\Sigma}^{-1} \otimes \mathbf{I}_{N_1}$.

\subsection{Parameter estimation}\label{sec:estimation} In the case of a latent Gaussian model, \eqref{eq:field_def} has parameters $\kappa_1,\kappa_2,\sigma_1,\sigma_2,\rho$ for the latent field, regression parameters $\beta_{1,1},\ldots, \beta_{1,K}$ and $\beta_{2,1},\ldots, \beta_{2,K}$ for the mean, and parameters $\sigma_{\varepsilon,1},\sigma_{\varepsilon,2},\rho_{\varepsilon}$ for the measurement noise. In the non-Gaussian case, the model additionally has the parameters $\eta_1,\eta_2,\mu_1,\mu_2$ for the latent field. 
In both cases, let $\mathbf{\Theta}$ denote the vector of all parameters that have to be estimated, and let $\mathbf{Y}$ denote all the observations.

In the Gaussian case, we estimate the parameter through numerical optimization of the log-likelihood of the data,
\begin{equation}\label{eq:likelihood}
\begin{aligned}
\log \pi( \mathbf{Y} \mid \boldsymbol{\Theta} ) =\ & - N_1 \log (2\pi) + \tfrac{1}{2} \log \left| \mathbf{Q}_x \right| + \tfrac{1}{2} \log \left| \mathbf{Q}_{\varepsilon} \right| - \tfrac{1}{2} \log \left| \mathbf{Q}_{x \mid y} \right| \\
& - \tfrac{1}{2} \boldsymbol{\mu}_{x \mid y}^\top \mathbf{Q}_x \boldsymbol{\mu}_{x \mid y} - \tfrac{1}{2} \left( \mathbf{Y} - \mathbf{A} \boldsymbol{\mu}_{x \mid y} \right)^\top \mathbf{Q}_{\varepsilon} \left( \mathbf{Y} - \mathbf{A} \boldsymbol{\mu}_{x \mid y} \right),
\end{aligned}
\end{equation}
where $\boldsymbol{\mu}_{x \mid y} = \mathbf{Q}_{x \mid y}^{-1} \mathbf{A}^\top \mathbf{Q}_{\varepsilon} \mathbf{Y}$, $\mathbf{Q}_{x \mid y} = \mathbf{Q}_x + \mathbf{A}^\top \mathbf{Q}_{\varepsilon} \mathbf{A}$ and $\mathbf{Q}_x =\mathbf{K}^\top\mathbf{C}^{-1}\mathbf{K}$.

In the non-Gaussian case, an explicit likelihood is not available. However, by applying Fisher's identity and the Rao--Blackwellization procedure (see Appendix~\ref{sec:like_NIG} for details),  we can express the gradient of the log-likelihood as
% &= E_{ \mathbf{v},\, \mathbf{w} } \left[ \nabla_{\boldsymbol{\Theta}} \log \pi( \mathbf{v}, \mathbf{w} \mid \mathbf{Y}, \boldsymbol{\Theta} ) \mid \mathbf{Y}, \boldsymbol{\Theta} \right] \\
\begin{equation}
\begin{aligned}
\nabla_{\boldsymbol{\Theta}} \log \pi(\mathbf{Y}\mid\boldsymbol{\Theta})  
&= E_{\mathbf{v}}\!\Bigl[E_{\mathbf{w}}\!\bigl[\nabla_{\boldsymbol{\Theta}}\log\pi(\mathbf{v},\mathbf{w};\mathbf{Y},\boldsymbol{\Theta})\mid\mathbf{v},\mathbf{Y},\boldsymbol{\Theta}\bigr]\mid\mathbf{Y},\boldsymbol{\Theta}\Bigr] \\[2pt]
&= E_{\mathbf{v}}\!\bigl[\nabla_{\boldsymbol{\Theta}}\log\pi(\mathbf{v}\mid\mathbf{Y},\boldsymbol{\Theta})\mid\mathbf{Y},\boldsymbol{\Theta}\bigr].
\end{aligned}
\end{equation}
Let $\mathbf{v} =(\mathbf{v}_1,\mathbf{v}_2)^{\!\top}$. Although \( \nabla_{\boldsymbol{\Theta}} \log \pi( \mathbf{v} \mid \mathbf{Y}, \boldsymbol{\Theta} ) \) is available in closed form, its expected value is not.  We therefore draw Gibbs samples of \( \mathbf{v} \) and approximate this expectation by
\begin{equation}\label{eq:likgrad}
G(\boldsymbol{\Theta})
  = \frac{1}{k}\sum_{j=1}^{k}
    \nabla_{\boldsymbol{\Theta}}
    \log \pi\!\bigl(\mathbf{v}^{(j)}\mid\mathbf{Y},\boldsymbol{\Theta}\bigr),
\end{equation}
where $\mathbf{v}^{(j)}$ are samples from distribution $\pi\!\bigl(\mathbf{v}\mid\mathbf{Y},\boldsymbol{\Theta}\bigr)$.
The samples are obtained using a Gibbs sampler that samples \( \pi( \mathbf{w} \mid \mathbf{Y}, \mathbf{v}, \boldsymbol{\Theta} ) \) and \( \pi( \mathbf{v} \mid \mathbf{Y}, \mathbf{w}, \boldsymbol{\Theta} ) \) (see Algorithm 1 in the Appendix). 
This step is computationally feasible and efficient because
$\mathbf{w}\mid\mathbf{Y},\mathbf{v}^{(j)},\boldsymbol{\Theta}
\sim
N\!\bigl(\hat{\boldsymbol{\xi}}^{(j)},(\hat{\mathbf{Q}}^{(j)})^{-1}\bigr)$
is a Gaussian Markov random field with
\begin{equation}\label{eq:gibbs_w}
\begin{aligned}
\hat{\mathbf{Q}}^{(j)} &=
\mathbf{K}^{\top}\operatorname{diag}\!\bigl(\mathbf{v}^{(j)}\bigr)^{-1}\mathbf{K}
+ \mathbf{A}^{\top}\mathbf{Q}_{\varepsilon}\mathbf{A},\\
\hat{\boldsymbol{\xi}}^{(j)} &=
\bigl(\hat{\mathbf{Q}}^{(j)}\bigr)^{-1}\!
\left(
\mathbf{A}^{\top}\mathbf{Q}_{\varepsilon}\bigl(\mathbf{y}-\mathbf{B}\boldsymbol{\beta}\bigr)
+ \mathbf{K}^{\top}\operatorname{diag}\!\bigl(\mathbf{v}^{(j)}\bigr)^{-1}
\bigl(\boldsymbol{\mu}\otimes\mathbf{I}_{N_1}\bigr)
\bigl(\mathbf{v}^{(j)}-\mathbf{h}\bigr)
\right),
\end{aligned}
\end{equation}
where $\boldsymbol{\beta}=(\beta_{1,1},\ldots,\beta_{1,K},\beta_{2,1},\ldots,\beta_{2,K})^{\top}$ are the regression parameters, and the matrix
$\mathbf{B}=[\boldsymbol{1}_{N}\;\; \mathbf{z}_1(\mathbf{s}) \ldots \mathbf{z}_K(\mathbf{s})]$ collects the covariates evaluated at the measurement locations.
Thus, this distribution can be sampled via sparse Cholesky factorization \parencite{Rue2000}. Further, \( \pi( \mathbf{v} \mid \mathbf{Y}, \mathbf{w}, \boldsymbol{\Theta} ) \) consists of vector-independent variables that can be sampled in parallel.

Based on this procedure, \eqref{eq:likgrad} provides a stochastic and unbiased estimate of the gradient of the likelihood. Therefore, we can use a stochastic gradient descent method to find maximum likelihood estimates of the parameters as proposed in \textcite{Asar2020} for non-Gaussian longitudinal models. Specifically, the iterative update is given by \( \boldsymbol{\Theta}^{(i)} = \lambda_i \cdot G\left( \boldsymbol{\Theta}^{(i-1)} \right) + \boldsymbol{\Theta}^{(i-1)} \), where \( \{ \lambda_i \} \) is a sequence of weights chosen to satisfy \( \sum \lambda_i \rightarrow \infty \) and \( \sum \lambda_i^2 < \infty \) in order to guarantee convergence \parencite{Andrieu2005}.

This estimation procedure is implemented in the \textit{ngme2} \textsf{R} package \parencite{ngme2}, which facilitates rapid and efficient computations for more general non-Gaussian latent models such as the autoregressive process of order one and the separable space-time model with NIG driving noise. The \textit{ngme2} package not only facilitates the estimation of our bivariate models, but also extends its utility to cross-validation and predictive performance evaluation which we briefly explain below. Further tutorials on the package and more technical details can be found at the package homepage \url{https://github.com/davidbolin/ngme2}.

\subsection{Spatial prediction and predictive performance}\label{sec:pred_eval}

A major interest in geostatistical applications lies in predicting latent fields at unobserved locations using irregularly sampled data. One way to quantify the predictive performance of a model is by evaluating the accuracy of point predictions by calculating, for example, the root mean squared error (RMSE) between the model's prediction and actual observation. However, for probabilistic models, one is often interested not only in pointwise accuracy but also in predictive uncertainty, which requires consideration of the whole predictive distribution. Formally, this can be written as \( \pi( X_k( \boldsymbol{s}_0 ) \mid \mathbf{Y}, \boldsymbol{\Theta} ) \), for the \( k \)th variable of the latent field at a given location \( \boldsymbol{s}_0 \). 

As for the estimation procedure, given the absence of a closed-form expression for the distribution of the non-Gaussian model, we resort to sampling methods to estimate these metrics. Let \( \mathbf{A} = [ \psi_1( \boldsymbol{s}_0 ),\ \ldots,\ \psi_n( \boldsymbol{s}_0 ) ] \) represent the basis functions from space discretization at \( \boldsymbol{s}_0 \). We then use samples \( \mathbf{v}^{(i)} \) from \( \pi( \mathbf{v} \mid \mathbf{Y}, \boldsymbol{\Theta} ) \), obtained via the Gibbs sampler, to approximate the expected value and variance of \( X_k \) at \( \boldsymbol{s}_0 \) as:
\[
E\left( X_k( \boldsymbol{s}_0 ) \mid \mathbf{Y}, \boldsymbol{\Theta} \right) \approx \frac{1}{M} \sum_{i=1}^{M} \mathbf{A} \hat{\boldsymbol{\xi}}^{(i)}, 
\quad 
\operatorname{Var}\left( X_k( \boldsymbol{s}_0 ) \mid \mathbf{Y}, \boldsymbol{\Theta} \right) \approx \frac{1}{M} \sum_{i=1}^{M} \mathbf{A}^\top \left( \hat{\mathbf{Q}}^{(i)} \right)^{-1} \mathbf{A},
\]
where $\hat{\boldsymbol{\xi}}^{(i)}$ and $\hat{\mathbf{Q}}^{(i)}$  are given in \eqref{eq:gibbs_w} and $M$ is a pre-specified number of samples.
To assess and compare the fit of our proposed models, we employ several metrics, including the continuous ranked probability score (CRPS) \parencite{Gneiting2007}, its scaled version (SCRPS) \parencite{Bolin2019}, root mean squared error (RMSE), and mean absolute error (MAE). The CRPS, a negatively oriented score, reflects the difference between the predicted and observed values, adjusted for the variability within the predictions themselves. CRPS is formally defined as:
\[
\operatorname{CRPS}( \mathbb{F},\ y ) = \mathrm{E}_{ \mathbb{F} } | X - y | - \tfrac{1}{2} \mathrm{E}_{ \mathbb{F} } \mathrm{E}_{ \mathbb{F} } \left| X - X' \right|,
\]
where \( X \) and \( X' \) are independent instances of a random variable with the cumulative distribution function \( \mathbb{F} \). For the Gaussian distribution, we can derive an analytical form. However, for the non-Gaussian fields approximate the expected values using the Gibbs samples. In particular, we use Proposition 5 in \textcite{Bolin2020}, which gives a Rao--Blackwellized version of the CRPS estimator.
We also use the scaled CRPS (SCRPS) introduced in \textcite{Bolin2019}, which is another proper scoring rule suitable for datasets with significant spatial and temporal variability in terms of predictability. The (negatively oriented) SCRPS  is
\begin{equation}\label{eq:SCRPS}
\operatorname{SCRPS}( \mathbb{F},\ y ) = \frac{ \mathrm{E}_{ \mathbb{F} } | X - y | }{ \mathrm{E}_{ \mathbb{F} } \mathrm{E}_{ \mathbb{F} } \left| X - X' \right| } + \tfrac{1}{2} \log\left( \mathrm{E}_{ \mathbb{F} } \mathrm{E}_{ \mathbb{F} } \left| X - X' \right| \right).
\end{equation}
Also for this, we use a Rao--Blackwellized estimator based on Gibbs sampling, as shown in the Appendix~\ref{sec:supp_theory}.

\section{Simulation Study}\label{sec:simulation}

In this section, a simulation study is conducted to investigate the impact of measurement noise correlation on model parameter estimates. The bivariate SPDE model in equation~\eqref{eq:multiSPDE} with Gaussian noise is tested for seven different configurations of the dependence parameter, \( \rho = \{ -0.7,\ -0.2,\ -0.05,\ 0,\ 0.05,\ 0.2,\ 0.7 \} \), representing weak, medium, and strong correlation values. For all models, we compare the model fit for cases with the general dependence and the independent measurement noise. To better understand how noise correlation affects parameter estimation, we test each model under six different conditions of noise correlation, \( \rho_\varepsilon = \{ -0.8,\ -0.4,\ -0.1,\ 0.1,\ 0.4,\ 0.8 \} \).
This resulted in a total of 42 different configurations for testing the models with Gaussian noise. We simulated a set of \( N = 1000 \) points for each configuration for each field, with ten independent replicates drawn from the bivariate SPDE model.

We used \textit{fmesher} R package \parencite{fmesher} to conduct spatial discretization of the domain, which facilitated the creation of mesh and projector matrices. Following this, we generated datasets using the constructed mesh, assuming a general noise structure with specified correlation parameters. These datasets were then analyzed using two different model structures. The first model was built on the assumption of i.i.d.\ measurement noise and the second model incorporated a general dependence structure, as proposed in our study.
\begin{figure}[htbp]
    \centering
    \begin{subfigure}[b]{0.48\linewidth}
        \centering
        \includegraphics[width=\linewidth]{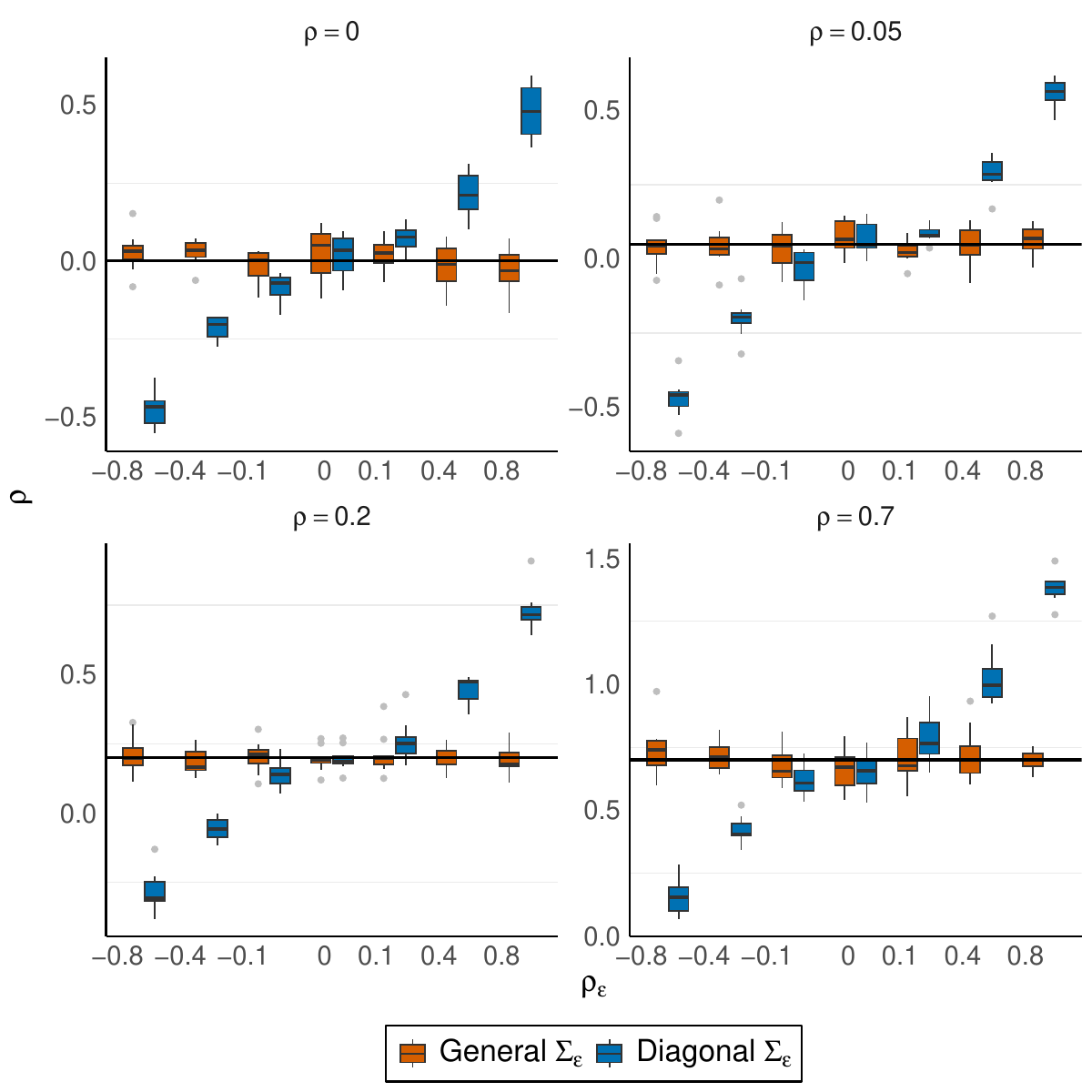}
        \caption{Estimates of \( \rho \).}
        \label{fig:boxplot_simulation_results_gauss}
    \end{subfigure}
    \hfill
    \begin{subfigure}[b]{0.48\linewidth}
        \centering
        \includegraphics[width=\linewidth]{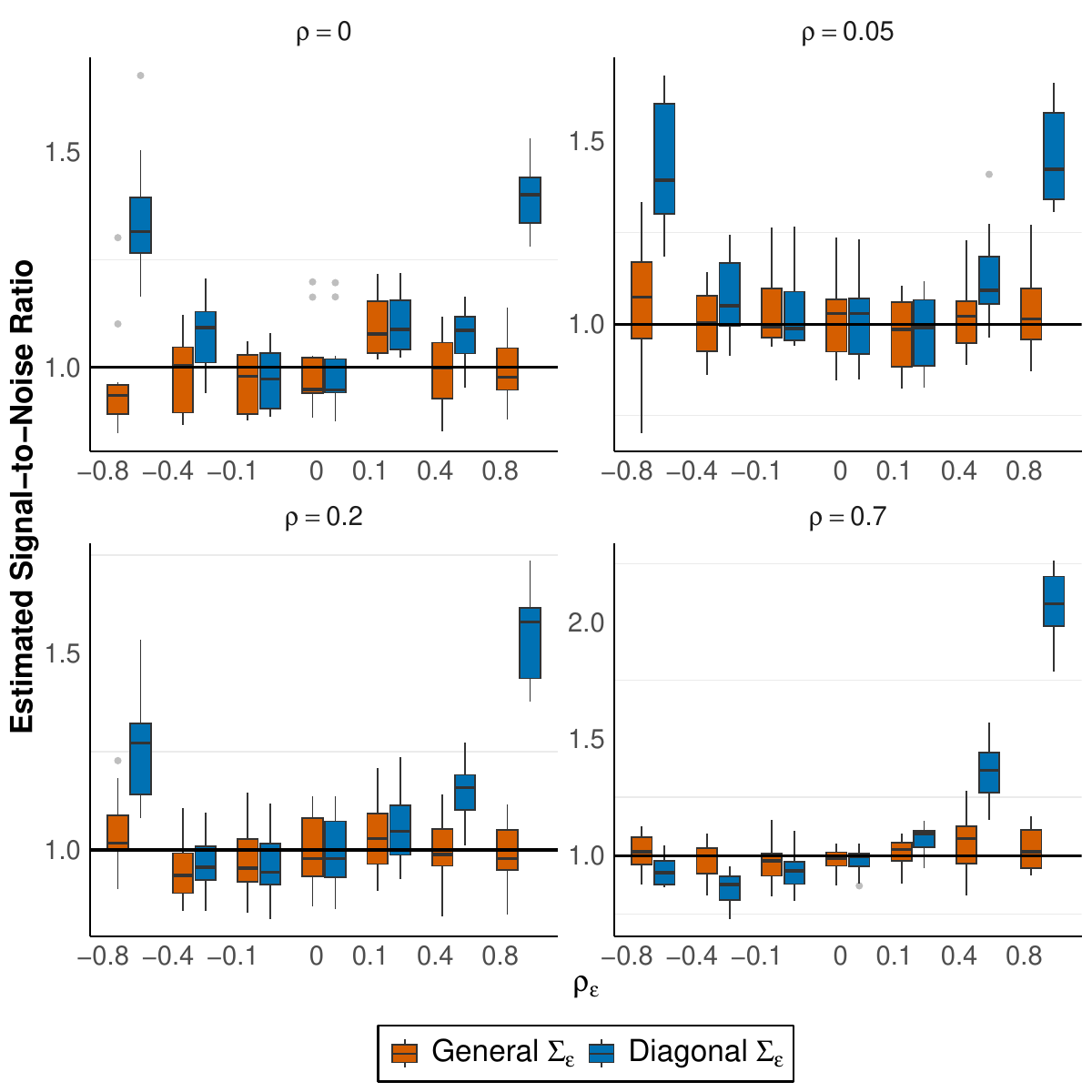}
        \caption{Estimates of signal-to-noise ratio.}
        \label{fig:boxplot_simulation_results_gauss_snr}
    \end{subfigure}
    \caption{Simulation results. (a) We compare the two Gaussian bivariate SPDE models based on the covariance matrix of measurement noise, \( \boldsymbol{\Sigma}_\varepsilon \). (b) Estimated signal-to-noise ratio, \( \sigma_1^2 / \sigma_{\varepsilon,1}^2 \), from the simulation results. The horizontal line shows the true value.}
    \label{fig:side_by_side_boxplots}
\end{figure}
Figure~\ref{fig:boxplot_simulation_results_gauss} presents a boxplot comparison of our extended model, which incorporates an additional correlation parameter (orange), and a traditional model that neglects correlation in the nugget effect (blue). The extended model displays narrower boxes, indicating reduced estimate variability and bias, and thereby more accurately reflects the true underlying field correlations. One can note that when the fields are uncorrelated (\( \rho= 0 \)) and a non-zero measurement noise correlation (\( \rho_\varepsilon \)) is present, ignoring this measurement noise correlation can cause substantial bias of the estimate of the true dependency structure. For instance, with \( \rho_\varepsilon = -0.8 \), the estimated correlation increases to \( \rho = -0.5 \), in contrast to the true value of \( 0 \). Thus, standard multivariate models can lead to under- or overestimation of dependence and correlation parameters, influenced by the sign of \( \rho_\varepsilon \).

The corresponding signal-to-noise ratio, \( \sigma_1^2 / \sigma_{\varepsilon,1}^2 \), was computed for each set of simulation results and is shown in Figure~\ref{fig:boxplot_simulation_results_gauss_snr}. While the ratio is generally well captured, deviations occur for extreme values of \( \rho_\varepsilon \) (e.g., \( | \rho_\varepsilon | = 0.8 \)), where disturbances from measurement noise are most pronounced. This is a critical observation, as discussed further in Section~\ref{sec:results}, because the application data exhibit high \( \rho_\varepsilon \) values, often exceeding \( 0.7 \) and reaching as high as \( 0.9 \). These findings highlight the importance of accurately accounting for noise correlations in scenarios with strong dependency structures.

%When \( \rho_\varepsilon \) is non-zero, it becomes necessary to extend the traditional models to accurately capture dependencies. The accuracy of estimating correlation \( \rho \) is severely compromised, especially when its true value is low, and further influenced by the sign of \( \rho_\varepsilon \). Incorporating i.i.d.\ measurement noise into hierarchical multivariate models does not fully address this issue, as the noise variance estimation remains unaffected by the different values of \( \rho_\varepsilon \). This inability can mask potential hidden correlations. Our proposed model mitigates this problem by introducing an additional correlation parameter in the noise, which facilitates the identification of the source of dependence, and thus leads to more accurate estimates of \( \rho \), \( \rho_\varepsilon \), \( \sigma_1 \), and \( \sigma_2 \). See Figure~\ref{fig:results_300_combined}.

%The additional parameter does not affect the estimation of the spatial scale parameter \( \kappa \) and the covariates \( \beta \). 
Similar boxplots for the estimates of the different parameters in the model are presented in the Appendix. There it can be noted that the proposed model and the traditional model without measurement noise correlation yield comparable results for the spatial scale parameters \( \kappa \). Thus, the main effect of not modeling the measurement noise correlation is a bias in the estimate of the correlation structure in the latent field. 
%which confirms that our model retains the strengths of conventional models while also providing a more comprehensive understanding of dependencies within multivariate spatial data.

%Our simulation study provides vital insights into the estimation of dependencies in multivariate models, particularly under varying conditions of measurement noise correlation. We demonstrate that neglecting the correlation in measurement noise can lead to significant biases when interpreting the true dependencies between fields. This is crucial in environmental applications, where understanding the relationships between ecological variables is key. Our findings reveal that correlated measurement errors can cause an overestimation of field dependencies and potentially lead to misinterpretations of natural phenomena. The proposed model, with its additional correlation parameter, effectively addresses these issues by accurately estimating dependencies and better representing the underlying processes. Overall, this work highlights the importance of considering measurement noise correlation in multivariate environmental models to ensure accurate interpretation and analysis.

\section{Application to Argo Profiling Float Data}\label{sec:application}

In this section, our proposed model is applied to real-world data from the Argo project. %and the statistical model detailed in Section~\ref{sec:BivariateModel} is adapted to fit the global ocean data. 
%This international project has been collecting seawater temperature and salinity data in the upper 2000\,m of the global ocean since 1999 at fine spatial and temporal resolutions. The in-situ observations are collected by fleets of sensors drifting in the ocean and collecting data at various depths. 
The inherent statistical challenge involves predicting or interpolating the field at unobserved spatial locations, using sparsely and irregularly sampled data, as demonstrated in Figure~\ref{fig:map_data_example}. 
%By doing so, we aim to enhance our understanding of oceanic conditions and trends, which may help to inform climate studies, marine biology research, and other scientific endeavors. 
%For more information about the floats and data collection process, please refer to \textcite{Argo} and the recent 20-year anniversary paper by \textcite{Wong2020}.

\subsection{Data}\label{sec:DataDescription}
\begin{figure}[b]
    \centering
    \includegraphics[width=0.45\linewidth]{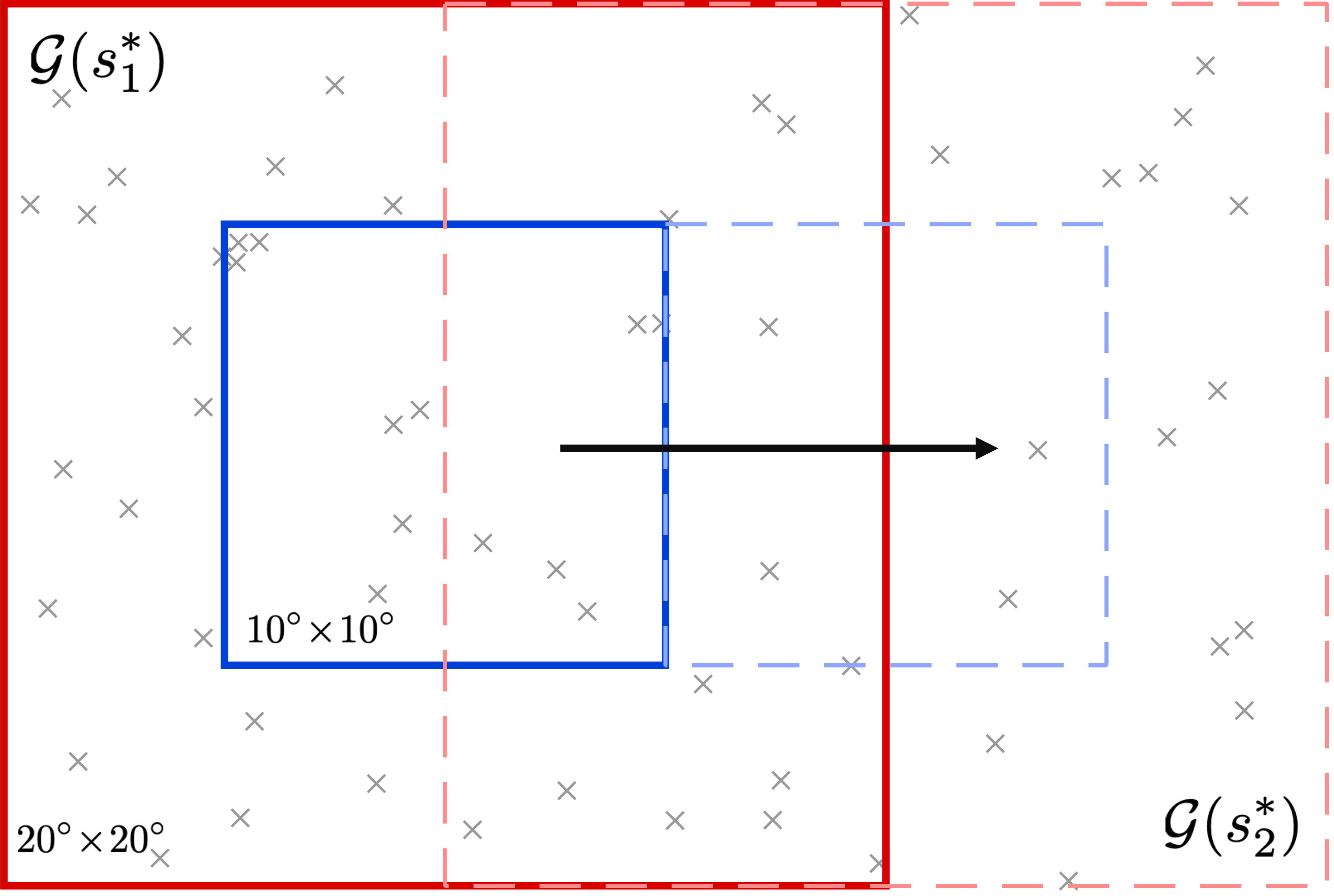}
    \caption{Schematic representation of the moving-window technique applied in bivariate SPDE model fitting. Central to the illustration is a \( 10^\circ \) by \( 10^\circ \) reference grid box, positioned at the equator, which is extended by an additional \( 5^\circ \) margin to construct a \( 20^\circ \) by \( 20^\circ \) window that includes the data used for parameter estimation.}
    \label{fig:moving_window}
\end{figure}
Our analysis, influenced by the approach of \textcite{Kuusela2018}, examines the Argo float data from January 2007 to 2020 obtained from an August 2023 snapshot of the Argo Global Data Assembly Center (GDAC, \textcite{ArgoData}). We limited our study to January to reduce seasonal fluctuations, as our model assumes temporal stationarity. However, to compute robust residuals for January,  the mean field is estimated by fitting a local polynomial regression using the complete dataset (January to December) for each year. 
Following Kuusela's preprocessing steps, we used expertly vetted delayed-mode data by applying rigorous quality control criteria: accepting only profiles with `good' or `probably good' quality flags, rejecting profiles with significant gaps or unrealistic values, and ensuring data consistency for temperature and salinity. Detailed quality control criteria are described in the Appendix~\ref{sec:overview}.
Using MATLAB R2022a \parencite{MATLAB}, we pre-processed the data, focusing on the adjusted variables of pressure, temperature, and practical salinity. After preprocessing, a total of 1,349,863 profiles spanning the entire annual cycle over 14 years were analyzed to interpolate specific pressure levels. The mean field was calculated using polynomial regression in equation~\eqref{eq:meanRG}, as established by \textcite{Roemmich2009}, and we subtracted this mean from the raw data to obtain the January residuals as explained below. These residuals, which represent deviations from the expected mean, were then used as input for our bivariate model fits.

\subsection{Moving-Window modeling}
\textcite{Kuusela2018} highlighted the importance of considering subtle variability in salinity at greater depths and suggested incorporating correlations between temperature and salinity to enhance model accuracy. Inspired by this methodology, we adopted the Matérn-SPDE fields in a moving-window approach to ensure smooth transitions in parameter estimation. This method utilizes data points near the desired prediction area, which are more informative for covariance function estimates \parencite{Haas1990}.

\begin{figure}[t]
    \centering
    \includegraphics[width=0.85\linewidth,trim=3cm 2cm 2cm 2cm,clip]{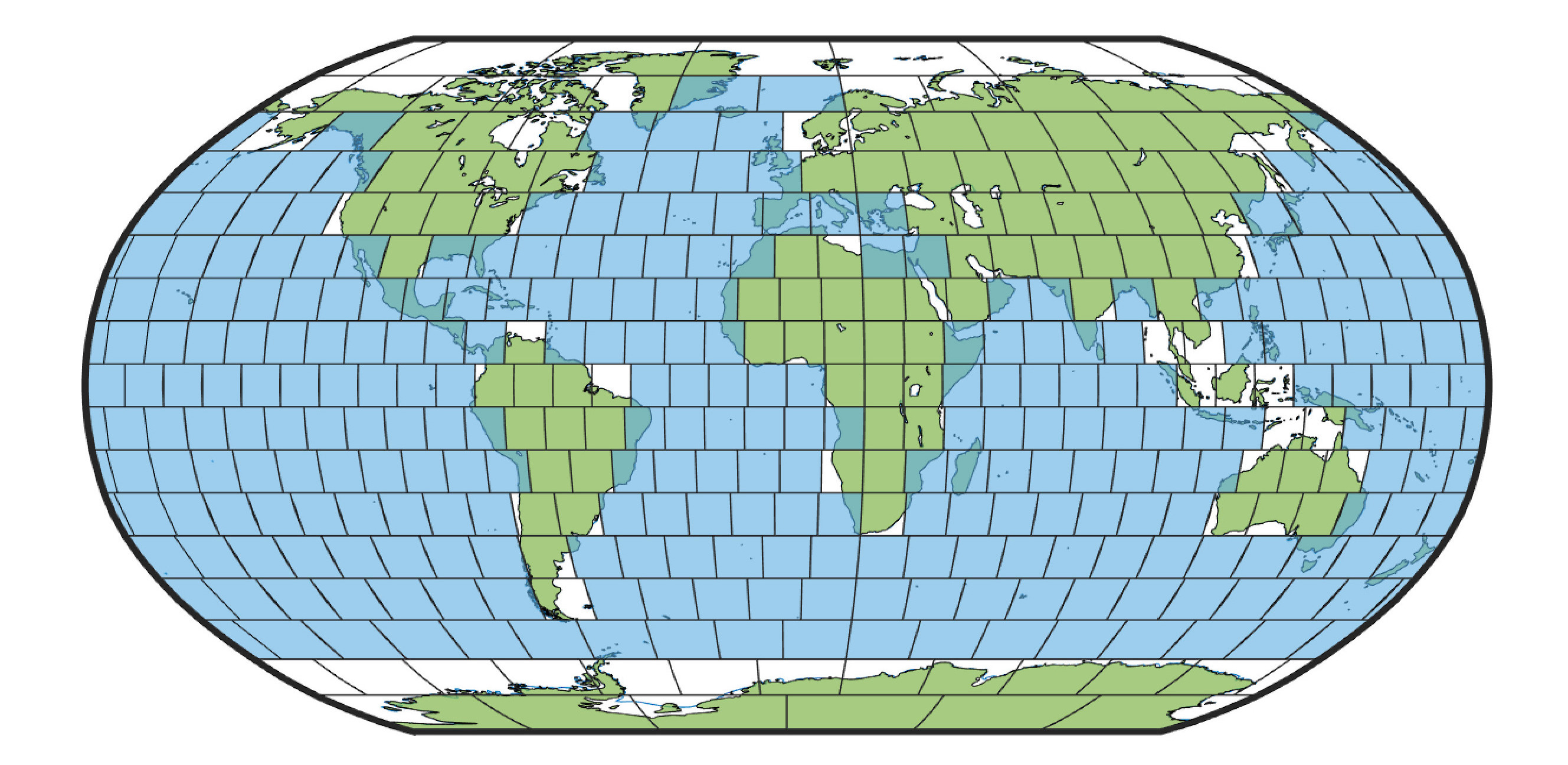}
    \caption{Grid used for model fitting for 300 dbar. All grid boxes have approximately equal surface areas.}
    \label{fig:grid}
\end{figure}

Our approach involves defining a prediction grid, \( \mathcal{G}( s^* ) \), determined by the minimum and maximum coordinates in latitude and longitude: \( s^*_{\text{min}} = [ s^*_{\text{lat}_{\text{min}}},\ s^*_{\text{lon}_{\text{min}}} ]^\top \) and \( s^*_{\text{max}} = [ s^*_{\text{lat}_{\text{max}}},\ s^*_{\text{lon}_{\text{max}}} ]^\top \), which is then expanded by a margin, \( s_{\text{win}} \), to encompass a larger area for better parameter estimation.
Specifically, we divided the world into 404 grid boxes of approximately equal surface area, anchored by a reference area at the equatorial region with coordinates \( [0^\circ\text{E},\ 5^\circ\text{S}] \times [10^\circ\text{E},\ 5^\circ\text{N}] \). Each grid box maintains a consistent height of \( 10^\circ \), but the width is adjusted at each latitude to ensure approximately equal surface area. This gridding strategy facilitates the use of a moving-window approach, applied from \( 0^\circ \) to \( 360^\circ \) east longitude, and includes overlaps between windows for consistent variation in model parameter estimates.
Our predictive model targets a reference \( 10^\circ \times 10^\circ \) box, subsequently enlarged by \( 5^\circ \) in all directions to form a \( 20^\circ \times 20^\circ \) window, as illustrated in Figure~\ref{fig:moving_window}. The actual width in degrees of each grid box increases closer to the poles, ensuring that each grid covers an approximately equal area. 
To ensure accurate parameter estimates, we require a minimum of 100 data points for each of the two fields within a grid box. Should this requirement remain unfulfilled or the box is empty (typically over areas of land), we omit the grid due to the excessive uncertainty of noise parameters. For the pressure level 300 dbar, there are 404 grid boxes; 128 grid boxes were omitted due to insufficient data. The remaining boxes are shown in blue in Figure~\ref{fig:grid}.

To obtain the residuals to model spatially, the mean value function $\mathbf{m}(\mathbf{s})$ in \eqref{eq:Xmodel} is specified as 
\begin{equation}\label{eq:meanRG}
\begin{aligned}
    m_i(\mathbf{s}) &= \beta_{i,0} + \beta_{i,x} x_c + \beta_{i,y} y_c + \beta_{i,xy} x_c y_c + \beta_{i,x^2} x_c^2 + \beta_{i,y^2} y_c^2 \\
    &\quad + \sum_{k=1}^K \left[ \beta_{i,c_k} \cos\left( \frac{2\pi k t}{365} \right) + \beta_{i,s_k} \sin\left( \frac{2\pi k t}{365} \right) \right],
\end{aligned}
\end{equation}
\(\mathbf{s} = (x_i, y_i)\) (with \(x\) and \(y\) corresponding to longitude and latitude, respectively), 
 \( x_c := x - x^* \) and \( y_c := y - y^* \) are spatial coordinates centered around \( x^* \) and \( y^* \), and \( K \) is a predefined maximum number of harmonics. The first line in \eqref{eq:meanRG} captures the local spatial structure of the mean field, and the second line models the seasonal cycle within the window. This regression model with \( K = 6 \) has been successfully adopted in the oceanographic literature to model the mean field of Argo observations \parencite{Roemmich2009}.

We fit the mean value for each grid cell, pressure level, and variable using local least-squares regression. For this, we have used all the data for 14 years from January to December so we can remove seasonality from the data. We then subtract the estimated mean from the observations and model the residuals through \eqref{eq:field_def}  with $\mathbf{m}_N = 0$. The final residual dataset used in the modeling contains 109,186 data points at 10 dbar, 113,996 at 300 dbar, and 99,999 at 1000 dbar for each field.

Treating data from different years as independent replicates, we denote \( \mathbf{Y}_1 \) and \( \mathbf{Y}_2 \) as temperature and salinity data, respectively. For each window, we apply four distinct models: bivariate Matérn-SPDE with Gaussian noise and either independent or general measurement noise, with NIG driving noise and either independent or general measurement noise. 
Model fitting was conducted using the the \textit{ngme2} \textsf{R} package. To analyze each grid, we employed 10,000 iterations of the gradient descent method; the Gaussian models were run with four parallel chains, while the NIG (non-Gaussian) models were run with two parallel chains. Numerical gradient methods and Rao--Blackwellization were utilized to enhance convergence speed and reduce estimate variance. Convergence control was achieved by monitoring the standard deviation across the four chains at designated checkpoints. Since we had 14 replicates in total, we could efficiently employ threads for parallel computing using OpenMP. All models were fitted on the IBEX computational cluster at King Abdullah University of Science and Technology, Thuwal, Saudi Arabia, \url{https://docs.hpc.kaust.edu.sa/systems/ibex/}, utilizing its 28 threads. This parallelization capability was pivotal to render our extensive moving-window computations computationally feasible.

\subsection{Results and Analysis}\label{sec:results}

\begin{figure}[h]
  \centering

  \begin{subfigure}[b]{\linewidth}
    \centering
    \includegraphics[width=\linewidth,trim=0cm 0.5cm 0cm 0cm,clip]{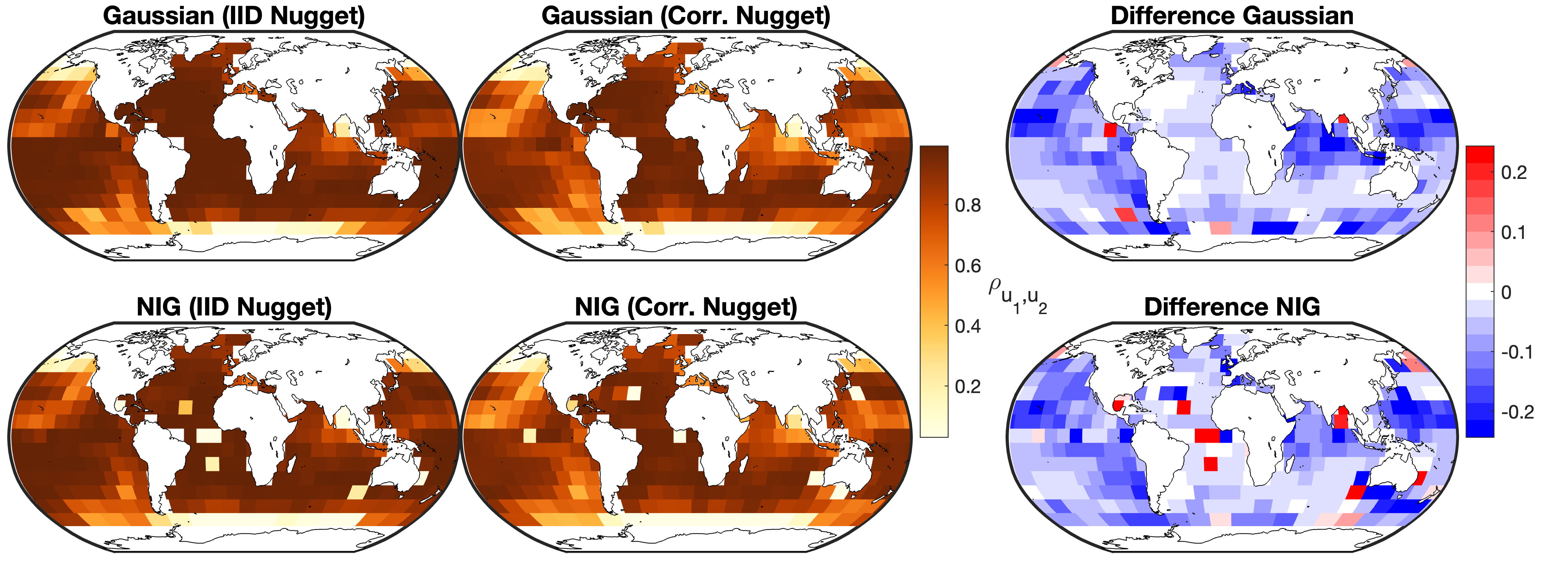}
    \caption{Correlation, \( \rho_{u_1,u_2} \) for model fits at 300 dbar pressure level (Columns: independent nugget, correlated nugget, difference = correlated – independent).}
    \label{fig:correlation_300}
  \end{subfigure}

  \vskip 0.5em  % Optional spacing between subfigures

  \begin{subfigure}[b]{0.8\linewidth}
    \centering
    \includegraphics[width=\linewidth]{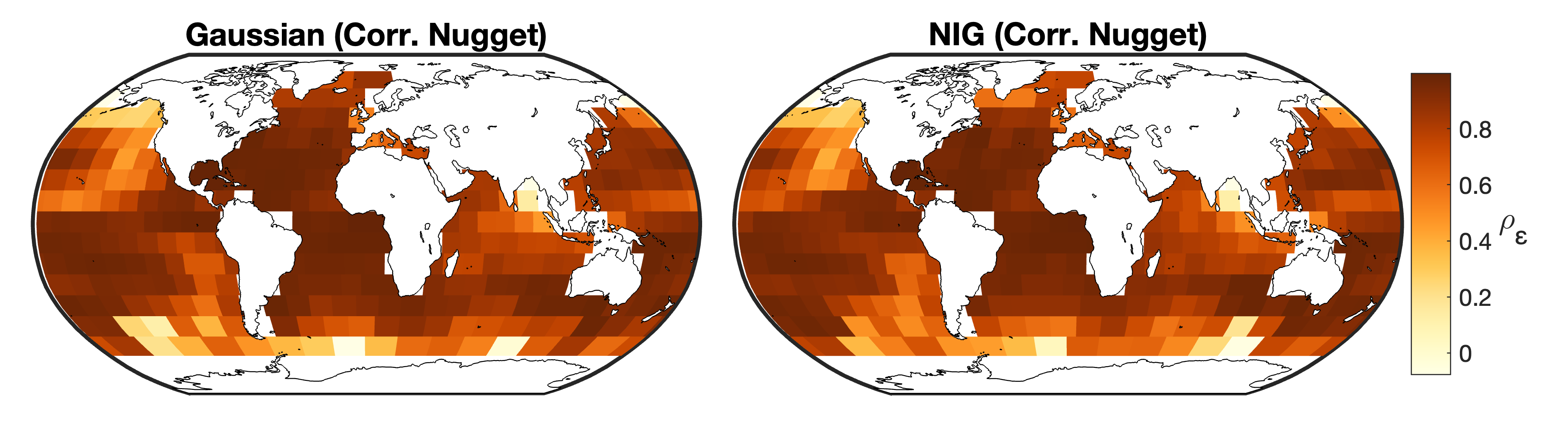}
    \caption{Measurement noise correlation, \( \rho_\varepsilon \) for proposed models at 300 dbar pressure level.}
    \label{fig:rho_epsilon_300}
  \end{subfigure}

  \caption{Results for model fits at 300 dbar pressure level. The models in the second column reveal the hidden dependence between the fields, indicating a possible underestimation of dependence in independent measurement noise models as shown in blue in the third column.}
  \label{fig:results_300_combined}
\end{figure}

In Figure~\ref{fig:results_300_combined}, part (a) illustrates a colormap of the Pearson correlation parameter computed with equation~\eqref{eq:PearsonCorrelation} for each model. The first column corresponds to the independent nugget effect model, the second column the correlated nugget model, and the third their difference (correlated - independent). Blue shades in the third column indicate decreases of up to -0.20, while red shades mark increases of up to +0.20.

\begin{figure}[t]
    \centering
    \includegraphics[width=\linewidth,trim=0cm 0.5cm 0cm 0cm,clip]{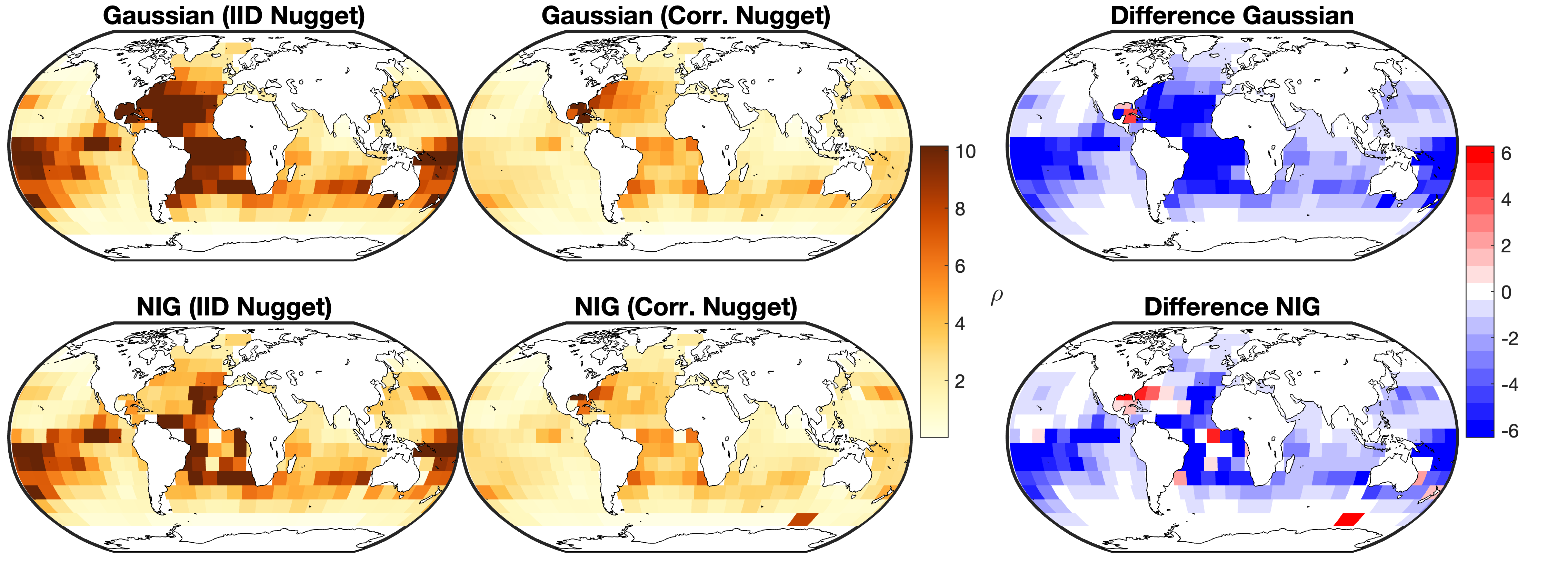}
    \caption{\( \rho \) parameter estimates at 300 dbar. }
    \label{fig:results_300_rho}
\end{figure}

Adding a correlated nugget lowers the median Pearson correlation from 0.95 to 0.87 in the Gaussian model and from 0.94 to 0.87 in the NIG model. The global average of the dependence parameter $\rho$ drops even more sharply: from 4.90 to 2.30 (Gaussian) and from 3.96 to 2.19 (NIG) (see Figure~\ref{fig:results_300_rho}). These numbers confirm that ignoring measurement-error correlation systematically overstates between-field dependence.

Part (b) of Figure~\ref{fig:results_300_combined} supports this interpretation, showing that the nugget correlation parameter $\rho_\varepsilon$ is uniformly large and positive, which points to widespread small-scale correlation. As shown in the simulation study of Section~\ref{sec:simulation}, such high $\rho_\varepsilon$ values increase the dependence unless they are estimated separately.
The apparent decrease in correlation between the latent processes suggests that previously estimated dependencies might originate from the measurement noise, specifically the nugget effect. The results suggest that the connections between the processes we are studying might not be as strong as previously believed, as correlated measurement errors seem to account for some of the correlations we observe.
 
The main reason for the difference in the estimated correlation between the fields is that the parameter estimates of \( \rho \) are affected by the correlation in the measurement noise. In the independent noise model, all dependence is captured by \( \rho \) because there is no additional parameter to account for dependence in the nugget effect. In contrast, the correlated noise model includes an extra parameter in the outer measurement noise that absorbs some of this dependence. Consequently, the estimated values of \( \rho \) are lower in the correlated model, as shown in the third column of Figure~\ref{fig:results_300_rho}. This negative difference indicates that the extra parameter in the correlated model reduces the dependence attributed to \( \rho \) compared to the independent model.

\begin{figure}[h]
    \centering

    \begin{subfigure}{\linewidth}
        \centering
        \includegraphics[width=\linewidth,trim=0cm 0.5cm 0cm 0cm,clip]{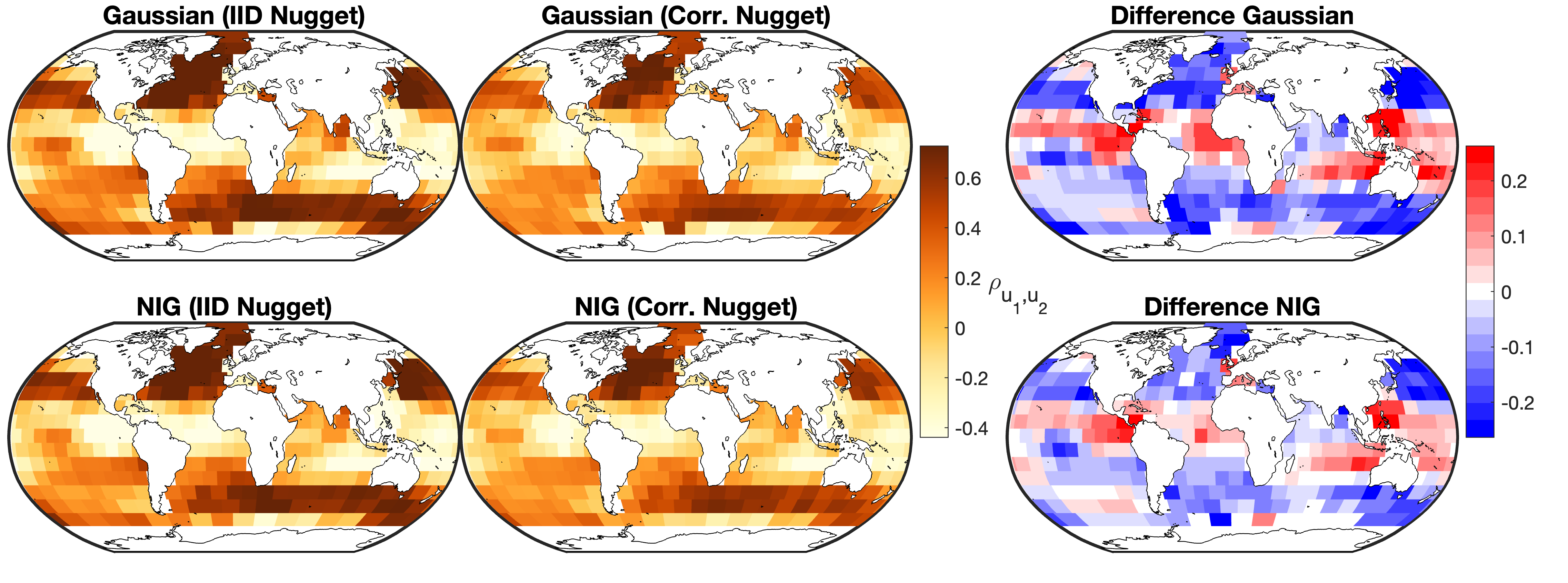}
        \caption{10 dbar}
        \label{fig:correlation_10}
    \end{subfigure}

    \vskip 0.5em

    \begin{subfigure}{\linewidth}
        \centering
        \includegraphics[width=\linewidth,trim=0cm 0.5cm 0cm 0cm,clip]{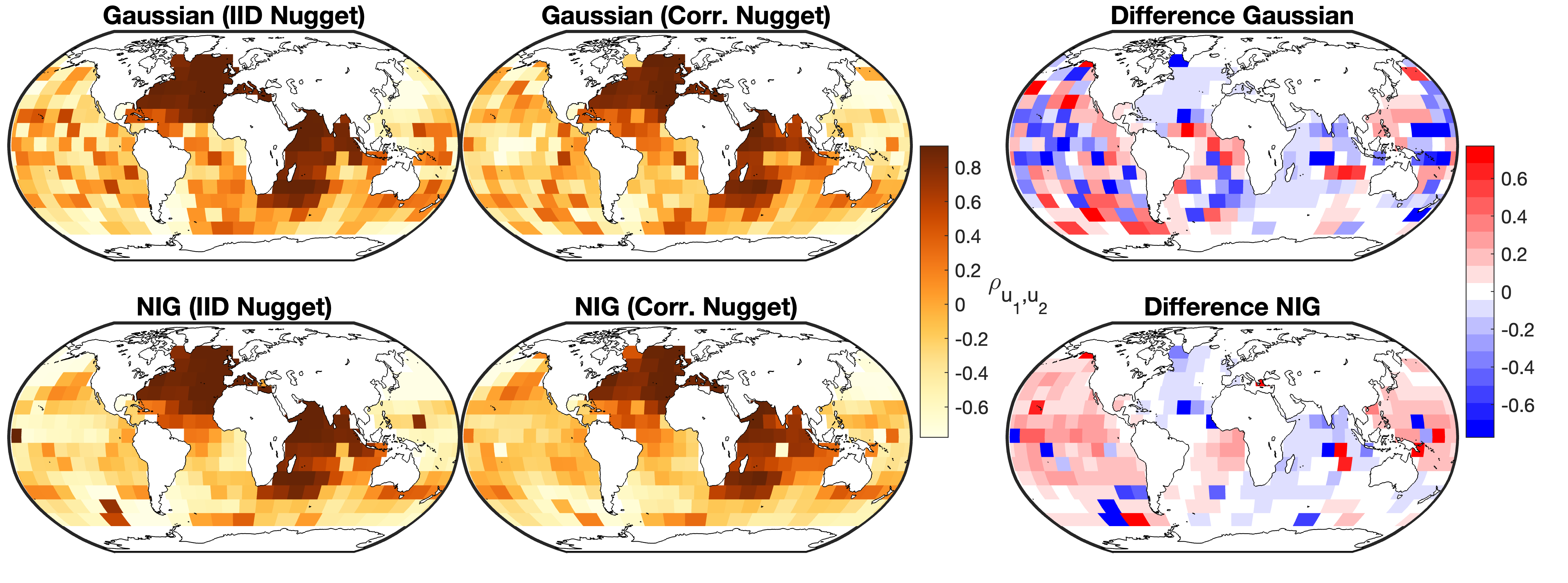}
        \caption{1000 dbar}
        \label{fig:correlation_1000}
    \end{subfigure}

    \caption{Correlation, \( \rho_{u_1,u_2} \), between two latent fields at (a) 10 dbar and (b) 1000 dbar. The third column in each panel shows the difference (correlated – independent).}
    \label{fig:correlation-10-1000}
\end{figure}

However, a different pattern for the estimated Pearson correlation can be observed at other pressure levels. The difference column in Figure~\ref{fig:correlation-10-1000} further highlights how ignoring measurement correlation error can generally overestimate true dependence. At 10 dbar (panel a), the correlated nugget model typically yields lower dependence, as indicated by the widespread blue areas. However, a notable exception is the region around the equator, where it exhibits the reverse behavior, with correlations up to 0.15 higher than in the independent model. By contrast, the 1000 dbar layer (panel b) is markedly more heterogeneous: while some regions show decreased correlation, most grid cells display a positive difference, and this effect appears strongest for the non-Gaussian models. This variability at depth possibly suggests that deep-water measurements are comparatively less contaminated by shared small-scale noise.

\begin{figure}[b]
    \centering
    \includegraphics[width=\linewidth,trim=0cm 0.5cm 0cm 0cm,clip]{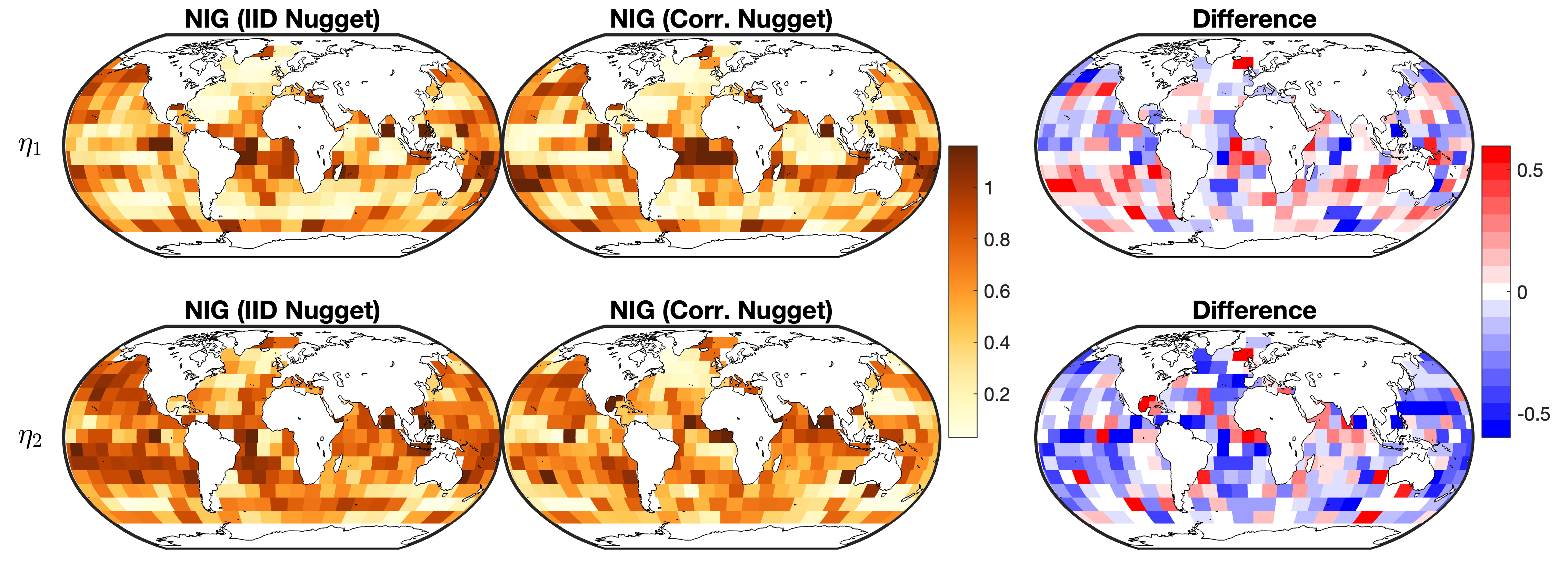}
    \caption{\( \eta \) parameter estimates at 300 dbar. \( \eta_1 \) corresponds to Temperature and \( \eta_2 \) to Salinity.}
    \label{fig:results_300_nu}
\end{figure}

For the non-Gaussian models, Figure~\ref{fig:results_300_nu} shows the estimated parameter $\eta$ which controls the tail of the distribution. 
The model approaches Gaussian behavior as \(\eta \to \infty\) and a heavy-tailed Cauchy distribution as \(\eta \to 0\).
We can, therefore, interpret the difference colormap as showing that the correlated nugget effect models are more non-Gaussian than the independent nugget case models when negative values (blue) are present. From the colormap of differences, it is clear that the correlated model for the temperature field becomes more non-Gaussian in most locations when using the correlated nugget effect. We also provide similar colormaps as in Figures~\ref{fig:results_300_combined}--\ref{fig:results_300_nu} for the other two pressure levels in the Appendix.

\subsection{Model diagnostics}
\begin{figure}[h]
    \centering

    \begin{subfigure}[b]{0.49\linewidth}
        \centering
        \includegraphics[width=\linewidth]{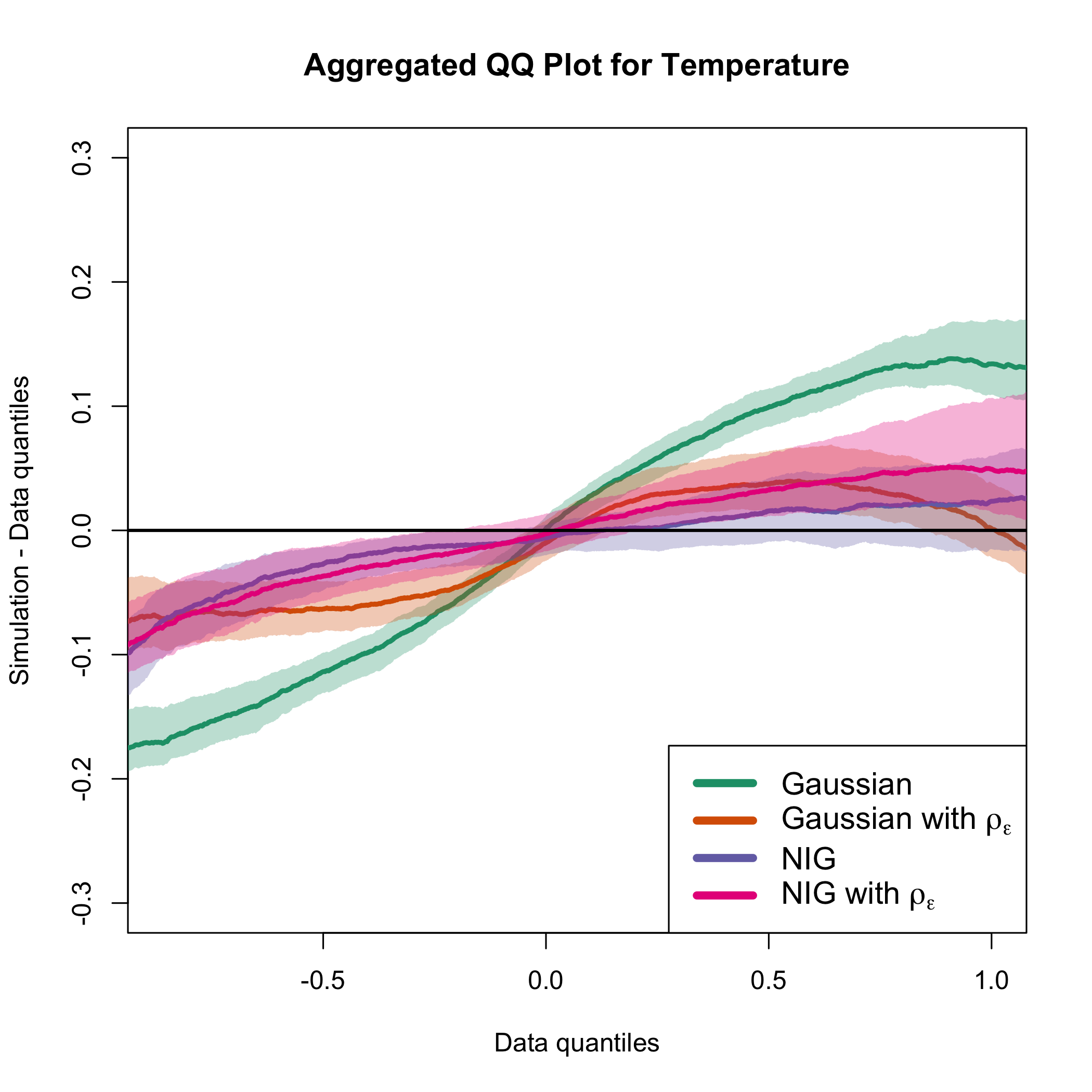}
        \caption{Temperature}
        \label{fig:agg_qq_temp}
    \end{subfigure}
    \hfill
    \begin{subfigure}[b]{0.49\linewidth}
        \centering
        \includegraphics[width=\linewidth]{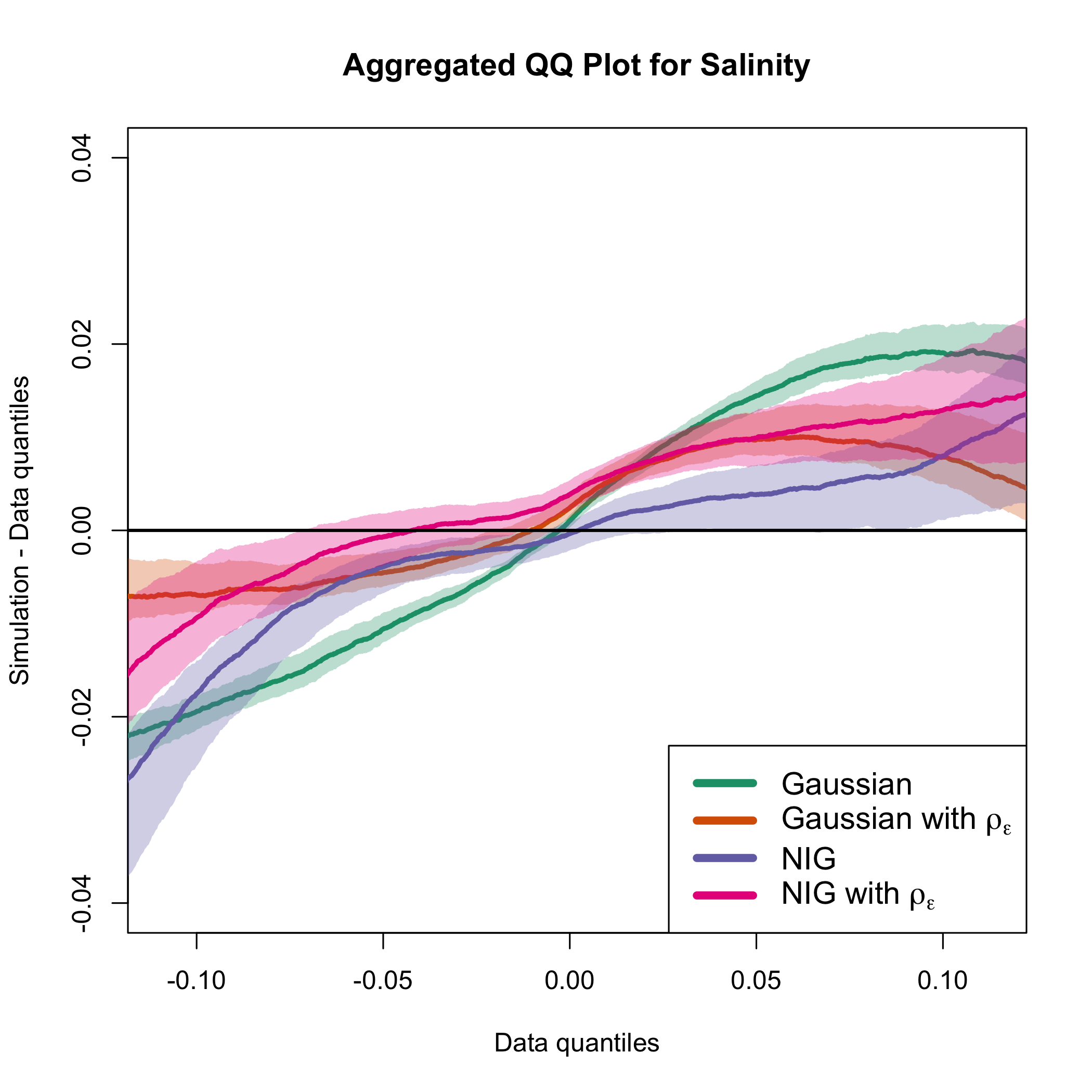}
        \caption{Salinity}
        \label{fig:agg_qq_psal}
    \end{subfigure}

    \caption{Aggregated QQ plots for 300 dbar with 20 simulations for each model. The solid colored line is the average of the 20 simulated QQ curves for each model, and the semi-transparent band around it is the point-wise 95\% simulation envelope. Predictive distributions are more accurate when the curves are close to horizontal straight lines at 0. The plots for the Gaussian and NIG models with correlated measurement noises overlap at the 0 line, which supports the results of cross-validation.}
    \label{fig:agg_qq_plots}
\end{figure}

One obvious question is whether the non-Gaussian models are preferable to the standard Gaussian models. Therefore, as a natural first step, we assess the suitability of the distributional assumptions for all models used. We exploit both the quantile-quantile (QQ) plots of the standardized marginal residuals and compare the different models. First, for the QQ plots for Gaussian models, one can examine its marginal residuals, which are standardized by their variances, and compare these against the theoretical quantiles of a standard normal distribution. However, this approach cannot easily be extended to the non-Gaussian models, since we do not necessarily know the true distribution of its residuals. To compare the validity of all models, Gaussian or otherwise, and compare them to each other, we exploit the adjusted version of the QQ plot; for each model, we predict the latent process from the data given the estimated parameters and compare it against the distribution of latent process simulated from the fitted model, as proposed in \textcite{Asar2020}. 
We replicated this process across 20 simulated datasets, creating a joint simulation envelope to evaluate the fit more formally. We perform simulations independently for each grid and then aggregate them to plot a single fit as shown in Figure~\ref{fig:agg_qq_plots}. For additional QQ plots and their stratified versions by latitude and oceans, please refer to the Appendix.

From Figure~\ref{fig:agg_qq_plots} we can note two things. First, for both the Gaussian and NIG models, the introduction of the correlation in the measurement noise improves the model fit. Second, and somewhat surprisingly, there is no big difference between the model fits of the Gaussian and NIG models with the correlation in the measurement noise. This suggests that it might not be necessary to use the non-Gaussian model as long as we include the correlation in the measurement noise.

\subsection{Evaluation of predictive performance}
As a final comparison of the models, we perform a leave-one-out cross-validation (LOOCV) for each grid to assess the predictive performance of the models, employing CRPS, SCRPS, MAE, and RMSE as metrics. The predictions were limited to the reference box and did not include the expanded 5-degree area. Global scores were calculated by taking the weighted averages of the data points in each grid box, with the weights corresponding to the number of data points in the box. The LOOCV results for temperature and salinity can be found in Tables~\ref{tab:CV_Argo_results_temperature} and~\ref{tab:CV_Argo_results_salinity}. In the tables, 
the lower the metric score, the better the model performance. The best model is indicated by bold values with the minimum score, while the second best model is shown by italic values.  %The results indicate that the best model in most metrics is the Gaussian model with correlated measurement noise, across all pressure levels, both for the temperature and salinity fields.
While the Gaussian model with correlated measurement noise is the clear winner at the intermediate 300 dbar level and often achieves the lowest RMSE, the non-Gaussian NIG model with correlated measurement noise attains the best scores at the surface (10 dbar) and frequently matches—or surpasses—the Gaussian model at 1000 dbar, particularly for CRPS and SCRPS.

As expected from the results of the QQ plots, the non-Gaussian NIG model with a general measurement noise structure exhibits strong performance, frequently securing the first or second-best scores --- especially at 10 dbar (three of four metrics) and for CRPS/SCRPS at 1000 dbar --- and consistently outperforming the traditional Gaussian model with an independent (diagonal) measurement noise structure. 
This indicates that --- while the Gaussian general model is the ultimately best model at 300 dbar --- incorporating non-Gaussian distributions such as the NIG, which allows for skewness and greater kurtosis than the Gaussian  can be superior at both the surface (10 dbar) and depth (1000 dbar), and demonstrates the advantages of coupling correlation with a more flexible marginal distribution and provide significant advantages over models that do not account for inter-variable correlations.
% ----------------------------------------------------
% Temperature
% ----------------------------------------------------
\begin{table}[H]
\centering
\caption{Cross-validation results of the moving-windows model---Temperature}
\label{tab:CV_Argo_results_temperature}
\begin{tabular}{@{}cllcccc@{}}
\toprule
\textbf{Pressure level} & \textbf{Model} & \textbf{Structure} & \textbf{RMSE} & \textbf{MAE} & \textbf{CRPS} & \textbf{SCRPS} \\ 
\midrule
10 & Gaussian & diagonal & \textit{0.7155} & \textit{0.4702} & 0.3421 & 0.7591 \\ 
   &          & general  & 0.7181 & 0.4733 & 0.3450 & 0.7642 \\ 
   & NIG      & diagonal & 0.7290 & 0.4731 & \textit{0.3408} & \textbf{0.7539} \\ 
   &          & general  & \textbf{0.7138} & \textbf{0.4701} & \textbf{0.3394} & \textit{0.7550} \\ 
\midrule
300 & Gaussian & diagonal & \textit{0.6557} & 0.3789 & 0.2886 & 0.6542 \\ 
    &          & general  & \textbf{0.6142} & \textbf{0.3561} & \textbf{0.2717} & \textbf{0.6189} \\ 
    & NIG      & diagonal & 0.7498 & 0.3970 & 0.2987 & 0.6580 \\ 
    &          & general  & 0.6597 & \textit{0.3725} & \textit{0.2805} & \textit{0.6204} \\ 
\midrule
1000 & Gaussian & diagonal & \textit{0.2135} & \textbf{0.1169} & \textit{0.0873} & 0.0048 \\ 
     &          & general  & \textbf{0.2116} & \textit{0.1175} & 0.0876 & 0.0142 \\ 
     & NIG      & diagonal & 0.2327 & 0.1214 & 0.0894 & \textit{0.0017} \\ 
     &          & general  & 0.2153 & 0.1178 & \textbf{0.0867} & \textbf{-0.0045} \\ 
\bottomrule
\end{tabular}
\end{table}

% ----------------------------------------------------
% Salinity
% ----------------------------------------------------
\begin{table}[ht]
\centering
\caption{Cross-validation results of the moving-windows model---Salinity}
\label{tab:CV_Argo_results_salinity}
\begin{tabular}{@{}cllcccc@{}}
\toprule
\textbf{Pressure level} & \textbf{Model} & \textbf{Structure} & \textbf{RMSE} & \textbf{MAE} & \textbf{CRPS} & \textbf{SCRPS} \\ 
\midrule
10 & Gaussian & diagonal & \textit{0.1806} & 0.1010 & 0.0784 & -0.0103 \\ 
   &          & general  & \textbf{0.1792} & \textit{0.1009} & 0.0793 & -0.0031 \\ 
   & NIG      & diagonal & 0.1878 & 0.1015 & \textit{0.0762} & \textbf{-0.0385} \\ 
   &          & general  & 0.1816 & \textbf{0.1004} & \textbf{0.0755} & \textit{-0.0351} \\ 
\midrule
300 & Gaussian & diagonal & 0.0844 & 0.0464 & 0.0358 & -0.4077 \\ 
    &          & general  & \textbf{0.0789} & \textbf{0.0435} & \textbf{0.0344} & \textit{-0.4402} \\ 
    & NIG      & diagonal & 0.0953 & 0.0481 & 0.0366 & -0.4119 \\ 
    &          & general  & \textit{0.0838} & \textit{0.0453} & \textit{0.0344} & \textbf{-0.4481} \\ 
\midrule
1000 & Gaussian & diagonal & 0.0258 & 0.0138 & 0.0106 & -1.0589 \\ 
     &          & general  & \textbf{0.0250} & \textit{0.0133} & 0.0109 & -1.1056 \\ 
     & NIG      & diagonal & 0.0278 & 0.0136 & \textit{0.0101} & \textit{-1.1362} \\ 
     &          & general  & \textit{0.0252} & \textbf{0.0130} & \textbf{0.0097} & \textbf{-1.1461} \\ 
\bottomrule
\end{tabular}
\end{table}

At the intermediate depth of 300 dbar, the Gaussian model that incorporates correlated structures delivers the best performance across the most evaluated metrics. These results demonstrate the importance of accounting for inter-variable correlations in multivariate spatial modeling.
In this context, the Gaussian general model still has the lowest RMSE at 300 dbar and 1000 dbar, but the NIG general model outperforms it at 10 dbar, showing depth-dependent behavior.

\begin{figure}[tb]
\centering
    \begin{subfigure}{0.8\linewidth}
        \centering
        \includegraphics[width=\linewidth,trim=0cm 0cm 0cm 0cm,clip]{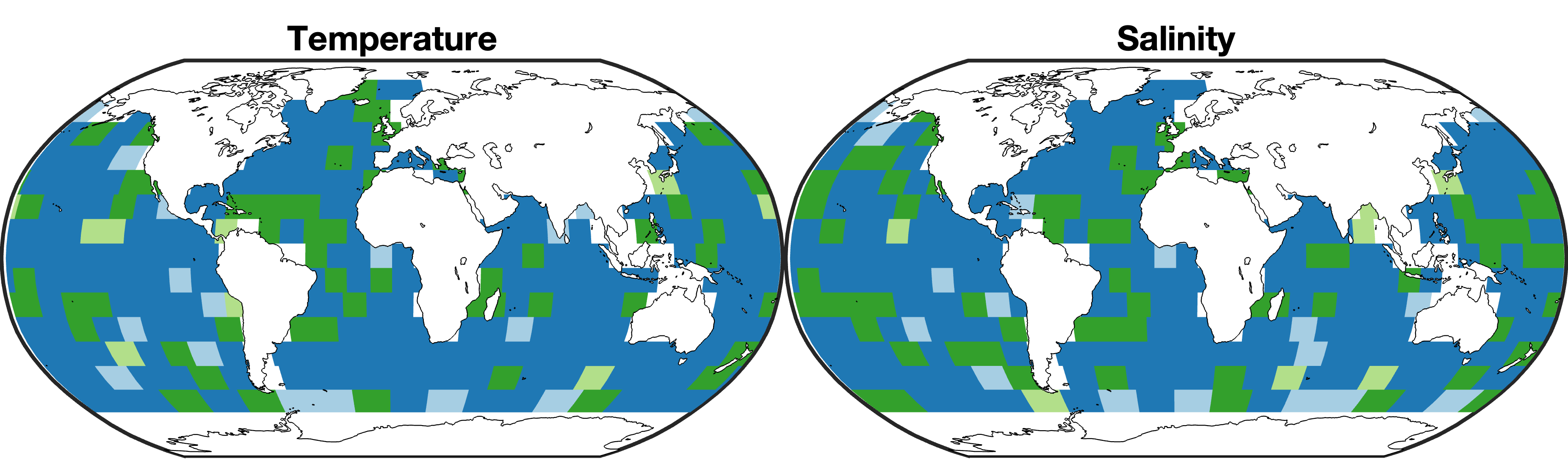}
        \caption{RMSE}
        \label{fig:best_model_rmse}
    \end{subfigure}

    \vskip 0.5em

    \begin{subfigure}{0.8\linewidth}
        \centering
        \includegraphics[width=\linewidth,trim=0cm 0cm 0cm 1.5cm,clip]{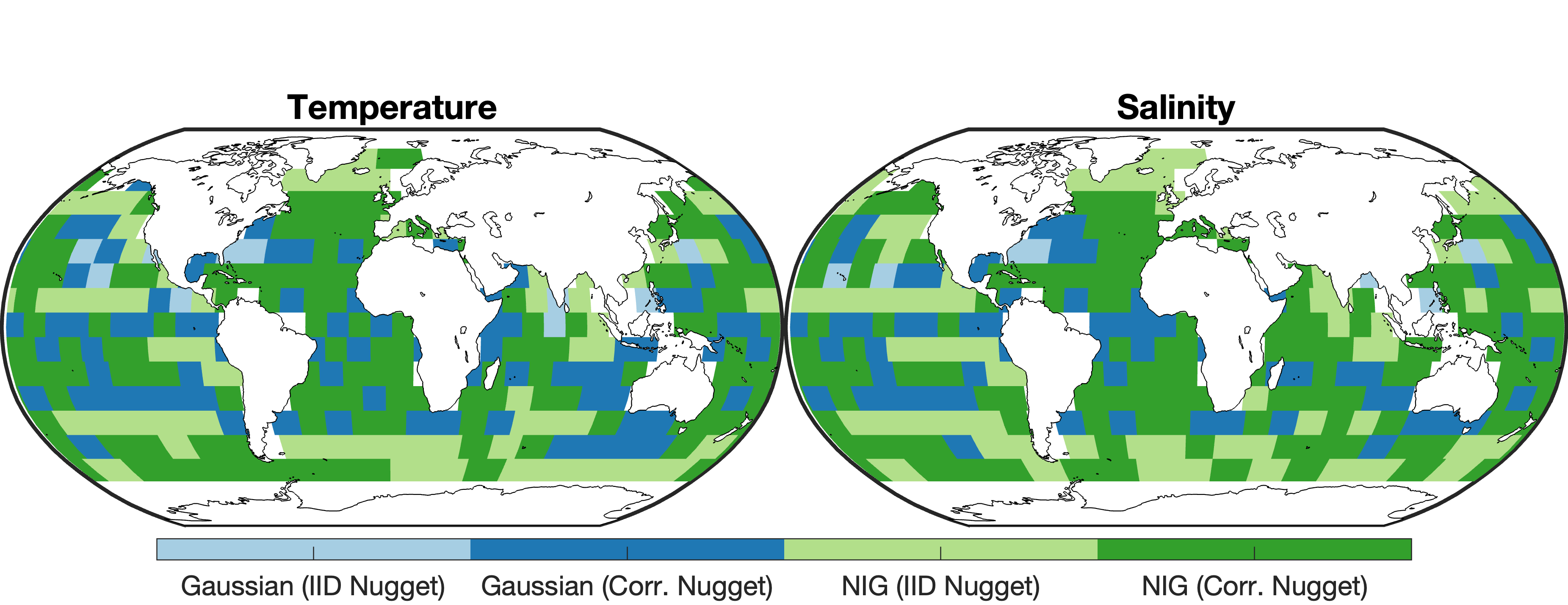}
        \caption{SCRPS}
        \label{fig:best_model_scrps}
    \end{subfigure}

    \caption{Best models for each grid for 300 dbar, according to (a) RMSE and (b) SCRPS.}
    \label{fig:best_model-300}
\end{figure}

Figure~\ref{fig:best_model-300} presents a color-coded map of the best-performing models for each field, as determined by RMSE and SCRPS. Table~\ref{tab:CV_Argo_results_temperature_best} summarizes the grid counts for which each model achieved the best performance for temperature and salinity, respectively.

In the salinity field, the mix of best-performing models in Table~\ref{tab:CV_Argo_results_salinity} suggests possible non-stationarity that needs to be addressed. At the shallowest level (10 dbar) the Gaussian model with a correlated nugget has the lowest RMSE, whereas CRPS and MAE favour the NIG model with a correlated nugget and SCRPS selects the NIG model with an independent nugget. This metric-dependent ranking mirrors the findings of \textcite{Fuglstad2015}, who encountered a similar divergence in model performance in their paper to forecast precipitation using RMSE and CRPS as evaluation metrics. Their conclusions suggest that this pattern of differences in the predictions of the best model could be due to a misspecification in the model, which may require separate model fits for different regions. 

\begin{table}[H]
\centering
\caption{Counts of grids with the best model performance---Temperature, Salinity}
\label{tab:CV_Argo_results_temperature_best}
\begin{tabular}{@{}cllcccc@{}}
\toprule
\textbf{Pressure level} & \textbf{Model} & \textbf{Structure} & \textbf{RMSE} & \textbf{MAE} & \textbf{CRPS} & \textbf{SCRPS}\\
\midrule
10 & Gaussian & diagonal & 64, 55 & 67, 62 & 60, 31 & 53, 16 \\
 &  & general & 72, 107 & 65, 63 & 55, 32 & 49, 24 \\
 & NIG & diagonal & 63, 49 & 57, 68 & 93, 126 & 113, 153 \\
 &  & general & 79, 67 & 89, 85 & 70, 89 & 63, 85 \\
\midrule
300 & Gaussian & diagonal & 16, 16 & 14, 16 & 21, 8 & 10, 7 \\
 &  & general & 198, 188 & 181, 167 & 105, 92 & 78, 62 \\
 & NIG & diagonal & 8, 7 & 12, 10 & 35, 43 & 63, 68 \\
 &  & general & 54, 65 & 69, 83 & 115, 133 & 125, 139 \\
\midrule
1000 & Gaussian & diagonal & 67, 37 & 65, 33 & 52, 20 & 48, 10 \\
 &  & general & 115, 126 & 102, 111 & 67, 54 & 48, 33 \\
 & NIG & diagonal & 27, 30 & 29, 39 & 55, 73 & 76, 109 \\
 &  & general & 61, 77 & 74, 87 & 96, 123 & 98, 118 \\
\bottomrule
\end{tabular}
\end{table}

\section{Conclusions}\label{sec:conclusion}

In this study, we extended traditional geostatistical frameworks by introducing a multivariate Matérn-SPDE model with correlated nugget effects to account for small-scale variability and measurement error correlations in Argo ocean data. Our simulation studies demonstrated that incorporating correlated nugget effects significantly enhances parameter estimation and spatial prediction accuracy, particularly in scenarios with strong measurement noise dependencies. By applying the proposed model to 14 years of Argo profiling float data, we found that existing approaches may overestimate cross-variable dependencies due to unaccounted-for correlated measurement errors. By relaxing the assumption of independent noise, our framework offers a more realistic representation of Argo data and thus enables better interpretation of fine-scale oceanic patterns and improved uncertainty quantification. %The integration of non-Gaussian noise using the NIG distribution further enhanced the model’s ability to capture heavy-tailed variability, as evidenced by diagnostic Q–Q plots. 
Our open-source implementation in the \textit{ngme2} R package will help to ensure reproducibility and facilitate broader adoption of the framework. 

The results for the Argo data indicate that the addition of the correlation in the measurement noise indeed improves the model performance for both Gaussian and non-Gaussian models. However, once this is accounted for, there seems to be little benefit of using a non-Gaussian model, at least of the types we considered here. 

While our model assumes spatial stationarity with fixed smoothness parameters (\(\alpha = 2\)), future work could explore the direct estimation of these parameters from the data. The rational SPDE method, as introduced by \textcite{Bolin2020rational} and refined by \textcite{Xiong2022}, could extend our model to locally non-stationary settings. It is also possible to expand the framework to fully three-dimensional or spatio-temporal contexts by addressing temporal autocorrelation in the Argo time series. Additionally, the challenge of modeling residuals in a non-Gaussian context remains an open question, as estimating the bivariate non-Gaussian measurement noise with extra correlation poses challenges to the stability of parameter estimates.

Looking forward, extending the spatial model to incorporate a third dimension, such as height or pressure, and refining the model to enable space-time multivariate analysis hold promise to provide more nuanced insights, especially for global non-stationary non-Gaussian models of subsurface temperatures in the upper layers of the ocean.  Furthermore, the enhanced temperature and salinity fields generated by our model could be used to predict oxygen concentrations using machine learning techniques, such as the random forest approach of \textcite{Giglio2018}, which has been demonstrated to yield highly accurate oxygen predictions at fixed pressure levels. Our proposed methodology could also be adapted to other environmental datasets with correlated measurement errors, including atmospheric and hydrological observations. 

\bigskip
\noindent\textbf{Funding.} This publication is based upon work supported by King Abdullah University of Science and Technology (KAUST) under Award No. ORFS-CRG11-2022-5015.

\appendix
\section*{Appendix}
In this Appendix, we provide details on data collection and quality control; the main results for the 10- and 1000-dbar pressure levels; and additional details for the simulation study. We also show results for cross-validation (CV) and uncertainty quantification performance at all pressure levels. Moreover, we provide details for the NIG‑SPDE model likelihood and Gibbs sampler. We prove Proposition~2.1 from the main text and provide the expression for the Rao--Blackwellized SCRPS. All code and processed data sets are archived at \url{https://github.com/d-saduakhas/Argo-SPDE}.

\section{Overview of Argo Data}\label{sec:overview}

Argo is a global network of profiling floats that provide high-resolution observations of the subsurface ocean. The network measures temperature (TEMP), salinity (PSAL), and pressure (PRES) up to a depth of 2,000 meters, offering an unparalleled resource for studying oceanographic phenomena. Each float follows a 10-day cycle of profiling and transmits data via satellite to centralized repositories. This dataset is crucial for spatio-temporal modeling due to its uniform spatial coverage and temporal resolution \parencite{ArgoData}.

The nominal accuracies of the Argo sensors are 0.005°C for temperature, 0.01 for salinity, and 2.5 dbar for pressure. Despite the reliability of temperature measurements, salinity and pressure data often require adjustments for sensor drift and calibration offsets. Only delayed-mode data verified by experts were used in our analysis to ensure the highest data quality \parencite{ArgoManual}.

Given the importance of Argo data in analyzing subsurface ocean dynamics, we tailored preprocessing methods to address its unique characteristics, such as variability in vertical resolution and sensor errors. Our quality control steps and interpolation to standardized pressure levels ensure a robust dataset for applying multivariate SPDE models.

\subsection{Quality Control}

We applied the quality‑control criteria proposed by \textcite{Kuusela2018Supplement} to process the raw data obtained from the GDAC website. Additional criteria specific to practical‑salinity data were also incorporated because of its small variability.

\begin{itemize}
  \item Data were restricted to January of each year, covering the period from 2007‑01‑01 to 2020‑01‑31.
  \item Quality‑control procedures included:
  \begin{itemize}
    \item Selecting delayed‑mode data verified by experts within 12~months.
    \item Using quality‑control flags with QC~$=1$ (good data).
    \item Utilising adjusted variables exclusively for PRES, TEMP and PSAL.
    \item Excluding profiles from problematic floats where \texttt{PRES\_ADJUSTED\_ERROR}~$\ge 20$.
  \end{itemize}
\end{itemize}

To further refine the data set, we first removed any anomalies in the input profile measurements. The data were then interpolated to standardized pressure levels of 10, 300 and 1000~dbar. Although there may be more advanced methodologies, we are confident that our criteria ensure a reliable and well‑structured data set suitable for subsequent analyzes.

\section{Likelihood and finite-element discretization}\label{sec:supp_theory}

In this section, we present the tools that enable
likelihood evaluation and inference. Section~\ref{sec:Qeps} introduces the
correlated-nugget precision matrix together with its log-scale
reparameterization.  Section~\ref{sec:like_NIG} presents the Gibbs
sampler for the NIG–SPDE model and the accompanying
Rao--Blackwellized expressions (Algorithm~\ref{gibbs}).  We conclude
with the CRPS/SCRPS formulas and the proof of Proposition~2.1.

%---------------------------------------------------------------------------
\subsection{Correlated-nugget precision}\label{sec:Qeps}

All measurement errors share the block precision 
\begin{equation}\label{eq:Qeps}
  \mathbf{\Sigma}^{-1}_{\varepsilon} = \mathbf{Q}_{\varepsilon} =
  \begin{bmatrix}
    \dfrac{1}{(1-\rho_{\varepsilon}^{2})\,\sigma_{\varepsilon_1}^2}\,\mathbf I_n &
    -\dfrac{\rho_{\varepsilon}}
           {(1-\rho_{\varepsilon}^{2})\,\sigma_{\varepsilon_1}\sigma_{\varepsilon_2}}
           \,\mathbf I_n\\[8pt]
    -\dfrac{\rho_{\varepsilon}}
           {(1-\rho_{\varepsilon}^{2})\,\sigma_{\varepsilon_1}\sigma_{\varepsilon_2}}
           \,\mathbf I_n &
    \dfrac{1}{(1-\rho_{\varepsilon}^{2})\,\sigma_{\varepsilon_2}^2}\,\mathbf I_n
  \end{bmatrix}_{m\times m},
\end{equation}
assuming the number of observations per field is $N_1 = N_2 = n$ and total observations $m= 2n$.
The following parameters were mapped onto a log scale—$\kappa_1, \kappa_2, \sigma_1, \sigma_2, \sigma_{\varepsilon_1}, \sigma_{\varepsilon_2}$ and $\rho_{\varepsilon}$—to enable the use of an unconstrained function minimizer. 
For unconstrained optimization we
map\footnote{Implemented in the \textsf{R} package \texttt{ngme2}
\parencite{ngme2}.}
\[
  \theta_{\sigma}=\log\sigma, \qquad
  \theta_{\rho_{\varepsilon}}=-\log\!\bigl((1-\rho_{\varepsilon})/
                                             (1+\rho_{\varepsilon})\bigr),
\]
which yields
\begin{equation}\label{eq:Qeps_mapped}
  \mathbf Q_{\varepsilon_t}=
  \begin{bmatrix}
     \frac14(e^{\theta_{\rho_{\varepsilon}}}+1)^2
            e^{-\theta_{\rho_{\varepsilon}}-2\theta_{\sigma_{\varepsilon_1}}}
            \mathbf I_n &
    -\frac14(e^{2\theta_{\rho_{\varepsilon}}}-1)
            e^{-\theta_{\rho_{\varepsilon}}-\theta_{\sigma_{\varepsilon_1}}
               -\theta_{\sigma_{\varepsilon_2}}}\mathbf I_n\\[6pt]
    -\frac14(e^{2\theta_{\rho_{\varepsilon}}}-1)
            e^{-\theta_{\rho_{\varepsilon}}-\theta_{\sigma_{\varepsilon_1}}
               -\theta_{\sigma_{\varepsilon_2}}}\mathbf I_n&
     \frac14(e^{\theta_{\rho_{\varepsilon}}}+1)^2
            e^{-\theta_{\rho_{\varepsilon}}-2\theta_{\sigma_{\varepsilon_2}}}
            \mathbf I_n
  \end{bmatrix},
\end{equation}
with log-determinant
\(
  \log|\mathbf Q_{\varepsilon_t}|
  =m\bigl[\log(e^{\theta_{\rho_{\varepsilon}}}+1)-\log 2
                   -\theta_{\sigma_{\varepsilon_1}}
                   -\theta_{\sigma_{\varepsilon_2}}
                   -\tfrac12\theta_{\rho_{\varepsilon}}\bigr].
\)

%---------------------------------------------------------------------------
\subsection{Inference for the NIG--SPDE model}
\label{sec:like_NIG}

The normal–inverse-Gaussian (NIG) driven SPDE yields an intractable
marginal likelihood. We therefore adopt the MCMC scheme of
\textcite{Bolin2020}, replacing their independent-nugget precision with the
correlated nugget matrix in Eq.~\eqref{eq:Qeps}. Gibbs updates,
Fisher-identity gradients and stochastic optimization are otherwise
unchanged.

\paragraph{Joint log-posterior.}
Up to an additive constant, the joint log-density of
\((\mathbf v,\mathbf w)\) given data \(\mathbf Y\) and parameters
\(\mathbf\Theta\) is
\begin{align}\label{eq:logpost}
\log \pi(\mathbf v,\mathbf w\mid \mathbf Y,\mathbf\Theta)=&
   -\tfrac12\log|\mathbf\Sigma_\varepsilon|
   -\tfrac12(\mathbf Y-\mathbf A\mathbf w-\mathbf B\boldsymbol\beta)^{\!\top}
            \mathbf Q_\varepsilon
            (\mathbf Y-\mathbf A\mathbf w-\mathbf B\boldsymbol\beta) \nonumber\\
 &-\tfrac12\bigl(\mathbf K\mathbf w-(\boldsymbol\mu\!\otimes\!\mathbf I_m)
                (\mathbf v-\mathbf h)\bigr)^{\!\top}
             \operatorname{diag}(\mathbf v)^{-1}
             \bigl(\mathbf K\mathbf w-(\boldsymbol\mu\!\otimes\!\mathbf I_m)
                (\mathbf v-\mathbf h)\bigr)                                   \nonumber\\
 &+\tfrac12\log|\mathbf K|
   -\mathbf 1^{\top}\!\log\mathbf v
   +\log\pi_{\Psi}(\mathbf v).
\end{align}

\paragraph{Rao--Blackwellization.}
Applying Fisher’s identity \parencite{Dempster1977} we can express the
score of the intractable marginal likelihood as
\begin{equation}\label{eq:RB_score}
\nabla_{\!\boldsymbol\Theta}\log\pi(\mathbf v\mid\mathbf Y,\boldsymbol\Theta)
  \;=\;
  E_{\mathbf w }
    \bigl[
      \nabla_{\!\boldsymbol\Theta}
      \log\pi(\mathbf v,\mathbf w\mid\mathbf Y,\boldsymbol\Theta)\mid \mathbf v,\mathbf Y,\boldsymbol\Theta
    \bigr].
\end{equation} 

A naive Monte-Carlo (MC) estimate of
\eqref{eq:RB_score} would draw \(\bigl(\mathbf v^{(j)},\mathbf
w^{(j)}\bigr)\) from the Gibbs sampler and average the
gradient
\(\nabla_{\!\boldsymbol\Theta}\log\pi(\mathbf v^{(j)},\mathbf
w^{(j)}\mid\mathbf Y,\boldsymbol\Theta)\).
Because \(\mathbf w\mid\mathbf Y,\mathbf v,\boldsymbol\Theta\) is
Gaussian with known mean and precision
(Eqs.\,\eqref{eq:Qtilde}–\eqref{eq:xitilde}) we can apply \eqref{eq:RB_score}
analytically, producing a Rao--Blackwellized estimator with lower
variance. For a fixed \(\mathbf v\) we have
\begin{align}
\widetilde{\mathbf Q} = \widetilde{\mathbf Q}(\mathbf v) &=
  \mathbf K^{\top}\!\operatorname{diag}(\mathbf v)^{-1}\mathbf K
  +\mathbf A^{\top}\mathbf Q_{\varepsilon}\mathbf A,
  \label{eq:Qtilde}\\
\widetilde{\boldsymbol\xi} =\widetilde{\boldsymbol\xi}(\mathbf v) &=
  \widetilde{\mathbf Q}(\mathbf v)^{-1}
  \Bigl(
     \mathbf A^{\top}\mathbf Q_{\varepsilon}
       \bigl(\mathbf Y-\mathbf B\boldsymbol\beta\bigr)
     +\mathbf K^{\top}\!\operatorname{diag}(\mathbf v)^{-1}
       (\boldsymbol\mu\!\otimes\!\mathbf I_{n})
       \bigl(\mathbf v-\mathbf h\bigr)
  \Bigr),                                            \label{eq:xitilde}
\end{align}
and thus, for any matrix $\mathbf{M}$,
\[
E_{\mathbf w}[\mathbf w \mid\mathbf Y,\mathbf v,\boldsymbol\Theta]
     =\widetilde{\boldsymbol\xi},\qquad
E_{\mathbf w}
      [\mathbf{w M}\mathbf w^{\top}\mid\mathbf Y,\mathbf v,\boldsymbol\Theta]
     = \mathrm{tr}(\mathbf{M }\widetilde{\mathbf Q}^{-1})+
     \widetilde{\boldsymbol\xi}\,\mathbf{M}\,\widetilde{\boldsymbol\xi}^{\top}
       .
\]

These moments yield closed-form expectations of the quadratic terms that
are present in \(\nabla_{\!\boldsymbol\Theta}\log\pi(\mathbf v,\mathbf w\mid\mathbf Y,\boldsymbol\Theta)\).
For instance, for any index \(i\) (where \(v_i\) denotes the realized
value of the variable \(\mathbf{v}_i\) drawn in the Gibbs step),
\[
E\!\Bigl[
  \mathbf w^{\top}\mathbf C^{\top}\!\operatorname{diag}\!\bigl(1/v_i\bigr)
  \mathbf K\mathbf w
  \;\Big|\; \mathbf v,\mathbf Y
\Bigr]
  =
  \mathrm{tr}\!\bigl(
      \mathbf C^{\top}\!\operatorname{diag}\!\bigl(1/v_i\bigr)\mathbf K
      \widetilde{\mathbf Q}^{-1}\bigr)
  +\widetilde{\boldsymbol\xi}^{\top}
     \mathbf C^{\top}\!\operatorname{diag}\!\bigl(1/v_i\bigr)\mathbf K
     \widetilde{\boldsymbol\xi}.
\]

We then can compute  Rao--Blackwellized gradients, for example,
\[
\nabla_{\!\boldsymbol\beta}
      \log\pi(\mathbf v\mid\mathbf Y,\boldsymbol\Theta)
  = \mathbf B^{\top}\mathbf Q_{\varepsilon}
     \bigl(\mathbf Y-\mathbf A\widetilde{\boldsymbol\xi}-\mathbf B\boldsymbol\beta\bigr),
\]
with analogous closed-form expressions for
\(\sigma_{\varepsilon_1}^2\), \(\sigma_{\varepsilon_2}^2\),
\(\kappa_1^2\), \(\kappa_2^2\), \(\rho\), \(\rho_\varepsilon\), etc.
These gradients and the corresponding expressions for the remaining parameters can be found in the Supplementary Material of \textcite{Bolin2020}.

\paragraph{Gibbs sampler.}
The sampler cycles over \(\mathbf w\) and \(\mathbf v\) as summarized in
Algorithm~\ref{gibbs}; only the correlated precision
\(\mathbf Q_\varepsilon\) distinguishes it from the algorithm in
\textcite{Bolin2020}. 

\begin{algorithm}[H]
\caption{Gibbs Sampler (Retrieved from \textcite{Bolin2020} and adapted)}\label{gibbs}
\begin{algorithmic}[1]
\Procedure{GIBBS}{$\mathbf y,\mathbf B,\mathbf v,\bm\Psi,\mathbf A_1,\mathbf A_2,\mathbf h$}
  \State $\mathbf K\gets\text{BuildOperator}(\bm\Psi)$
  \State $\hat{\mathbf Q}\gets\mathbf K^{\top}\operatorname{diag}(\mathbf v)^{-1}\mathbf K
                      +\mathbf A^{\top}\mathbf Q_\varepsilon\mathbf A$
  \State $\hat{\boldsymbol\xi}\gets\hat{\mathbf Q}^{-1}\!\bigl(
          \mathbf A^{\top}\mathbf Q_\varepsilon(\mathbf y-\mathbf B\boldsymbol\beta)
          +\mathbf K^{\top}\operatorname{diag}(\mathbf v)^{-1}
            (\boldsymbol\mu\!\otimes\!\mathbf I_{2n})(\mathbf v-\mathbf h)\bigr)$
  \State Sample $\mathbf w\sim\mathcal N(\hat{\boldsymbol\xi},\hat{\mathbf Q}^{-1})$
  \State $\bigl[\mathbf E_1^{\top},\mathbf E_2^{\top}\bigr]^{\top}\gets\mathbf K\mathbf w$
  \State Sample $\mathbf v\sim\pi(\mathbf v\mid\mathbf E_1,\mathbf E_2,\bm\Psi)$
  \State \textbf{return} $(\mathbf w,\mathbf v,\hat{\boldsymbol\xi},\hat{\mathbf Q})$
\EndProcedure
\end{algorithmic}
\end{algorithm}

\subsection{Proper scoring rules}
If the predictive distribution is Gaussian with mean $\mu$ and variance $\sigma^2$, the analytic expression for the CRPS \parencite{Gneiting2007} is
\begin{equation}
\operatorname{CRPS}\bigl(\mathcal{N}(\mu,\sigma^2),x\bigr)=\sigma\Bigl[\tfrac{1}{\sqrt{\pi}}-2\varphi\bigl(\tfrac{x-\mu}{\sigma}\bigr)-\tfrac{x-\mu}{\sigma}\bigl(2\Phi\bigl(\tfrac{x-\mu}{\sigma}\bigr)-1\bigr)\Bigr],
\end{equation}
where $\varphi$ and $\Phi$ denote the PDF and CDF of the standard Gaussian distribution, respectively. Similarly, for a Gaussian predictive distribution the analytic expression for the SCRPS \parencite{Bolin2019} is
\begin{equation}
\operatorname{SCRPS}\bigl(\mathcal{N}(\mu,\sigma^2),x\bigr)=-\sqrt{\pi}\,\varphi(z)-\frac{\sqrt{\pi}\,z}{2}\bigl(2\Phi(z)-1\bigr)-\frac{1}{2}\log\Bigl(\frac{2\sigma}{\sqrt{\pi}}\Bigr),
\end{equation}
where $z=(\mu-x)/\sigma$.
Assume that the random variable \( X \) is a normal–variance mixture with CDF
\[
F(x) = \int \Phi\left(\frac{x - \mu(v)}{\sigma(v)}\right) \, dF_v(v).
\]
Let \( V_j^{(i)} \), \( j = 1, 2 \), \( i = 1, \ldots, N \), be independent draws from the mixing distribution \( F_v \), and define
\( \mu_V = E(X \mid V) \), \( \sigma_V^2 = \operatorname{Var}(X \mid V) \)
% Let $V_j^{(i)},\ j=1,2,\ i=1,\ldots,N$, be independent draws from the mixing distribution $F_v$, and define $\mu_V=E(X\mid V)$, $\sigma_V^2=\operatorname{Var}(X\mid V)$ 
and
\[
M(\mu,\sigma^2)=2\sigma\,\varphi\bigl(\tfrac{\mu}{\sigma}\bigr)+\mu\bigl\{2\Phi\bigl(\tfrac{\mu}{\sigma}\bigr)-1\bigr\}.
\]
Using the delta method, one can show that the Rao--Blackwellized estimator
\begin{equation}\label{eq:SCRPS_proposition}
\begin{split}
\operatorname{SCRPS}_N^{\mathrm{RB}}(F,y)=&\;\frac{\frac{1}{N}\sum_{i=1}^{N}M\bigl(\mu_{V_1^{(i)}}-y,\ \sigma_{V_1^{(i)}}^2\bigr)}{\frac{1}{N}\sum_{i=1}^{N}M\bigl(\mu_{V_1^{(i)}}-\mu_{V_2^{(i)}},\ \sigma_{V_1^{(i)}}^2+\sigma_{V_2^{(i)}}^2\bigr)}\\[1ex]
&+\frac{1}{2}\log\Biggl(\frac{1}{N}\sum_{i=1}^{N}M\bigl(\mu_{V_1^{(i)}}-\mu_{V_2^{(i)}},\ \sigma_{V_1^{(i)}}^2+\sigma_{V_2^{(i)}}^2\bigr)\Biggr)
\end{split}
\end{equation}
is asymptotically more efficient than the standard SCRPS estimator. Consequently, we use $\operatorname{SCRPS}_N^{\mathrm{RB}}$ in this paper.

\subsection{Proof of Proposition 2.1 (Pearson correlation)}
\begin{proof}
Let $d=2$ and $\alpha_1=\alpha_2=2$, so that $\nu_i=\alpha_i-d/2=1$ for $i=1,2$. For the bivariate Mat\'ern–SPDE field the cross‑covariance of $u_1$ and $u_2$ (with $i\neq j$) is
\[
\operatorname{Cov}\{u_1(\mathbf{s}),u_2(\mathbf{t})\}=\frac{\rho}{c_1c_2\sqrt{1+\rho^{2}}}\,
   \mathcal{F}^{-1}\!\Bigl\{\frac{1}{(2\pi)^2(\kappa_1^{2}+\|\mathbf{k}\|^{2})(\kappa_2^{2}+\|\mathbf{k}\|^{2})}\Bigr\}\bigl(\|\mathbf{s}-\mathbf{t}\|\bigr),
\]
where $c_i=\sqrt{\Gamma(\nu_i)\bigl[(4\pi)^{d/2}\kappa_i^{2\nu_i}\sigma_i^{2}\Gamma(\alpha_i)\bigr]^{-1}}$. With $d=2$ and $\nu_i=1$ we obtain $c_1c_2=(4\pi\sigma_1\sigma_2\kappa_1\kappa_2)^{-1}$.

\paragraph{Case $\boldsymbol{\kappa_1\neq\kappa_2}$.}
At zero lag ($\mathbf{h}=\mathbf{s}-\mathbf{t}=0$) the inverse Fourier transform reduces to
\[\frac{1}{2\pi}\int_{\mathbb{R}^{2}}\frac{d\mathbf{k}}{(\kappa_1^{2}+\|\mathbf{k}\|^{2})(\kappa_2^{2}+\|\mathbf{k}\|^{2})}=\frac{1}{2\pi}\frac{\ln(\kappa_1/\kappa_2)}{\kappa_1^{2}-\kappa_2^{2}}.\]
Hence
\[\operatorname{Cov}\{u_1,u_2\}(0)=\frac{2\rho\,\sigma_1\sigma_2\kappa_1\kappa_2}{\sqrt{1+\rho^{2}}}\;\frac{\ln(\kappa_1/\kappa_2)}{\kappa_1^{2}-\kappa_2^{2}},\qquad
\kappa_1\neq\kappa_2.\]
Dividing by $\sigma_1\sigma_2$ gives the Pearson correlation at zero lag
\[\rho_{u_1,u_2}(0)=\frac{2\rho}{\sqrt{1+\rho^{2}}}\,\frac{\kappa_1\kappa_2\,\ln(\kappa_1/\kappa_2)}{\kappa_1^{2}-\kappa_2^{2}},\qquad
\kappa_1\neq\kappa_2.\]

\paragraph{Case $\boldsymbol{\kappa_1=\kappa_2\equiv\kappa}$.}
Both the numerator and the denominator in the expression above vanish
as $\kappa_2\to\kappa_1$, so we take the limit:
\[
\lim_{\kappa_2\to\kappa_1}
\frac{\kappa_1\kappa_2\,\ln(\kappa_1/\kappa_2)}
     {\kappa_1^{2}-\kappa_2^{2}}
  \;=\;
\frac12
\quad\text{(l’Hôpital’s rule)}.
\]
Substituting this limit gives
\[
\rho_{u_1,u_2}(0)
  =\frac{2\rho}{\sqrt{1+\rho^{2}}}\times\frac12
  =\frac{\rho}{\sqrt{1+\rho^{2}}},
  \qquad\kappa_1=\kappa_2.
\]
Combining the two cases completes the proof.
\end{proof}

\newpage

\section{Main results}
In this section, we show the global parameter and CV maps, QQ plots, and the simulation study results. The global parameter maps are presented in the order $\{\rho_{\varepsilon},\ \rho,\ \kappa_{1:2},\ \sigma_{1:2},\ \sigma_{\varepsilon_{1:2}},\ \eta,\ \mu\}$. 
At each depth (10, 300, and 1000 dbar) the panels are organized as follows: column 1 shows results under the independent-noise specification, column 2 shows the correlated-noise specification, and column 3 plots their difference (correlated – independent). 

\subsection{Parameter estimates}
% \begin{figure}[H]
%     \centering
%     \begin{subfigure}{\textwidth}
%         \centering
%          \includegraphics[width=\linewidth,trim=0cm 0.5cm 0cm 0cm,clip]{Figures/10/10_correlation_combined.png}
%         \caption{10 dbar}
%     \end{subfigure}
%     \begin{subfigure}{\textwidth}
%         \centering
%          \includegraphics[width=\linewidth,trim=0cm 0.5cm 0cm 0cm,clip]{Figures/300/300_correlation_combined.png}
%         \caption{300 dbar}
%     \end{subfigure}
%     \begin{subfigure}{\textwidth}
%         \centering
%         \includegraphics[width=\linewidth,trim=0cm 0.5cm 0cm 0cm,clip]{Figures/1000/1000_correlation_combined.png}
%         \caption{1000 dbar}
%     \end{subfigure}
%     \caption{Correlation, \( \rho_{u_1,u_2} \), between two latent fields (a) 10 dbar and (b) 300 dbar, (c) 1000 dbar. The difference is calculated as the correlated model - independent model values. Figure (b) is the same as Figure 5 (a) in the main paper.}
%     \label{fig:correlation-10-300}%\label{fig:correlation-1000}
% \end{figure}

\begin{figure}[H]
    \centering
    \begin{subfigure}[b]{\textwidth}
        \centering
        \includegraphics[width=\linewidth]{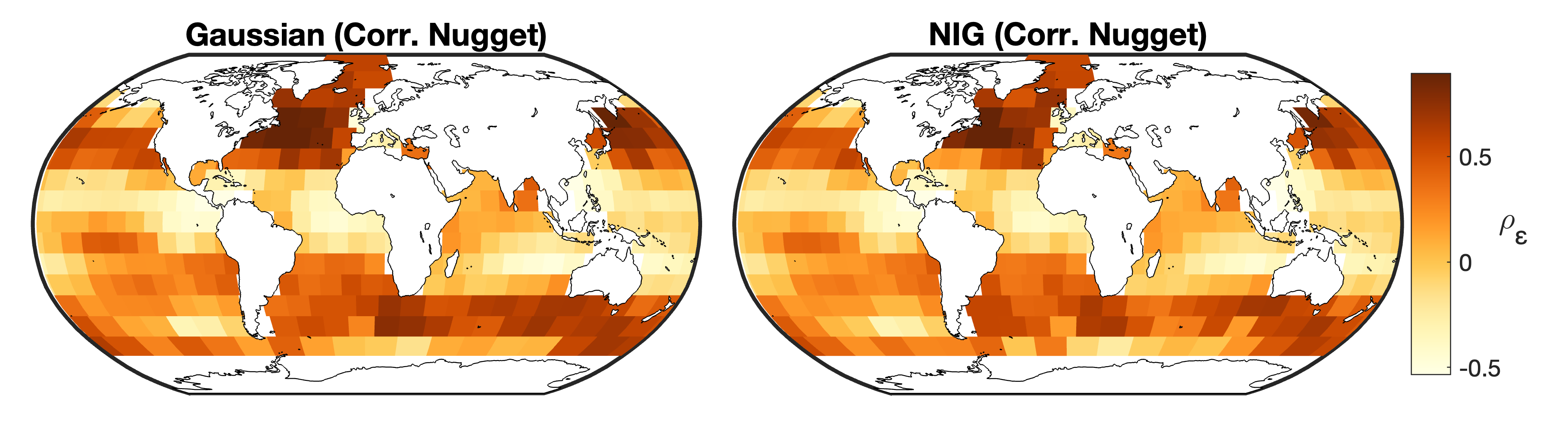}
        \caption{10 dbar}
    \end{subfigure}
    \hfill \vspace{5mm} 
    \begin{subfigure}[b]{\linewidth}
        \centering
        \includegraphics[width=\linewidth]{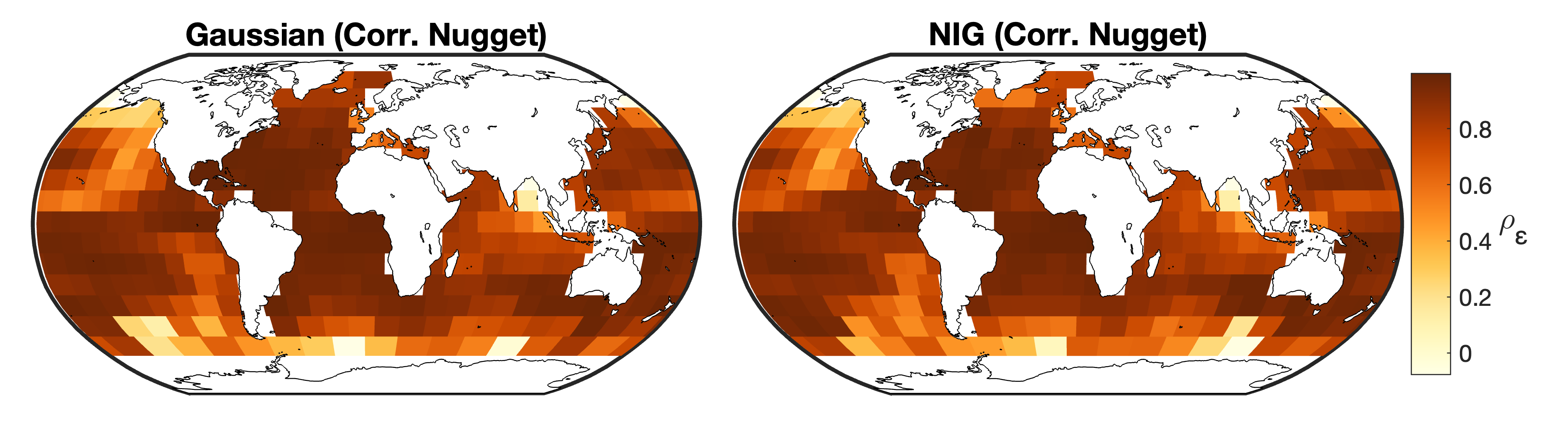}
        \caption{300 dbar}
    \end{subfigure}   
    \hfill
    \vspace{5mm} 
    \begin{subfigure}[b]{\linewidth}
        \centering
        \includegraphics[width=\linewidth]{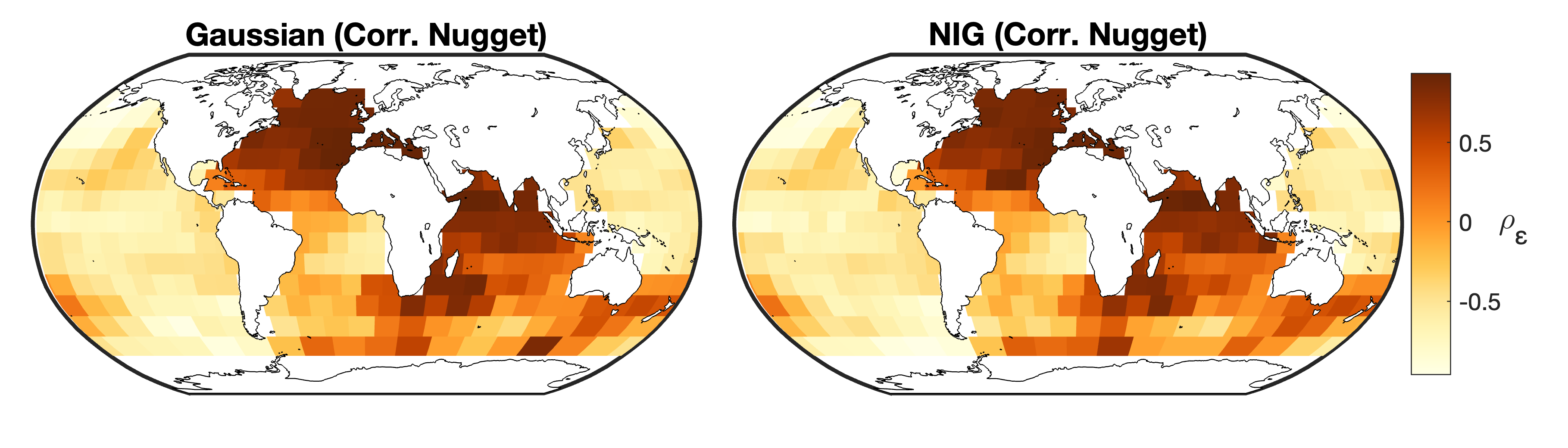}
        \caption{1000 dbar}
    \end{subfigure}   
    \caption{Measurement noise correlation, $\rho_\varepsilon$. Figure (b) is the same as Figure 5 (b) in the main paper.}
    \label{fig:rho_e-10-300}
\end{figure}

\newpage

\begin{figure}[H]
    \centering
    \begin{subfigure}[b]{\textwidth}
        \centering
        \includegraphics[width=\linewidth,trim=0cm 0.5cm 0cm 0cm,clip]{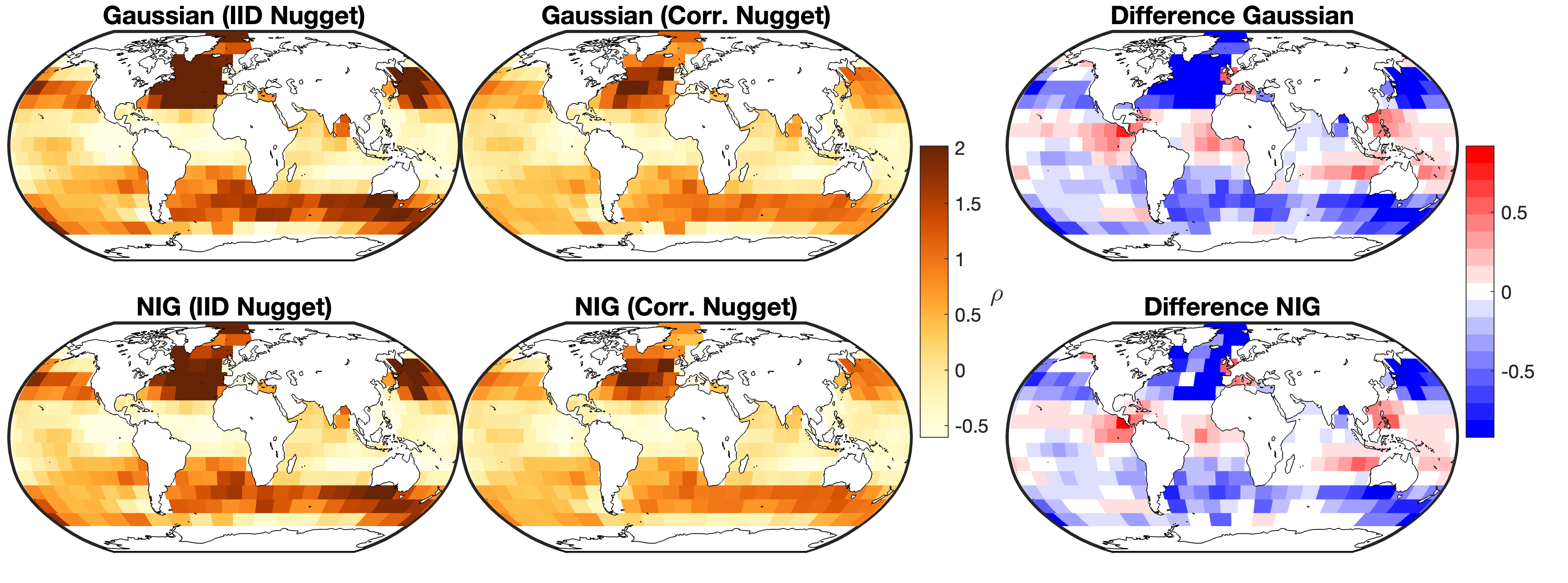}
        \caption{10 dbar}
    \end{subfigure}
     \hfill \vspace{5mm} 
    \begin{subfigure}[b]{\linewidth}
        \centering
        \includegraphics[width=\linewidth,trim=0cm 0.5cm 0cm 0cm,clip]{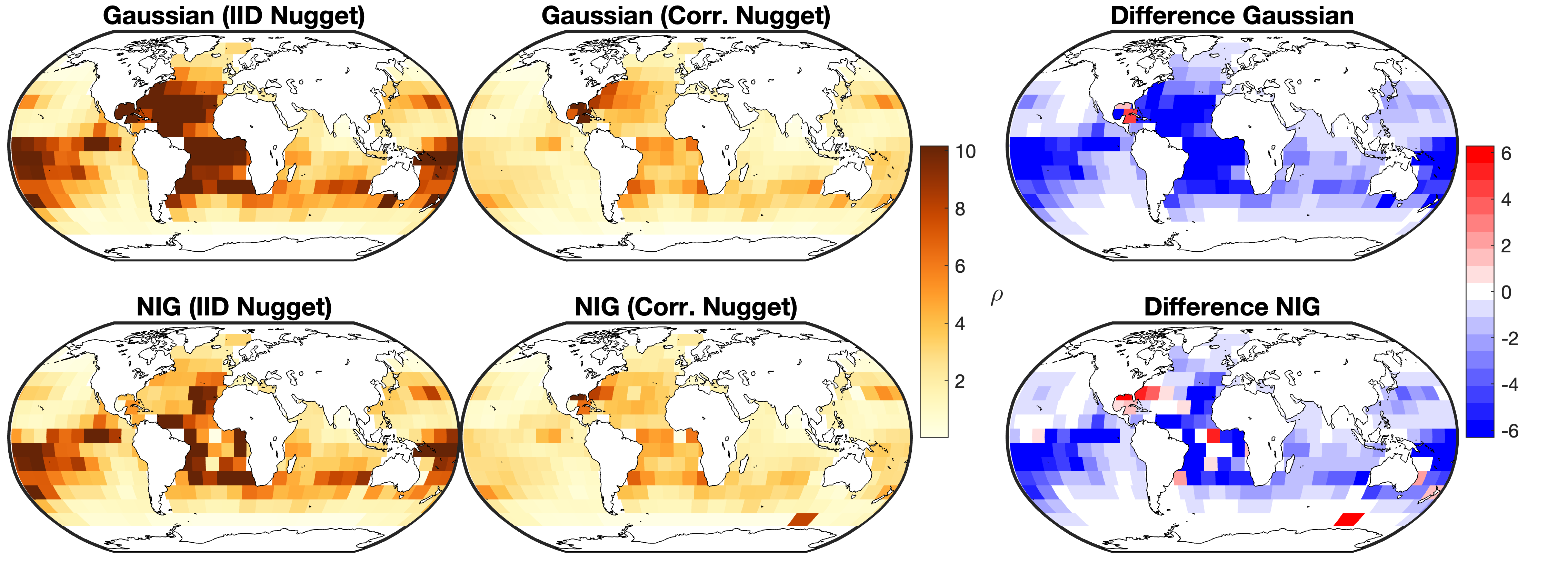}
        \caption{300 dbar}
    \end{subfigure}   
     \hfill \vspace{5mm} 
    \begin{subfigure}[b]{\linewidth}
        \centering
        \includegraphics[width=\linewidth,trim=0cm 0.5cm 0cm 0cm,clip]{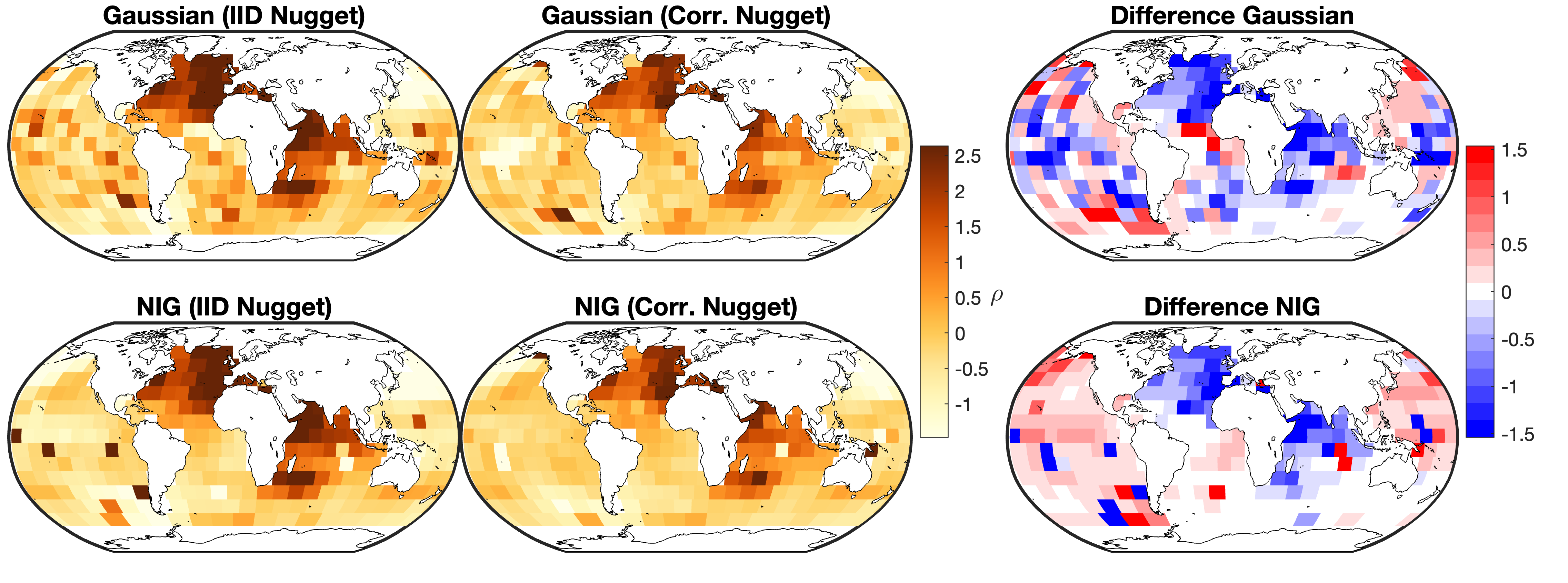}
        \caption{1000 dbar}
    \end{subfigure}   
    \caption{$\rho$}
\end{figure}

\newpage
\begin{figure}[H]
    \centering
    \begin{subfigure}[b]{\textwidth}
        \centering
        \includegraphics[width=\linewidth,trim=0cm 0.5cm 0cm 0cm,clip]{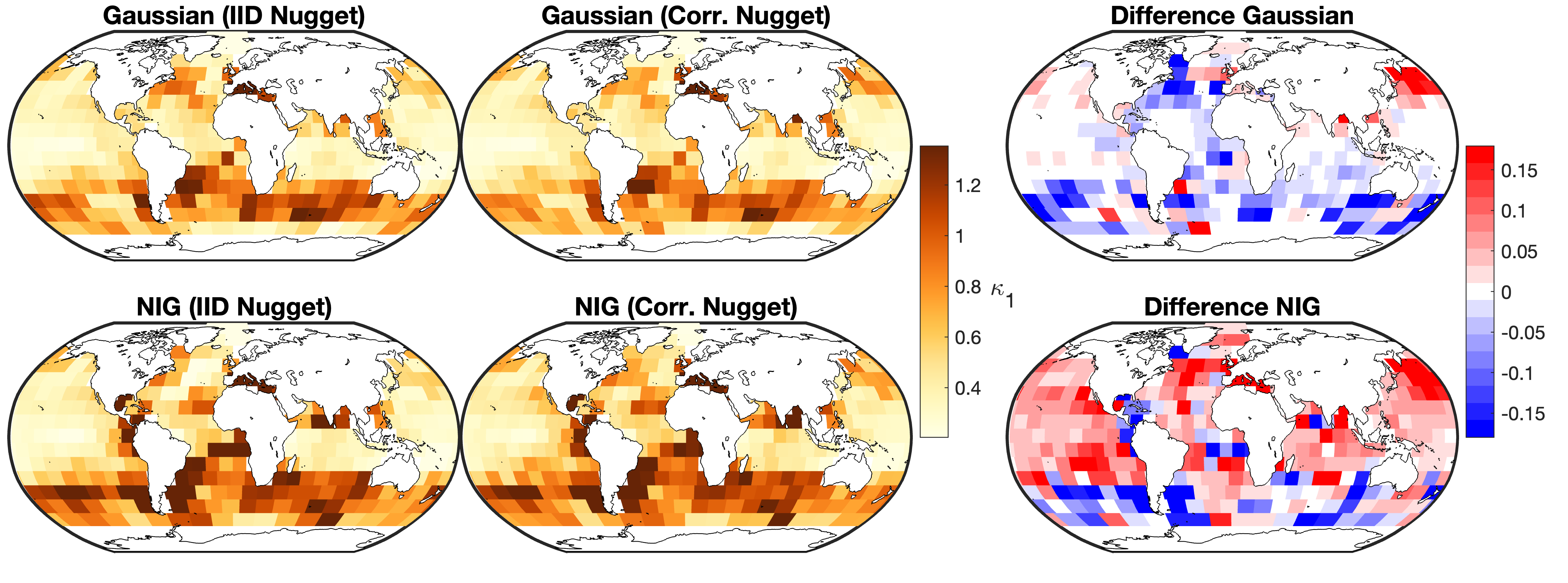}
        \caption{10 dbar}
    \end{subfigure}
     \hfill \vspace{5mm} 
    \begin{subfigure}[b]{\linewidth}
        \centering
        \includegraphics[width=\linewidth,trim=0cm 0.5cm 0cm 0cm,clip]{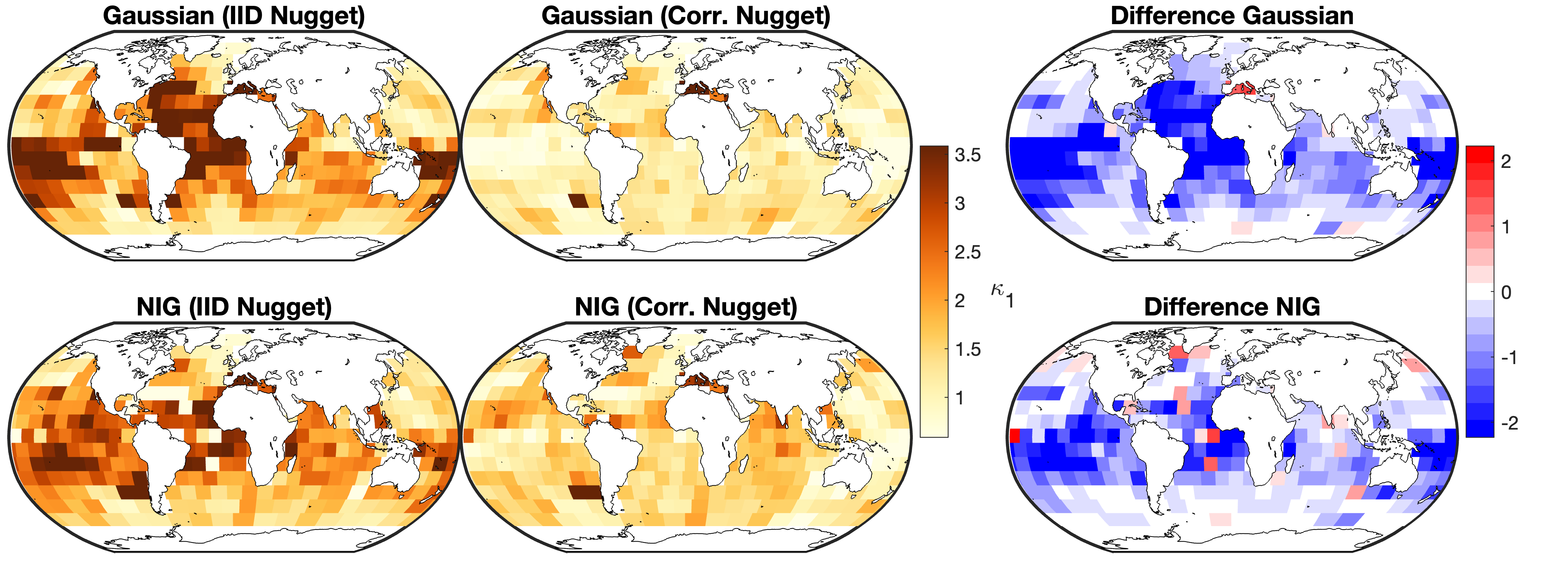}
        \caption{300 dbar}
    \end{subfigure}   
     \hfill \vspace{5mm} 
    \begin{subfigure}[b]{\linewidth}
        \centering
        \includegraphics[width=\linewidth,trim=0cm 0.5cm 0cm 0cm,clip]{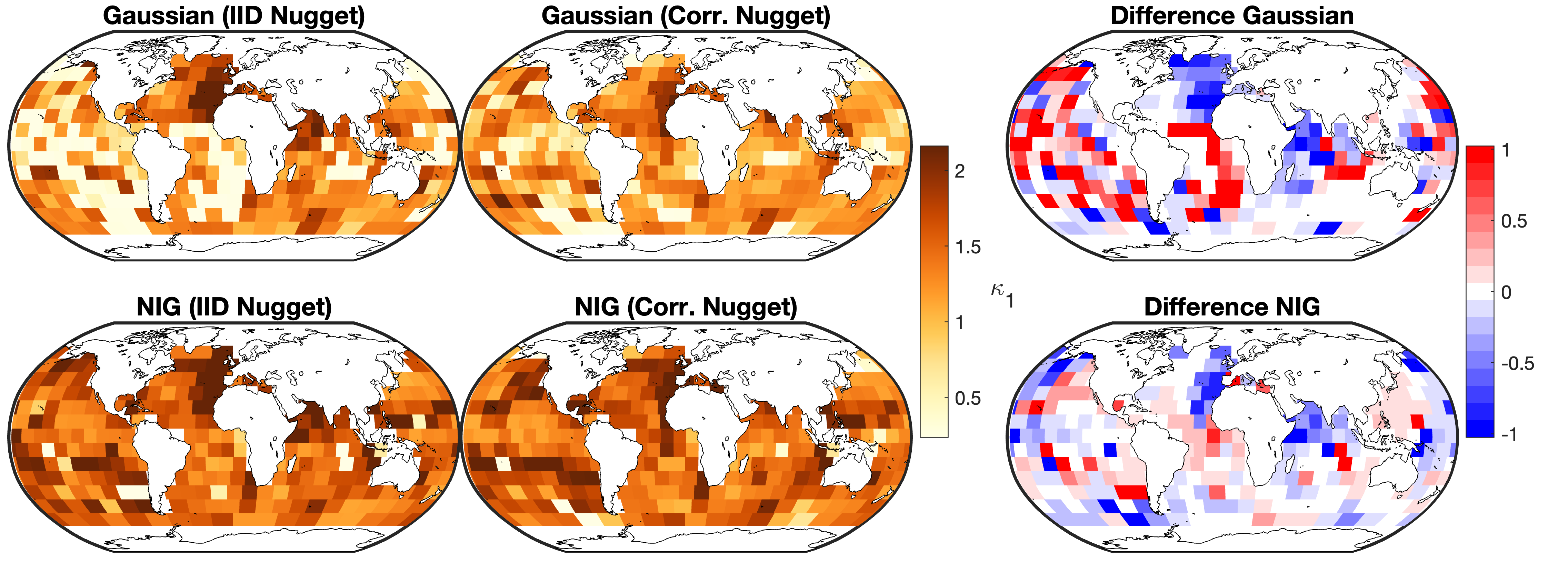}
        \caption{1000 dbar}
    \end{subfigure}   
    \caption{$\kappa_1$}
\end{figure}

\begin{figure}[H]
    \centering
    \begin{subfigure}[b]{\textwidth}
        \centering
        \includegraphics[width=\linewidth,trim=0cm 0.5cm 0cm 0cm,clip]{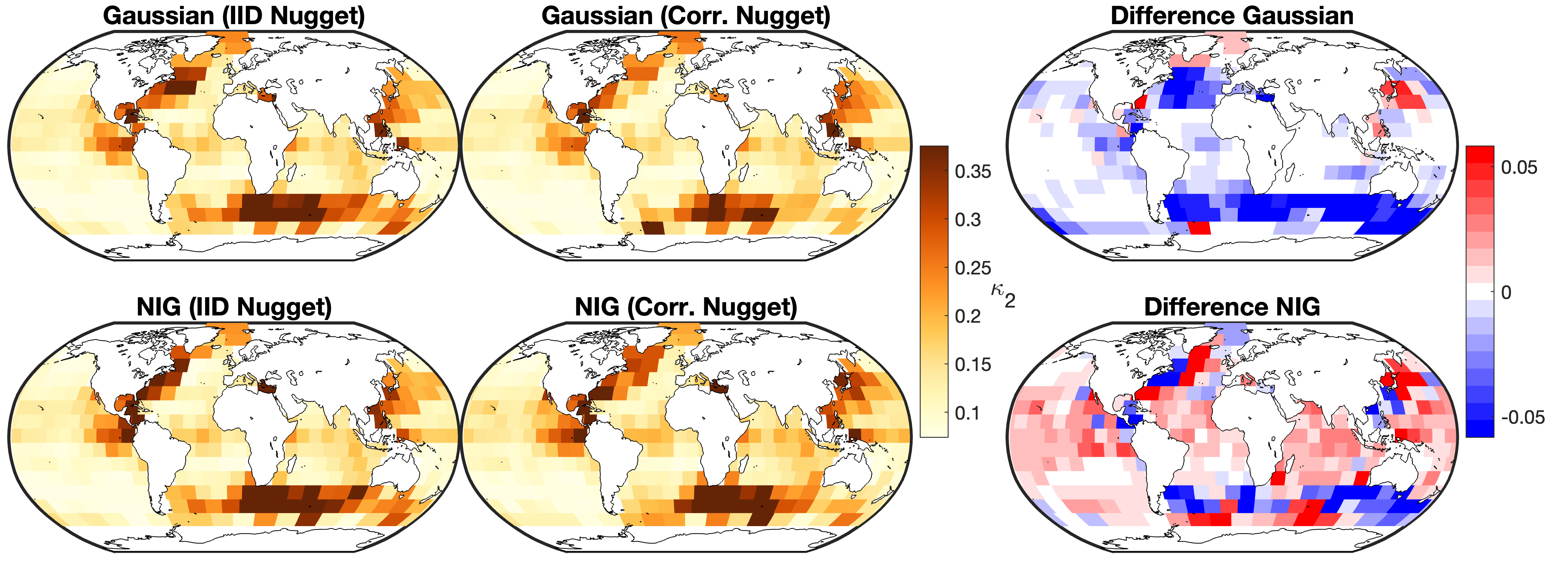}
        \caption{10 dbar}
    \end{subfigure}
     \hfill \vspace{5mm} 
    \begin{subfigure}[b]{\linewidth}
        \centering
        \includegraphics[width=\linewidth,trim=0cm 0.5cm 0cm 0cm,clip]{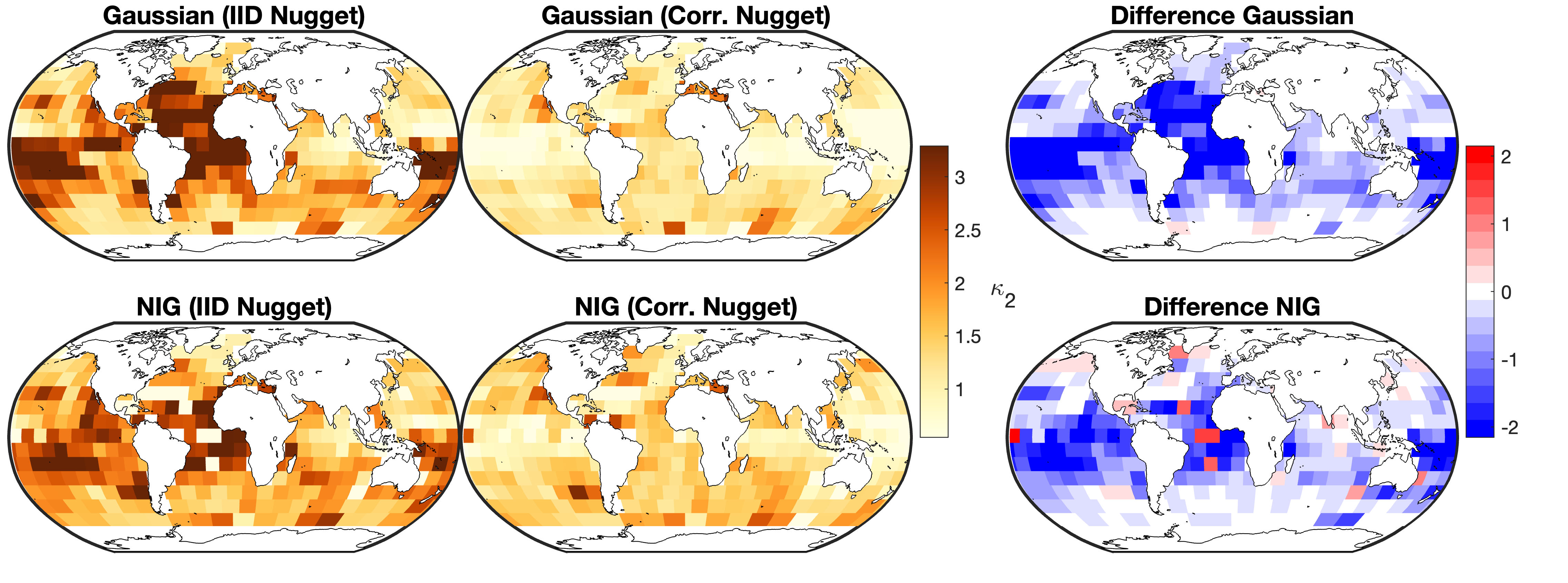}
        \caption{300 dbar}
    \end{subfigure}   
     \hfill \vspace{5mm} 
    \begin{subfigure}[b]{\linewidth}
        \centering
        \includegraphics[width=\linewidth,trim=0cm 0.5cm 0cm 0cm,clip]{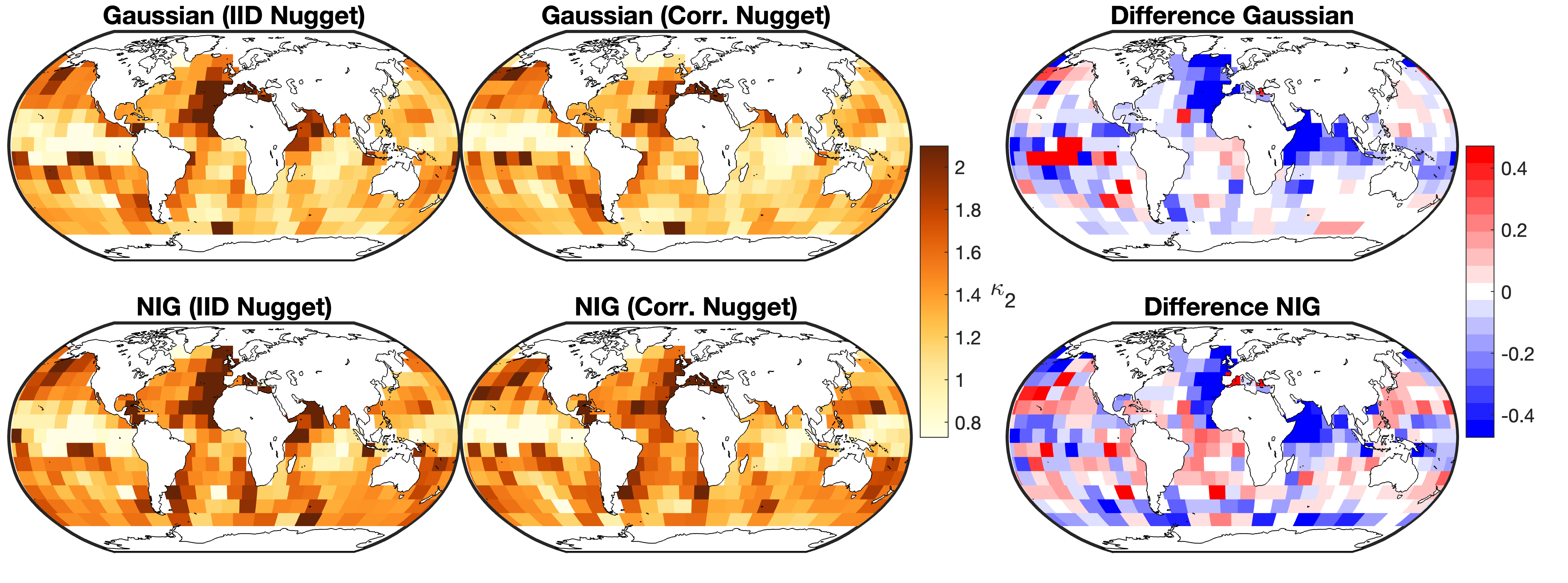}
        \caption{1000 dbar}
    \end{subfigure}   
    \caption{$\kappa_2$}
    \label{fig:kappa3}
\end{figure}

\newpage
\begin{figure}[H]
    \centering
    \begin{subfigure}[b]{\textwidth}
        \centering
        \includegraphics[width=\linewidth,trim=0cm 0.5cm 0cm 0cm,clip]{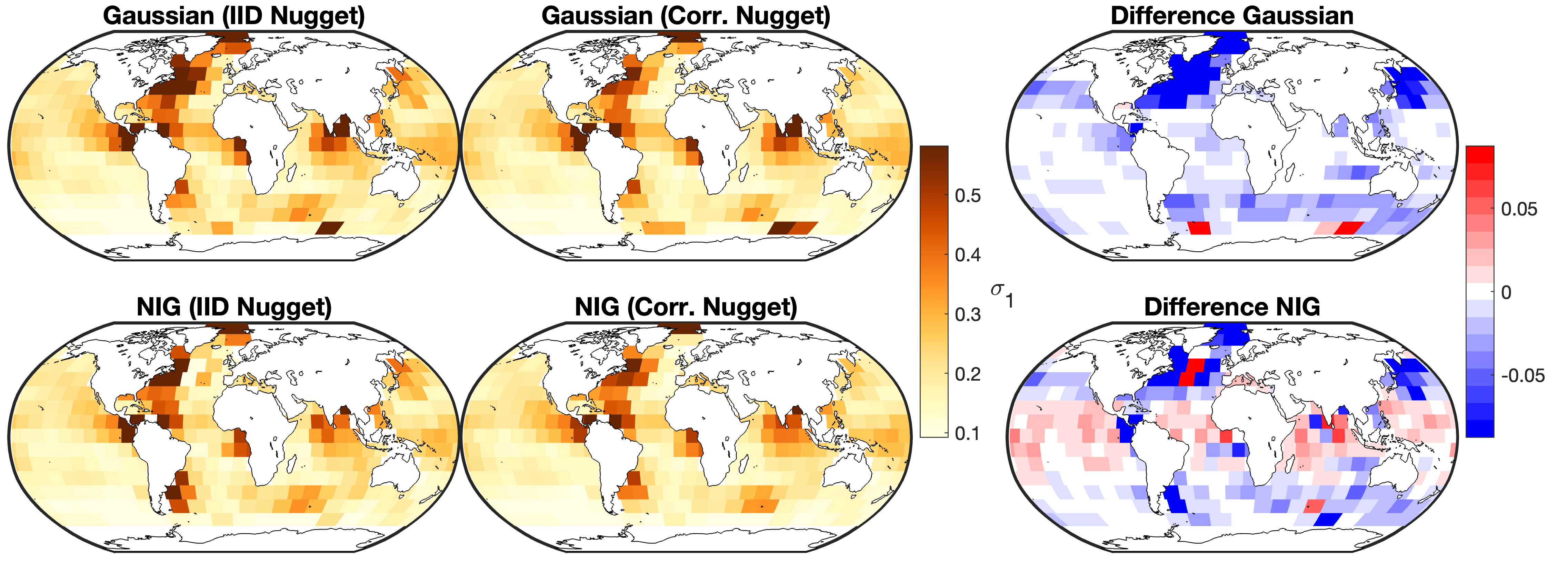}
        \caption{10 dbar}
    \end{subfigure}
     \hfill \vspace{5mm} 
    \begin{subfigure}[b]{\linewidth}
        \centering
        \includegraphics[width=\linewidth,trim=0cm 0.5cm 0cm 0cm,clip]{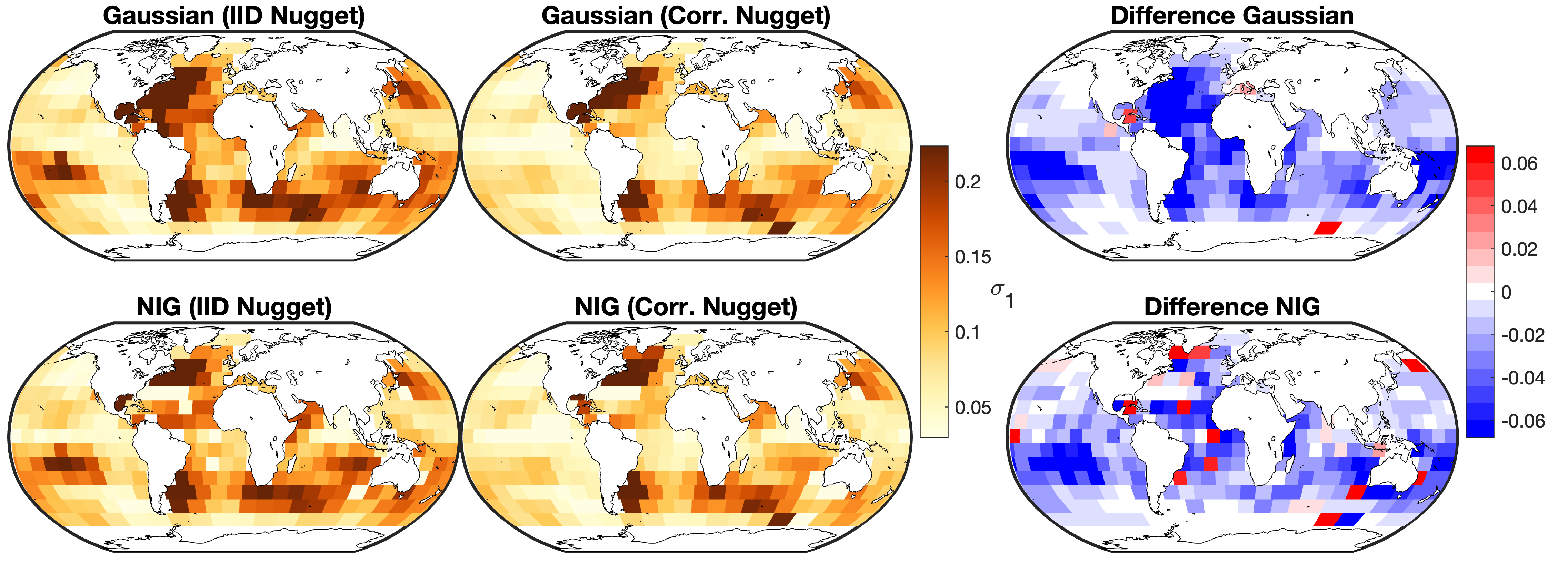}
        \caption{300 dbar}
    \end{subfigure}   
     \hfill \vspace{5mm} 
    \begin{subfigure}[b]{\linewidth}
        \centering
        \includegraphics[width=\linewidth,trim=0cm 0.5cm 0cm 0cm,clip]{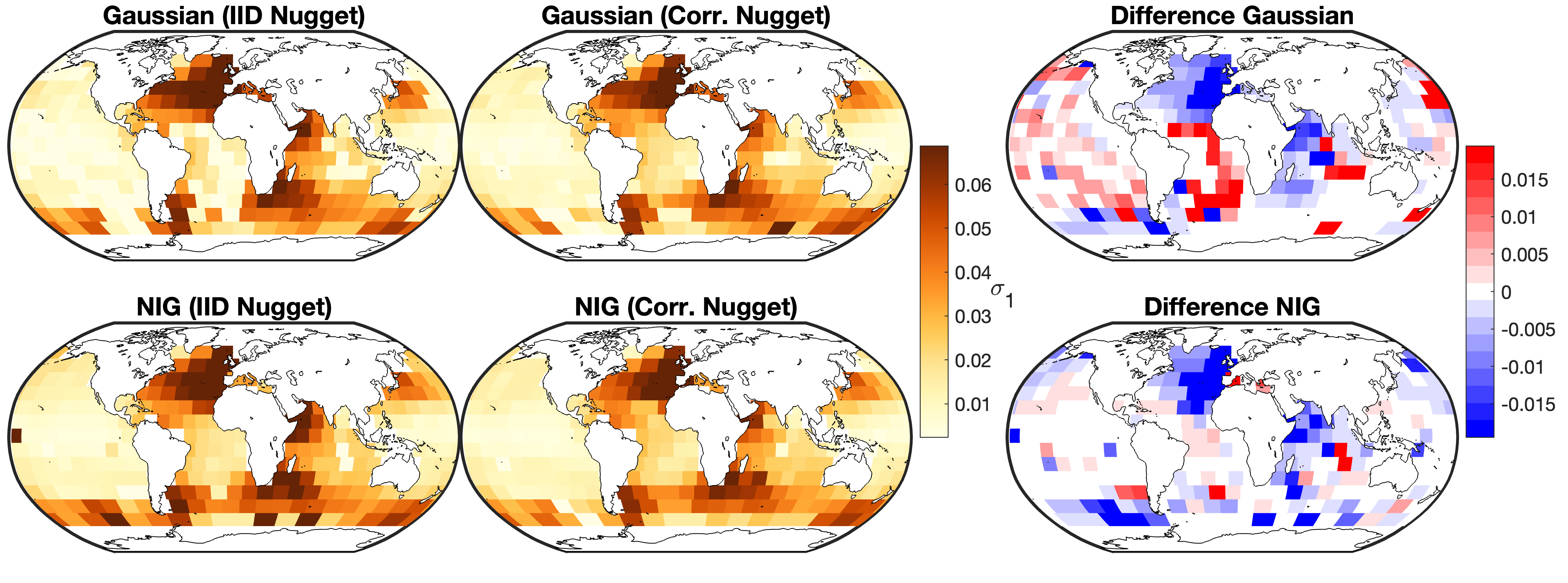}
        \caption{1000 dbar}
    \end{subfigure}   
    \caption{$\sigma_1$}
    \label{fig:sigma1}
\end{figure}

\begin{figure}[H]
    \centering
    \begin{subfigure}[b]{\textwidth}
        \centering
        \includegraphics[width=\linewidth,trim=0cm 0.5cm 0cm 0cm,clip]{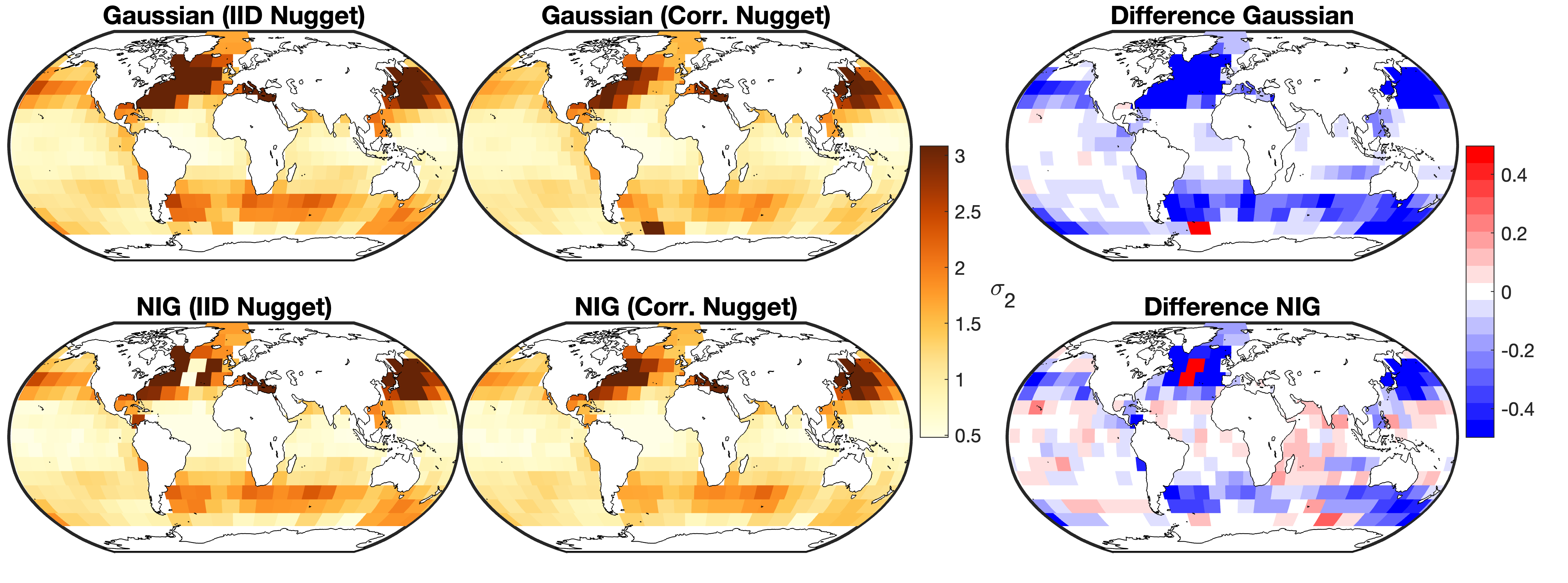}
        \caption{10 dbar}
    \end{subfigure}
     \hfill \vspace{5mm} 
    \begin{subfigure}[b]{\linewidth}
        \centering
        \includegraphics[width=\linewidth,trim=0cm 0.5cm 0cm 0cm,clip]{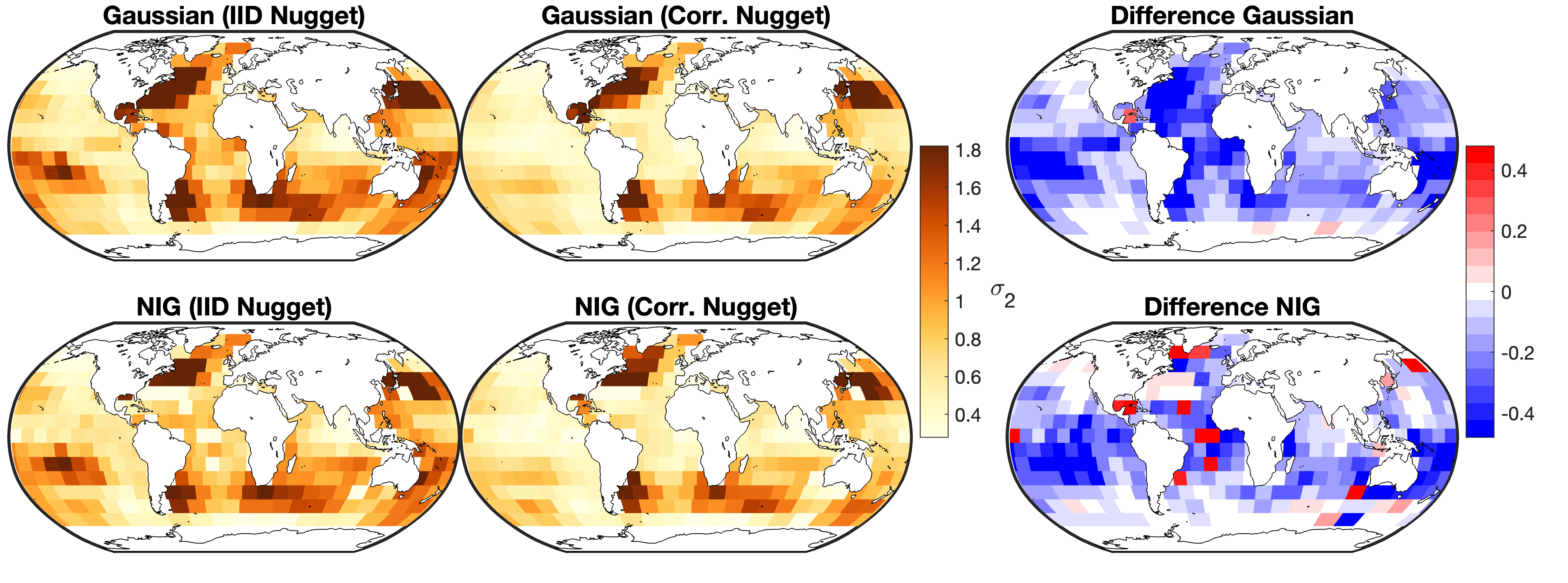}
        \caption{300 dbar}
    \end{subfigure}   
     \hfill \vspace{5mm} 
    \begin{subfigure}[b]{\linewidth}
        \centering
        \includegraphics[width=\linewidth,trim=0cm 0.5cm 0cm 0cm,clip]{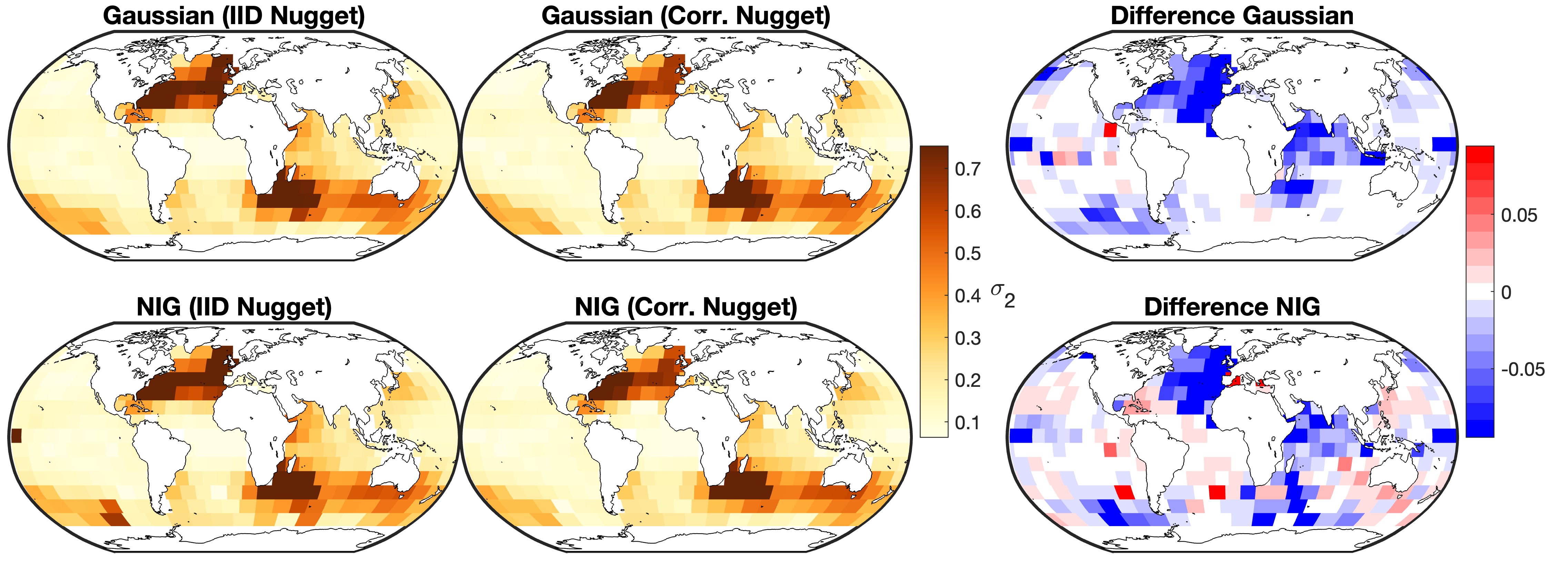}
        \caption{1000 dbar}
    \end{subfigure}   
    \caption{$\sigma_2$}
    \label{fig:sigma2}
\end{figure}

\newpage
\begin{figure}[H]
    \centering
    \begin{subfigure}[b]{\textwidth}
        \centering
        \includegraphics[width=\linewidth,trim=0cm 0.5cm 0cm  0cm,clip]{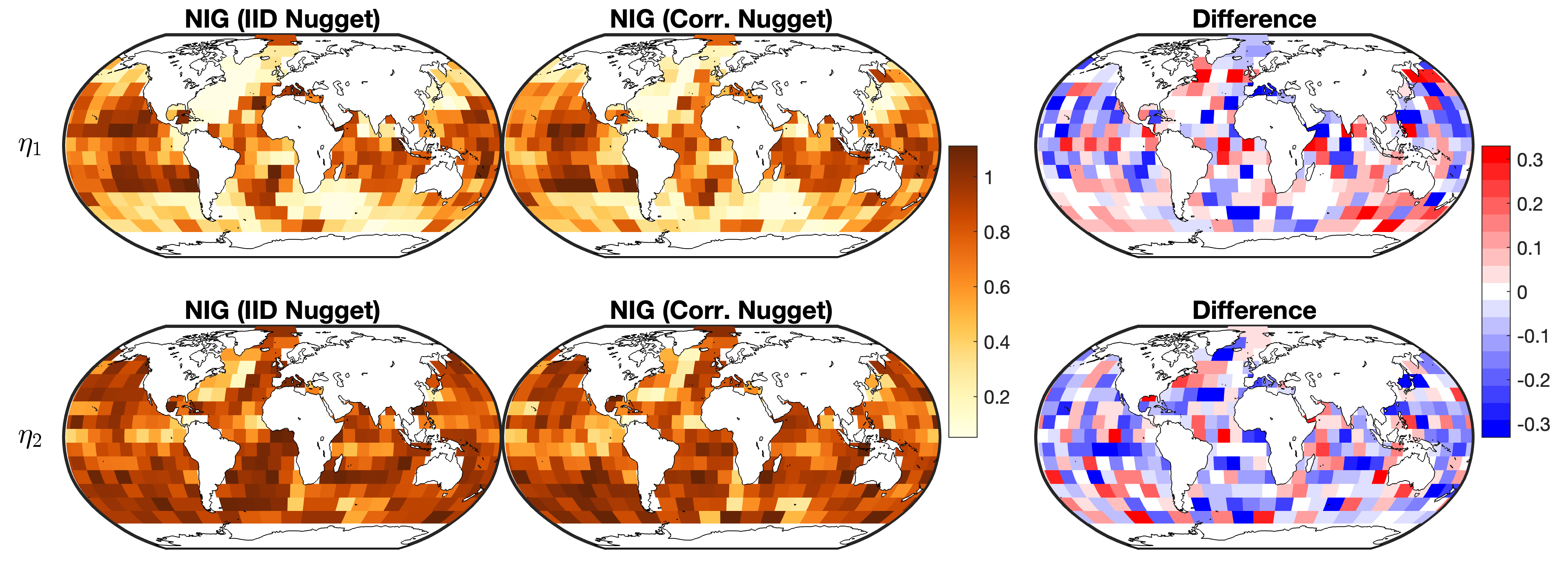}
        \caption{10 dbar}
    \end{subfigure}
    \hfill
     \vspace{10mm} 
    \begin{subfigure}[b]{\linewidth}
        \centering
        \includegraphics[width=\linewidth,trim=0cm 0.5cm 0cm  0cm,clip]{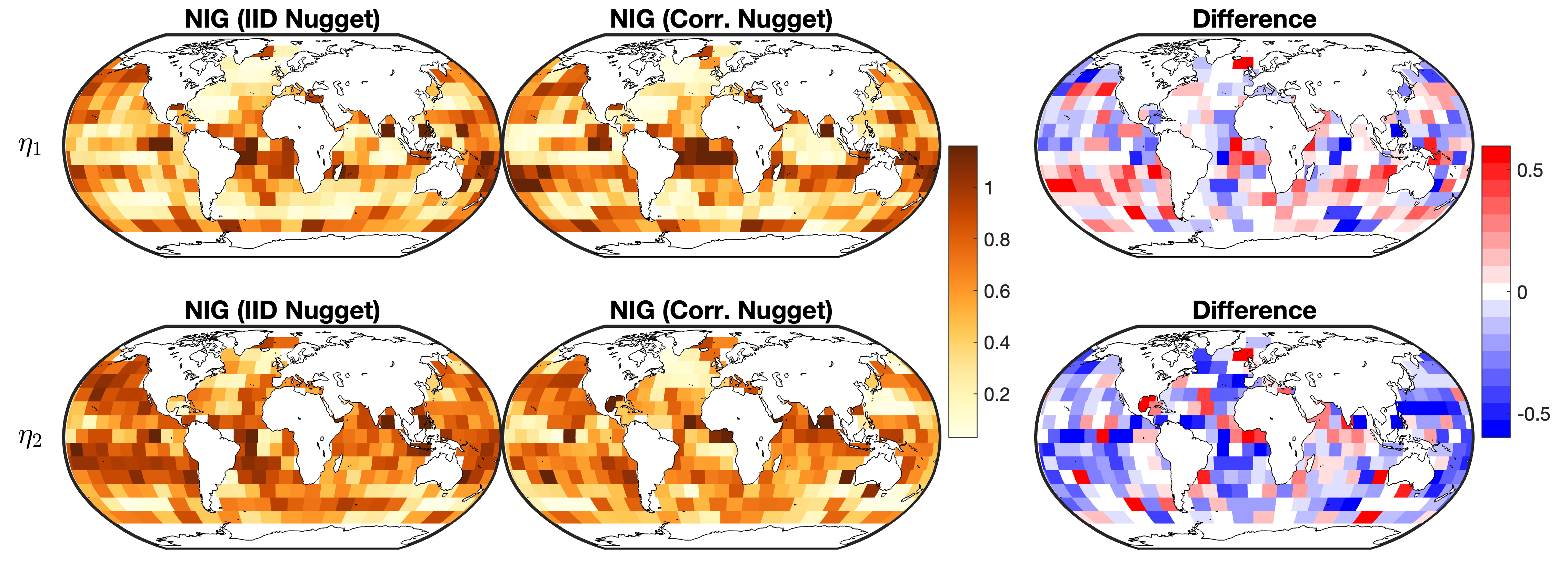}
        \caption{300 dbar}
    \end{subfigure}   
    \hfill
     \vspace{10mm} 
    \begin{subfigure}[b]{\textwidth}
        \centering
        \includegraphics[width=\linewidth,trim=0cm 0.5cm 0cm  0cm,clip]{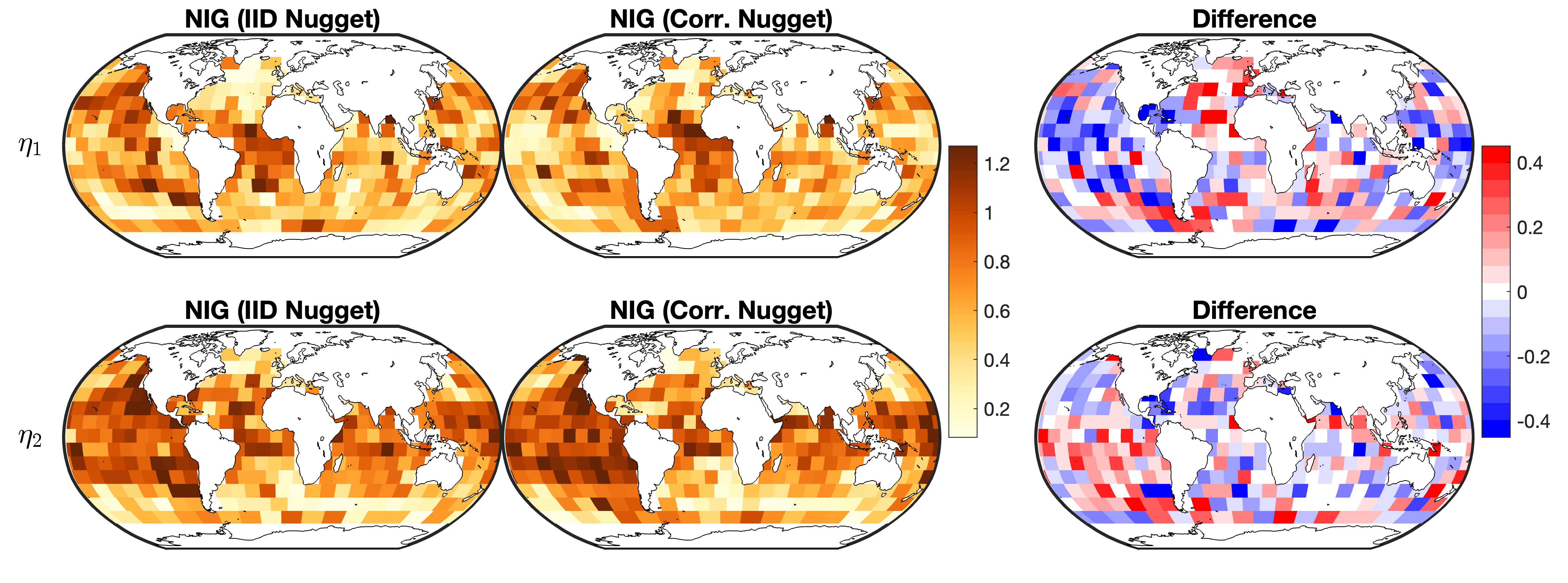}
        \caption{1000 dbar}
    \end{subfigure}
      \caption{$\eta$}
    \label{fig:nig_cor_nu}
\end{figure}

\newpage
\begin{figure}[H]
    \centering
    \begin{subfigure}[b]{\linewidth}
        \centering
         \includegraphics[width=\linewidth,trim=0cm 0.5cm 0cm  0cm,clip]{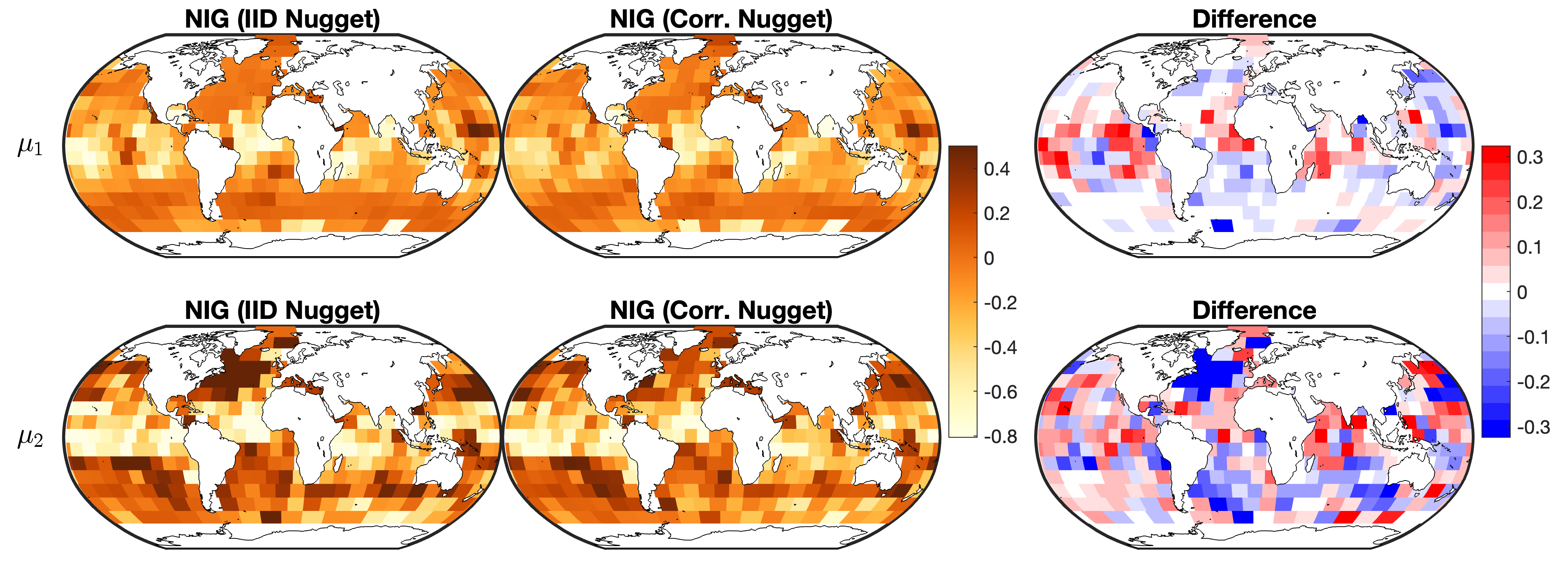}
        \caption{10 dbar}
    \end{subfigure}
    \hfill
    \vspace{10mm} 
    \begin{subfigure}[b]{\linewidth}
        \centering
         \includegraphics[width=\linewidth,trim=0cm 0.5cm 0cm  0cm,clip]{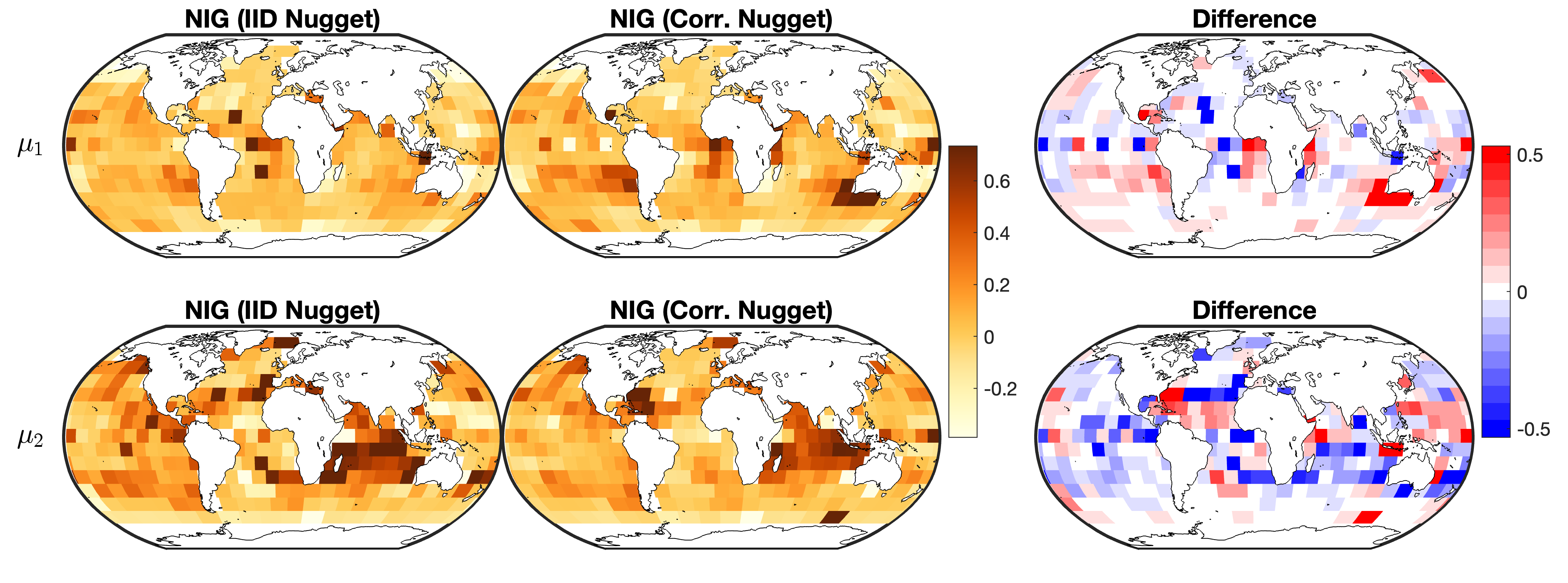}
        \caption{300 dbar}
    \end{subfigure}   
\hfill
\vspace{10mm} 
    \begin{subfigure}[b]{\textwidth}
        \centering
        \includegraphics[width=\linewidth,trim=0cm 0.5cm 0cm  0cm,clip]{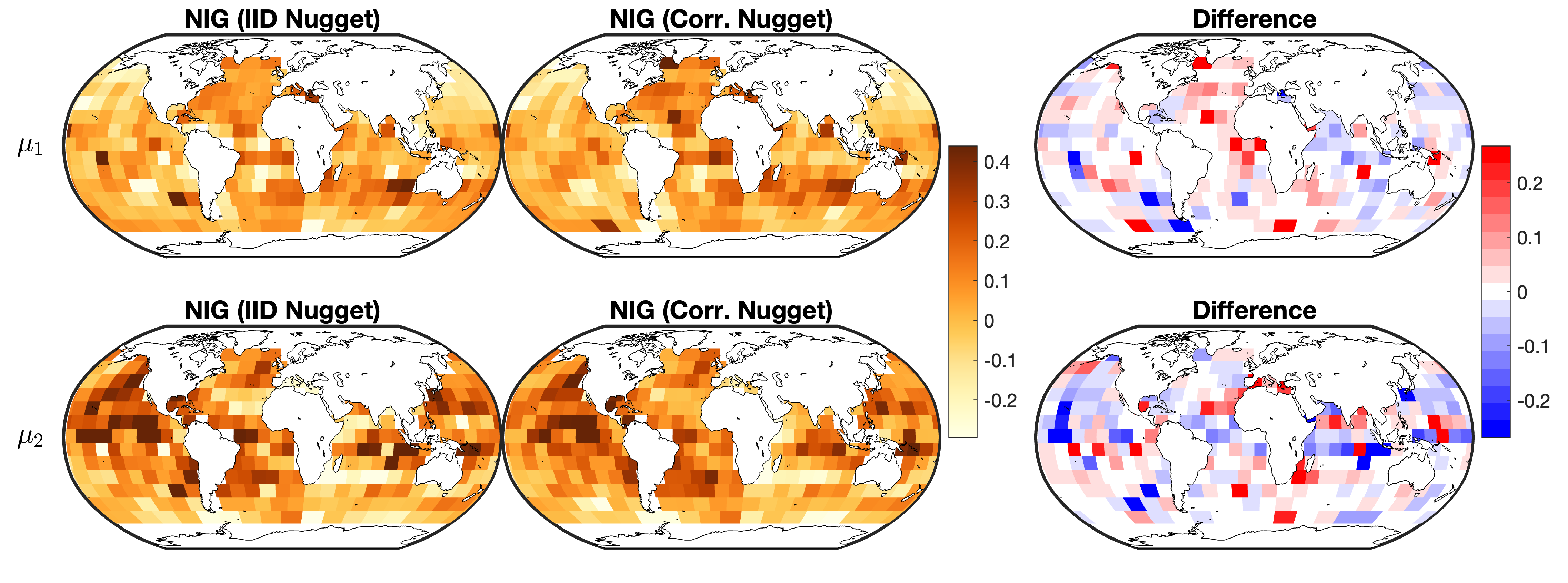}
        \caption{1000 dbar}
    \end{subfigure}
    \caption{$\mu$}
\end{figure}

\newpage
\subsection{CV results}
This section shows the best-performing model at each grid point according to all evaluated metrics.

\begin{figure}[H]
    \centering
    \begin{subfigure}[b]{\linewidth}
        \centering
         \includegraphics[width=\linewidth,trim=0cm 0cm 0cm 0cm,clip]{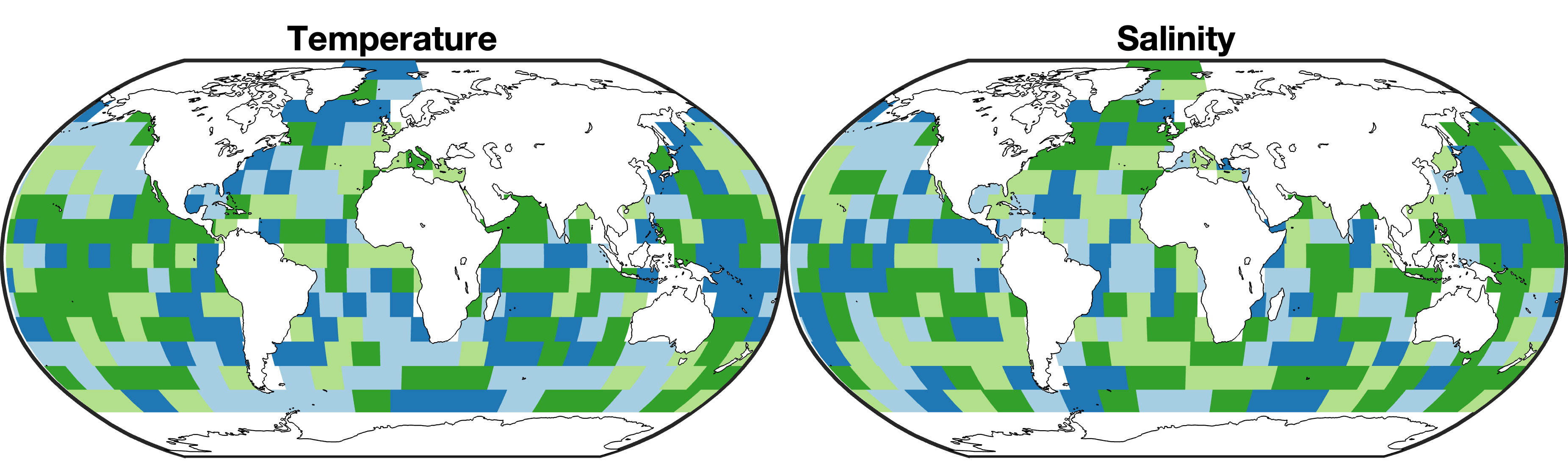}
        \caption{MAE}
    \end{subfigure}
    \hfill
    \begin{subfigure}[b]{\linewidth}
        \centering
         \includegraphics[width=\linewidth,trim=0cm 0cm 0cm 0cm,clip]{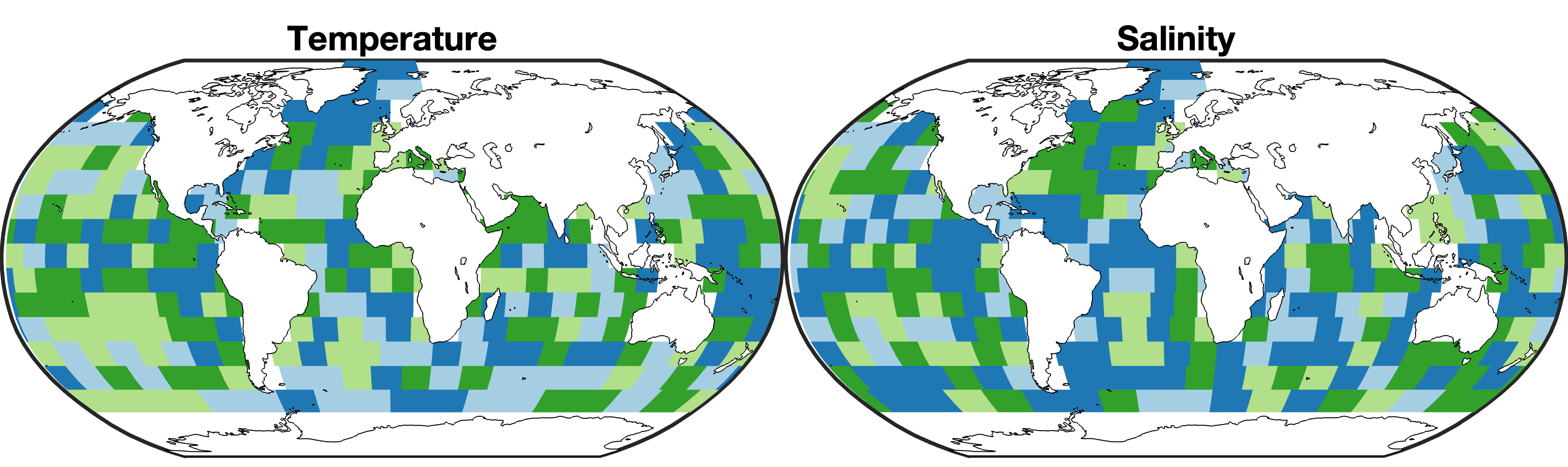}
        \caption{RMSE}
    \end{subfigure}   
\hfill 
    \begin{subfigure}[b]{\textwidth}
        \centering
        \includegraphics[width=\linewidth,trim=0cm 0cm 0cm 0cm,clip]{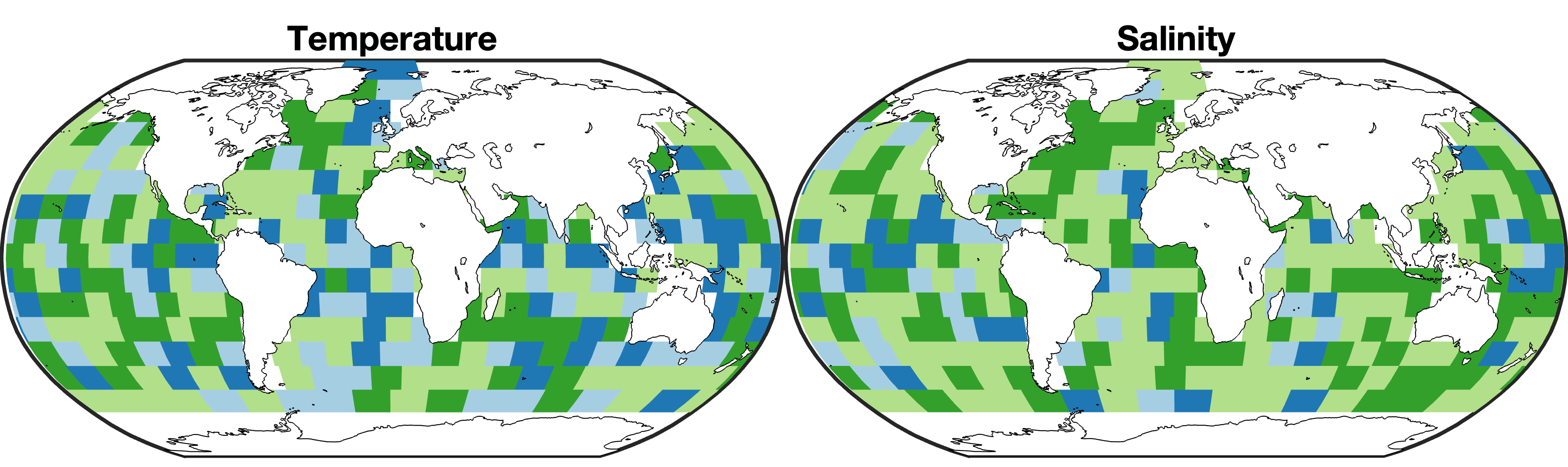}
        \caption{CRPS}
    \end{subfigure}
\begin{subfigure}[b]{\textwidth}
        \centering
        \includegraphics[width=\linewidth,trim=0cm 0cm 0cm 1.5cm,clip]{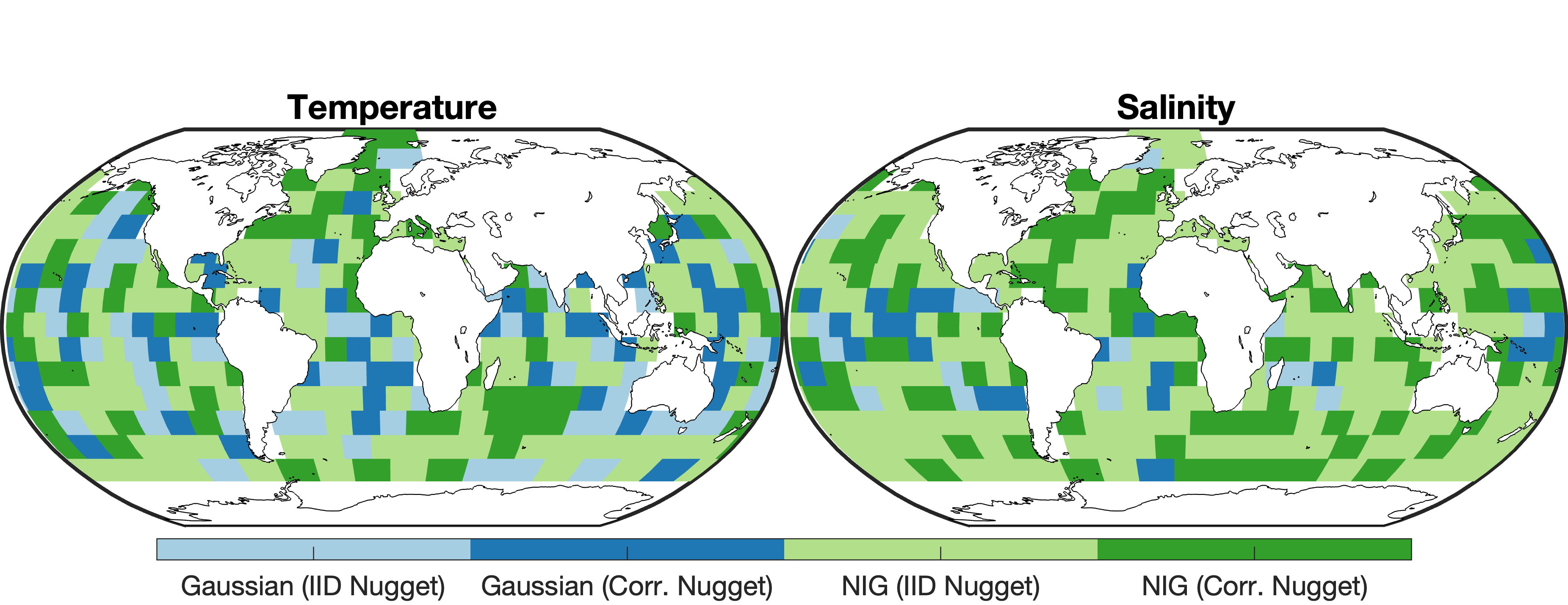}
        \caption{SCRPS}
    \end{subfigure}
    \caption{10 dbar}
\end{figure}

\begin{figure}[H]
    \centering
    \begin{subfigure}[b]{\linewidth}
        \centering
         \includegraphics[width=\linewidth,trim=0cm 0cm 0cm  0cm,clip]{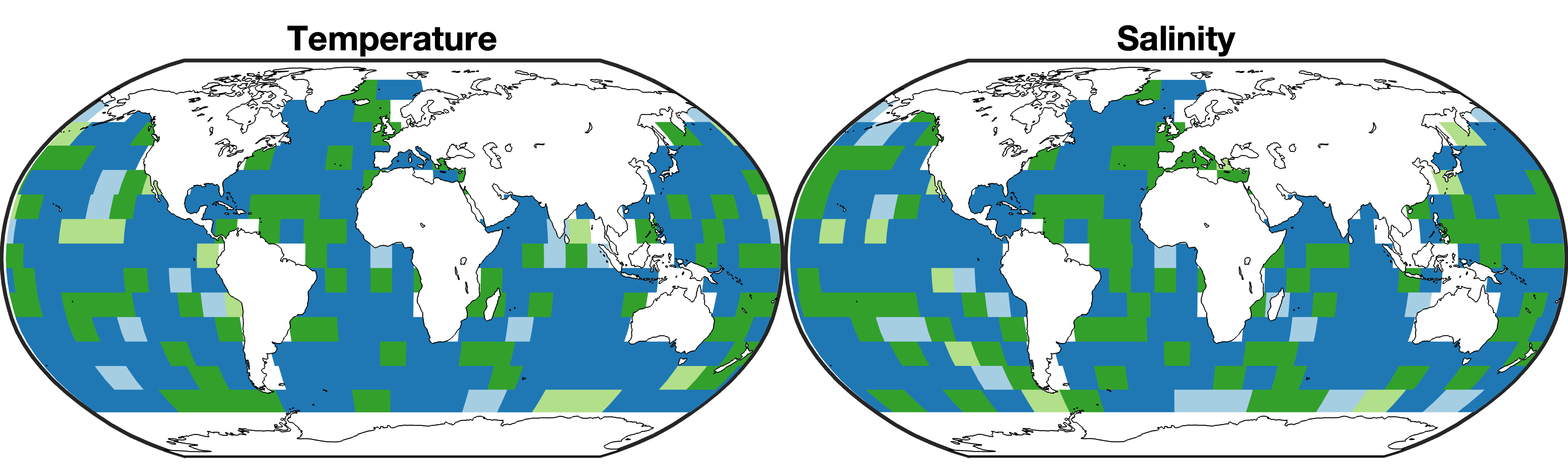}
        \caption{MAE}
    \end{subfigure}
    \hfill
    \begin{subfigure}[b]{\linewidth}
        \centering
         \includegraphics[width=\linewidth,trim=0cm 0cm 0cm  0cm,clip]{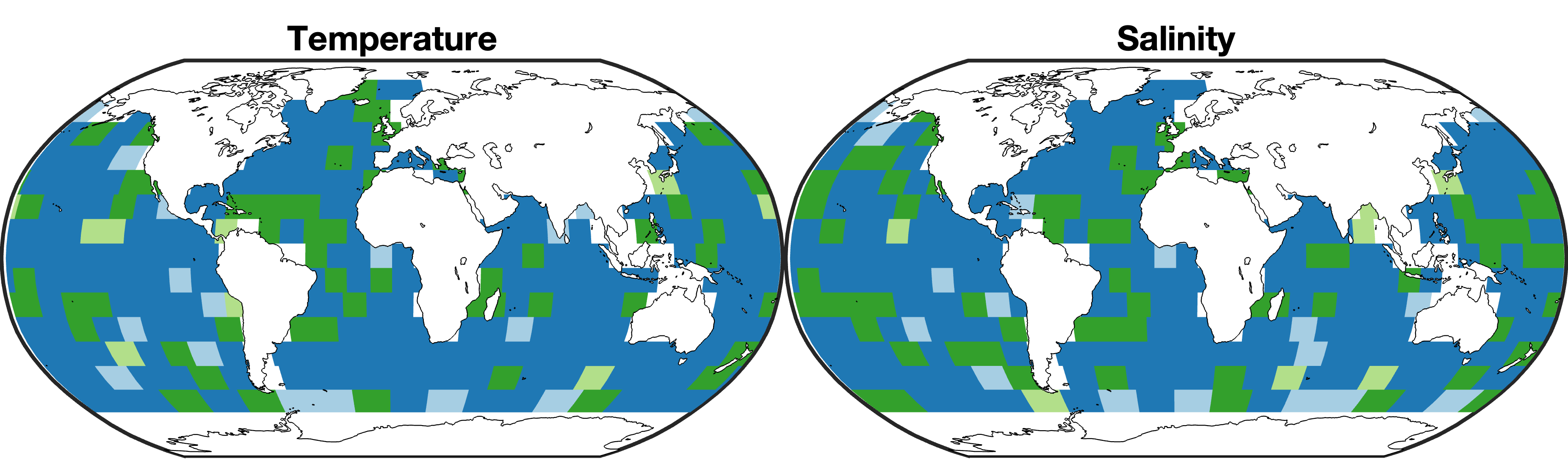}
        \caption{RMSE}
    \end{subfigure}   
\hfill 
    \begin{subfigure}[b]{\textwidth}
        \centering
        \includegraphics[width=\linewidth,trim=0cm 0cm 0cm 0cm,clip]{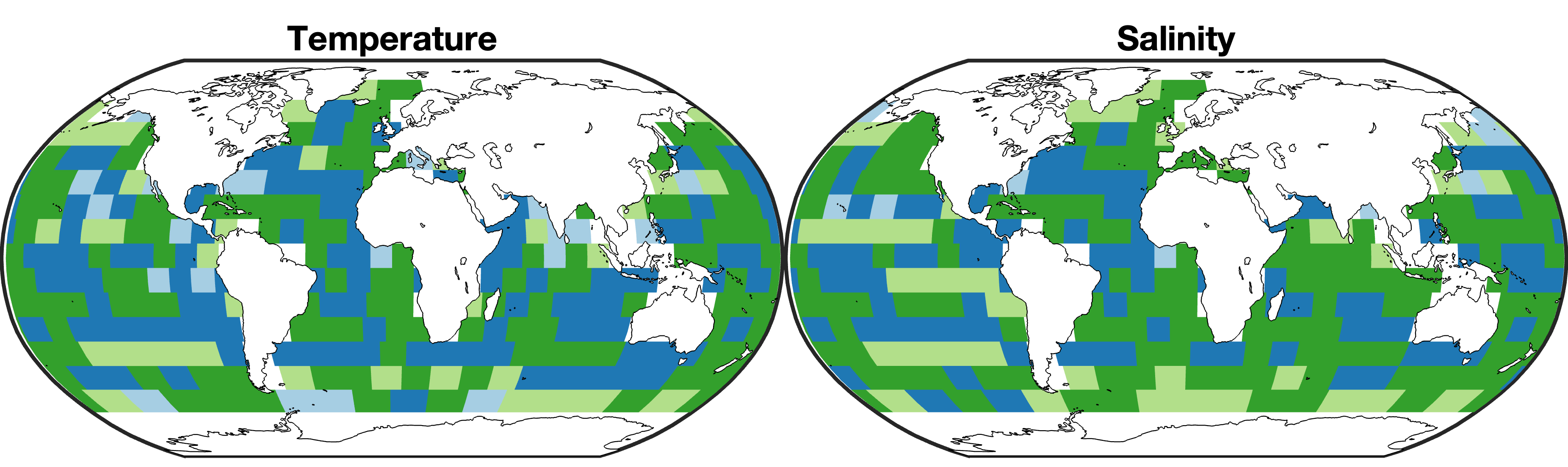}
        \caption{CRPS}
    \end{subfigure}
     \begin{subfigure}[b]{\textwidth}
        \centering
        \includegraphics[width=\linewidth,trim=0cm 0cm 0cm 1.5cm,clip]{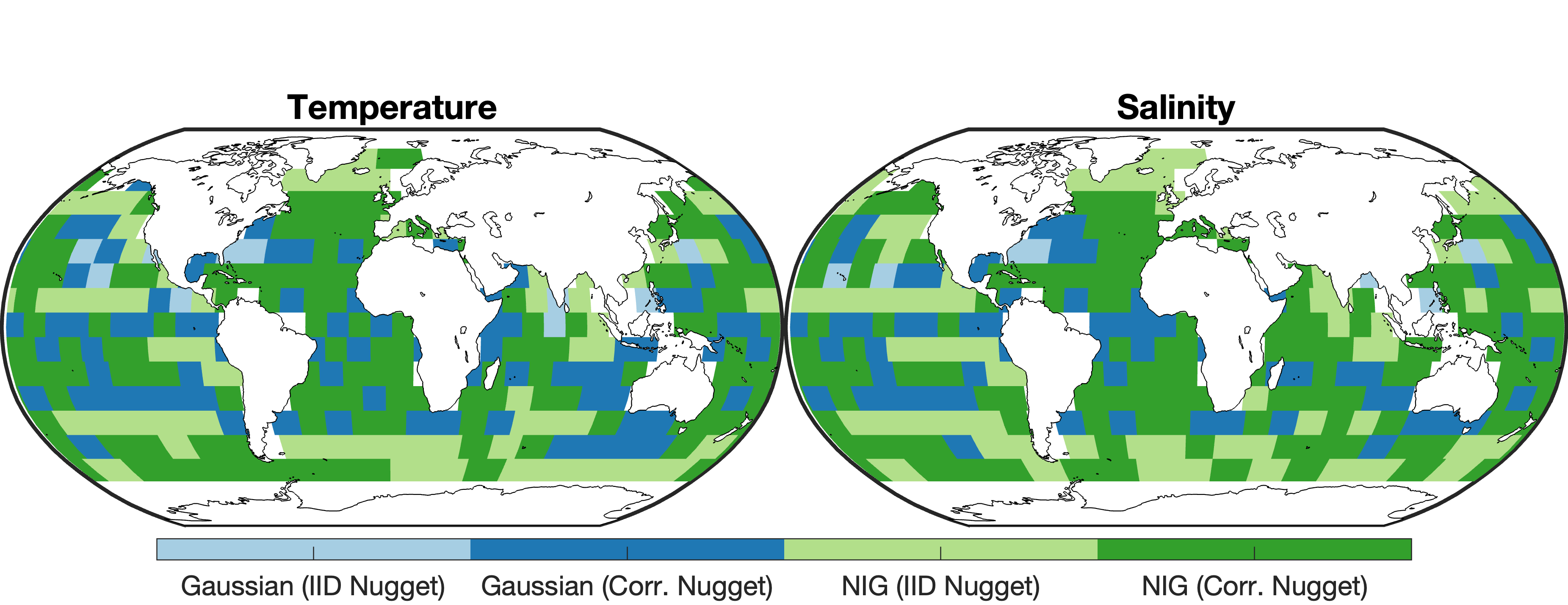}
        \caption{SCRPS}
    \end{subfigure}
    \caption{300 dbar}
\end{figure}

\begin{figure}[H]
    \centering
    \begin{subfigure}[b]{\linewidth}
        \centering
         \includegraphics[width=\linewidth,trim=0cm 0cm 0cm  0cm,clip]{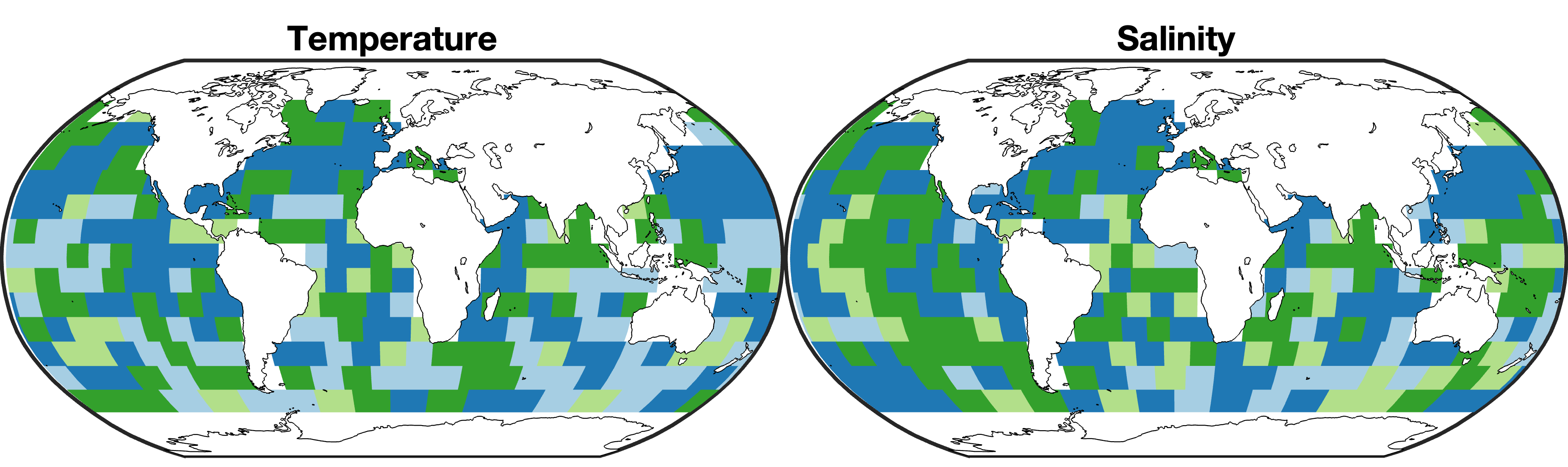}
        \caption{MAE}
    \end{subfigure}
    \hfill
    \begin{subfigure}[b]{\linewidth}
        \centering
         \includegraphics[width=\linewidth,trim=0cm 0cm 0cm  0cm,clip]{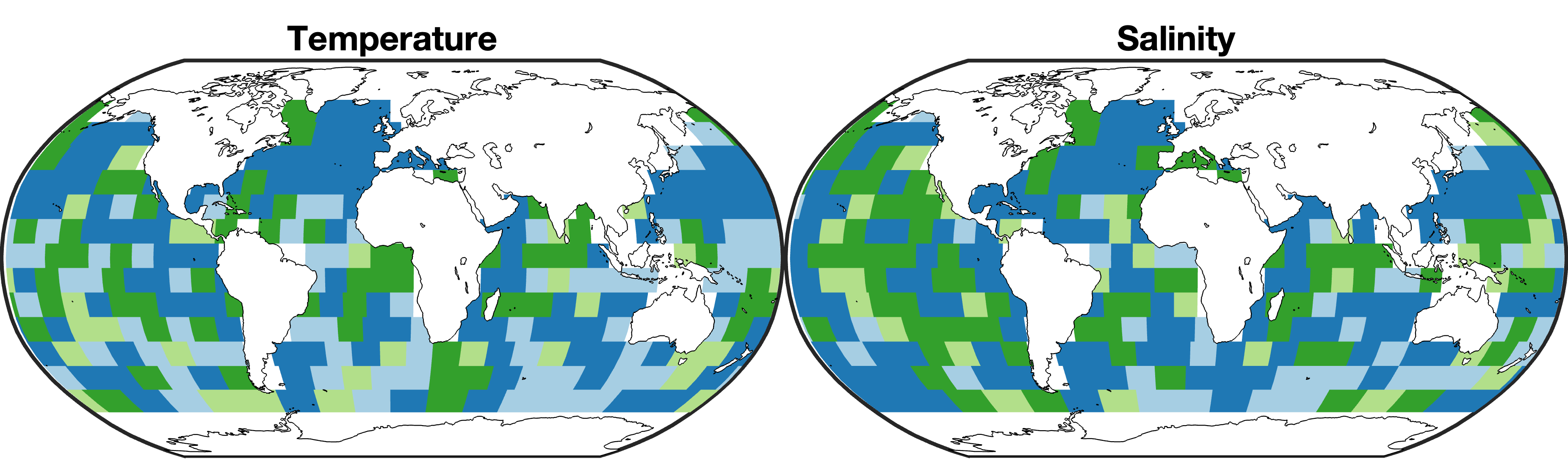}
        \caption{RMSE}
    \end{subfigure}   
\hfill 
    \begin{subfigure}[b]{\textwidth}
        \centering
        \includegraphics[width=\linewidth,trim=0cm 0cm 0cm  0cm,clip]{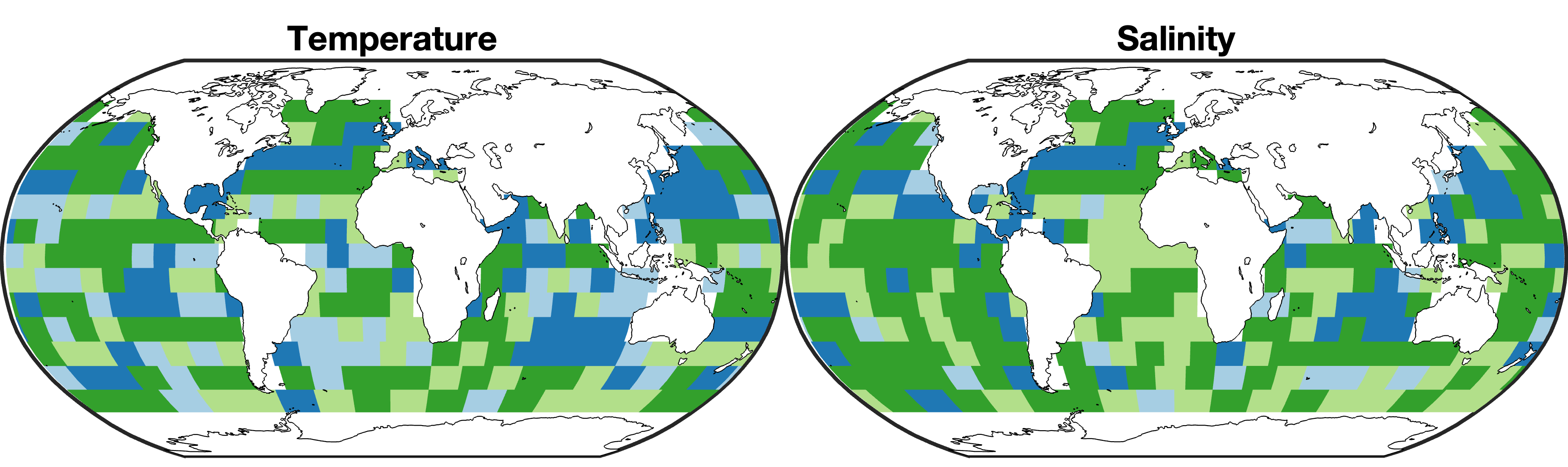}
        \caption{CRPS}
    \end{subfigure}
     \begin{subfigure}[b]{\textwidth}
        \centering
        \includegraphics[width=\linewidth,trim=0cm 0cm 0cm 1.5cm,clip]{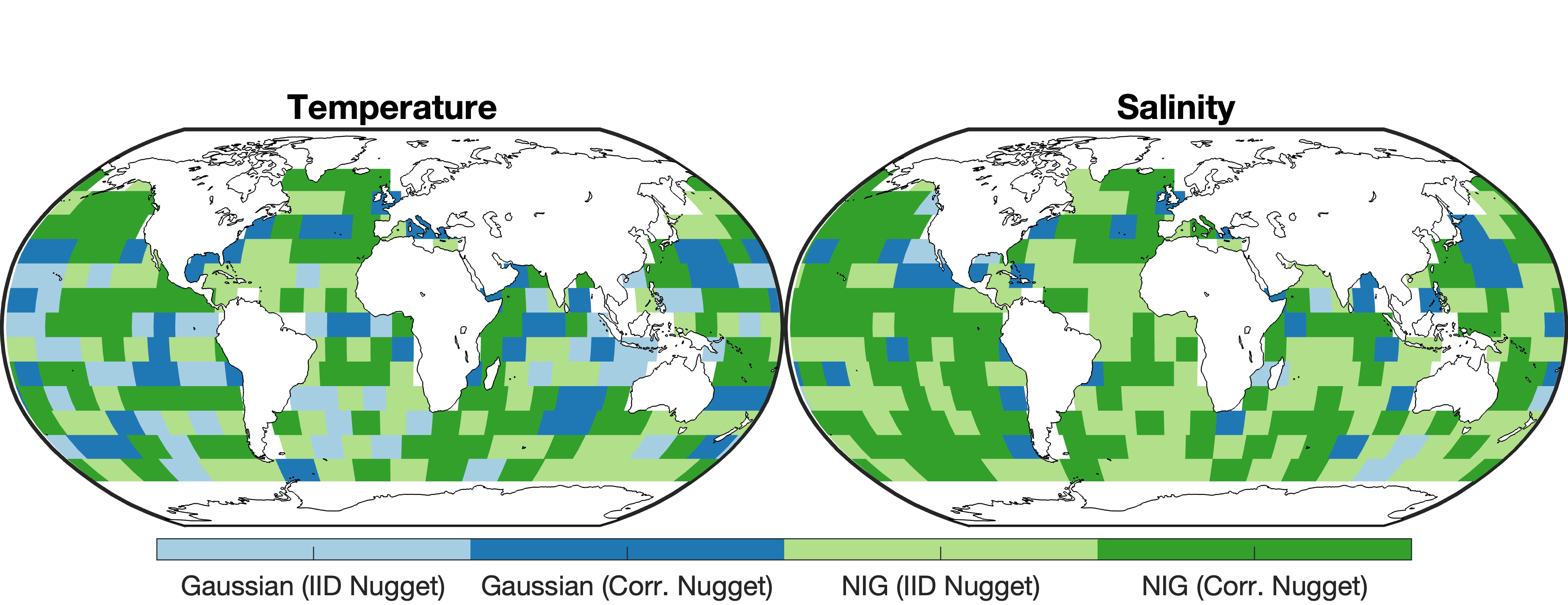}
        \caption{SCRPS}
    \end{subfigure}
    \caption{1000 dbar}
\end{figure}

\newpage

\subsection{Uncertainty quantification performance}
%Posterior predictive envelopes are evaluated via aggregated and latitude-stratified QQ plots.  Results are grouped by depth and variable (temperature or salinity).
\begin{figure}[H]
    \centering
    % First row: 10 dbar
    \begin{subfigure}[b]{0.4\linewidth}
        \centering
        \includegraphics[width=\linewidth]{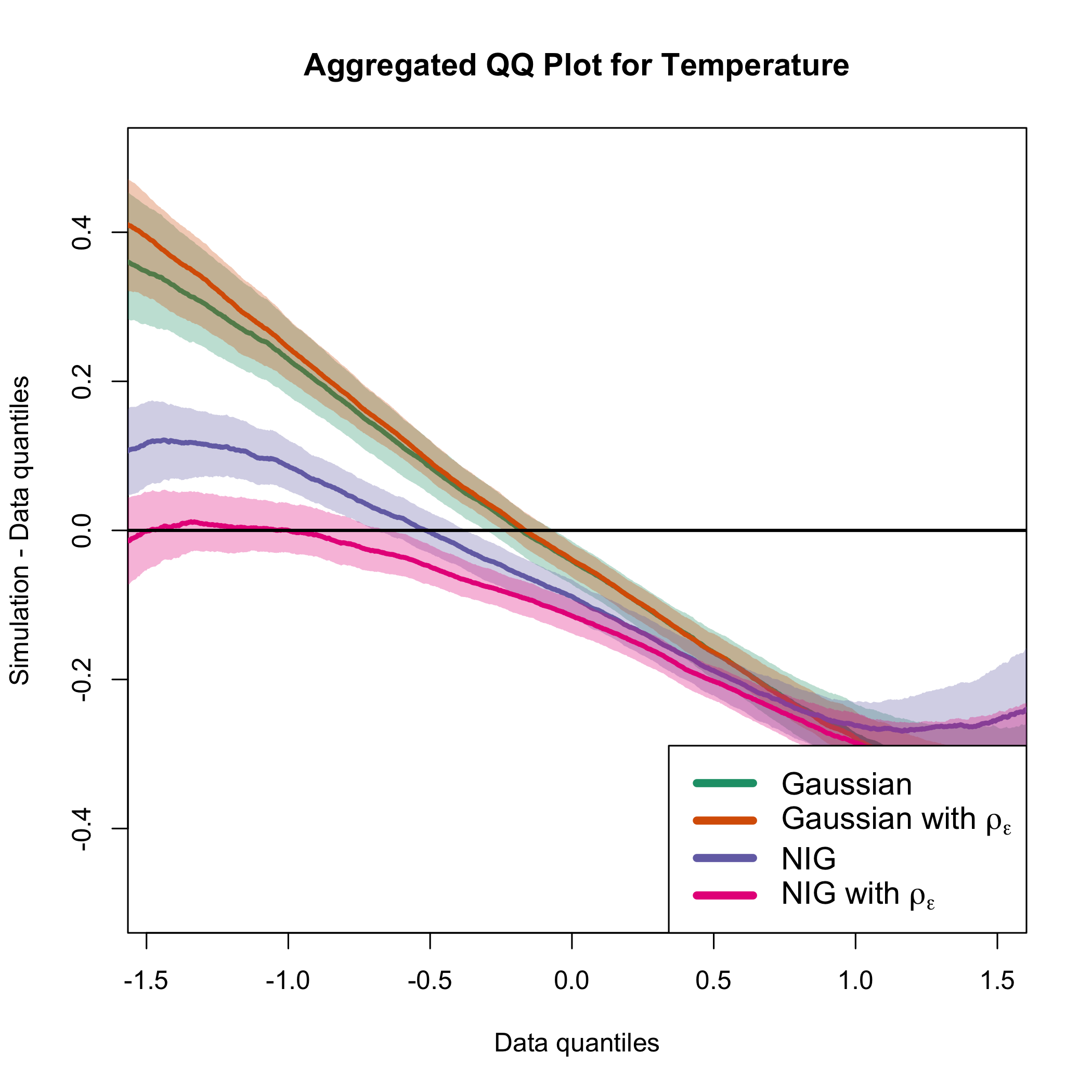}
        \caption{10 dbar Temperature}
        \label{fig:10_temp_qq}
    \end{subfigure}
    \hfill
    \begin{subfigure}[b]{0.4\linewidth}
        \centering
        \includegraphics[width=\linewidth]{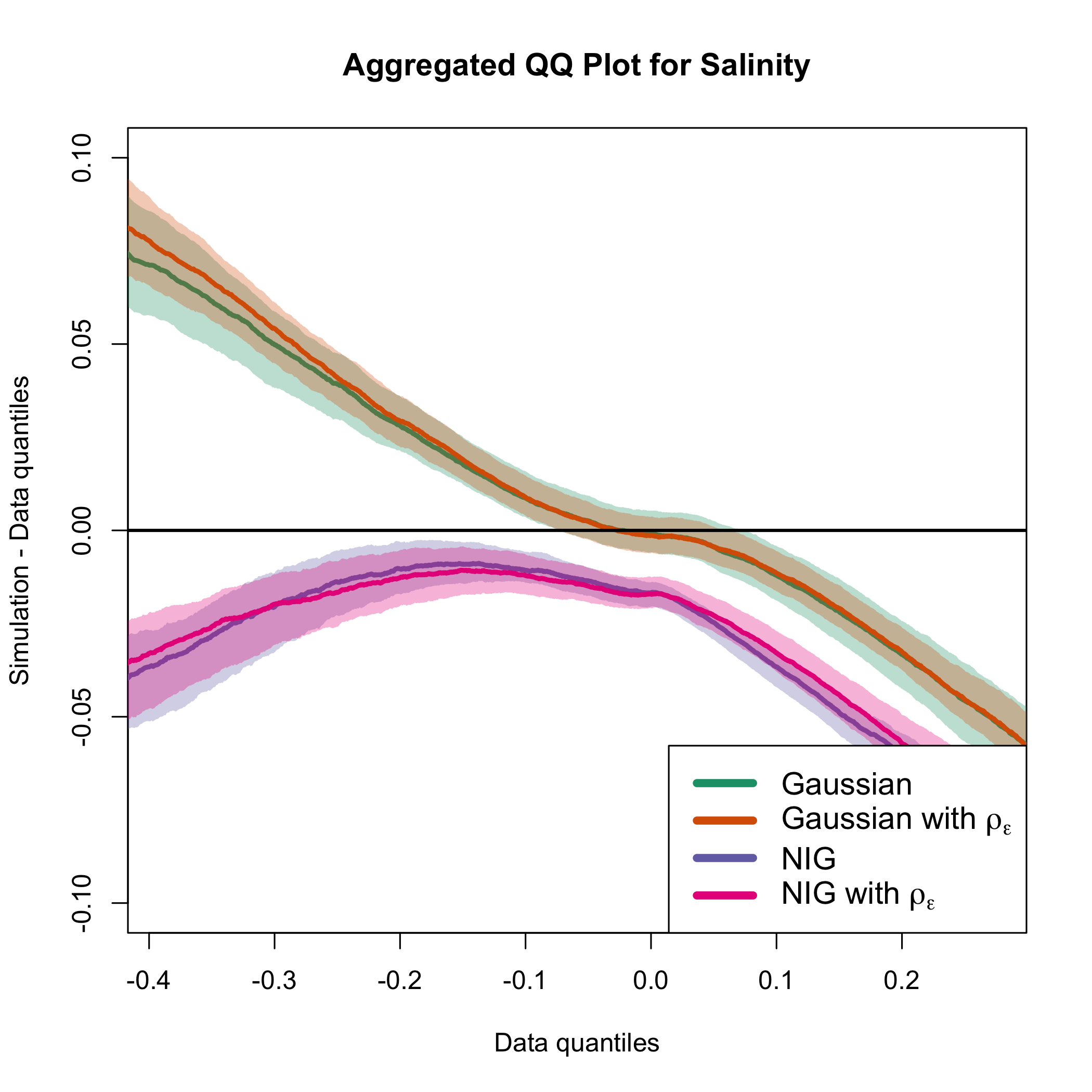}
        \caption{10 dbar Salinity}
        \label{fig:10_psal_qq}
    \end{subfigure}

    % Second row: 300 dbar
    \begin{subfigure}[b]{0.4\linewidth}
        \centering
        \includegraphics[width=\linewidth]{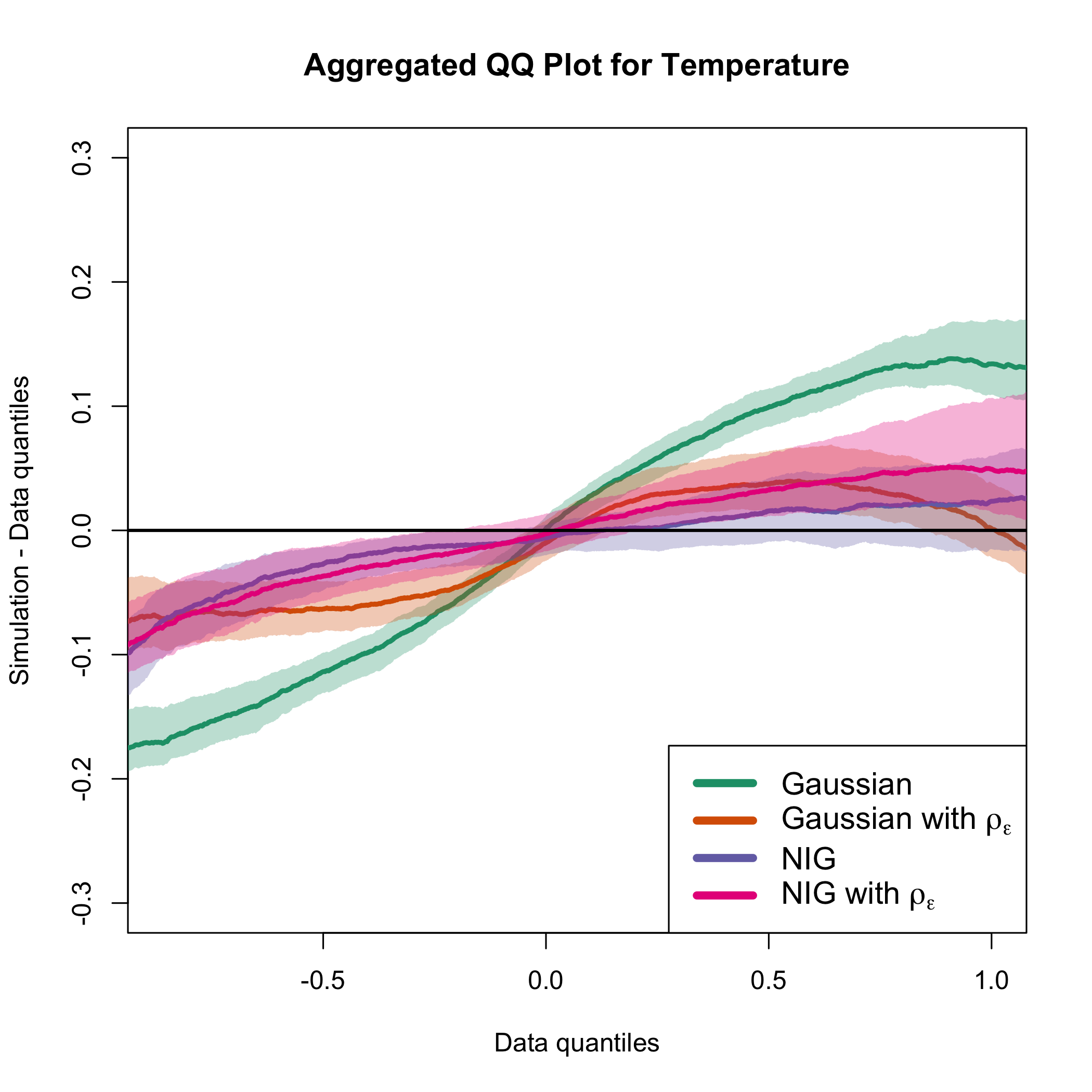}
        \caption{300 dbar Temperature}
        \label{fig:300_temp_qq}
    \end{subfigure}
    \hfill
    \begin{subfigure}[b]{0.4\textwidth}
        \centering
        \includegraphics[width=\textwidth]{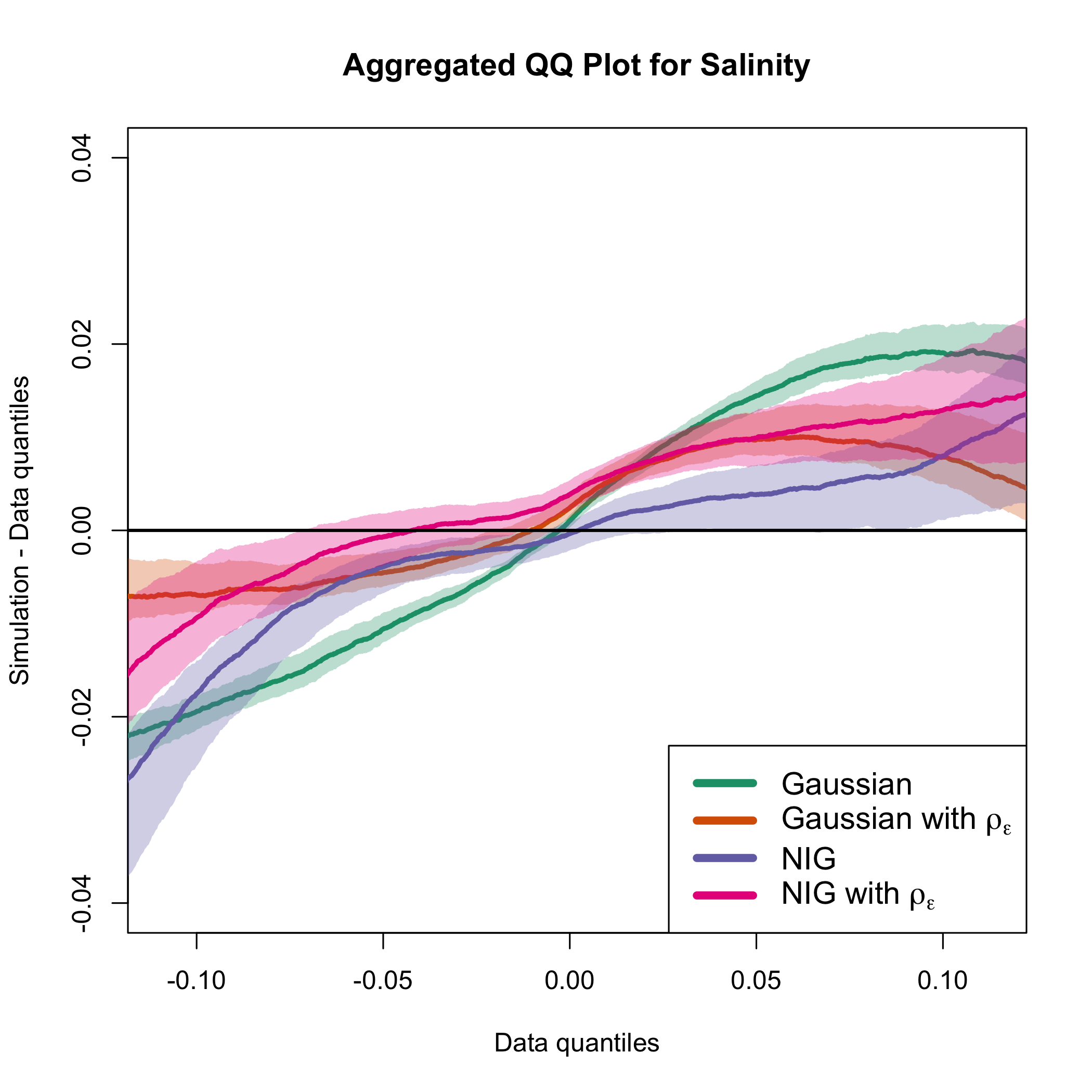}
        \caption{300 dbar Salinity}
        \label{fig:300_psal_qq}
    \end{subfigure}

    % Third row: 1000 dbar
    \begin{subfigure}[b]{0.4\textwidth}
        \centering
        \includegraphics[width=\textwidth]{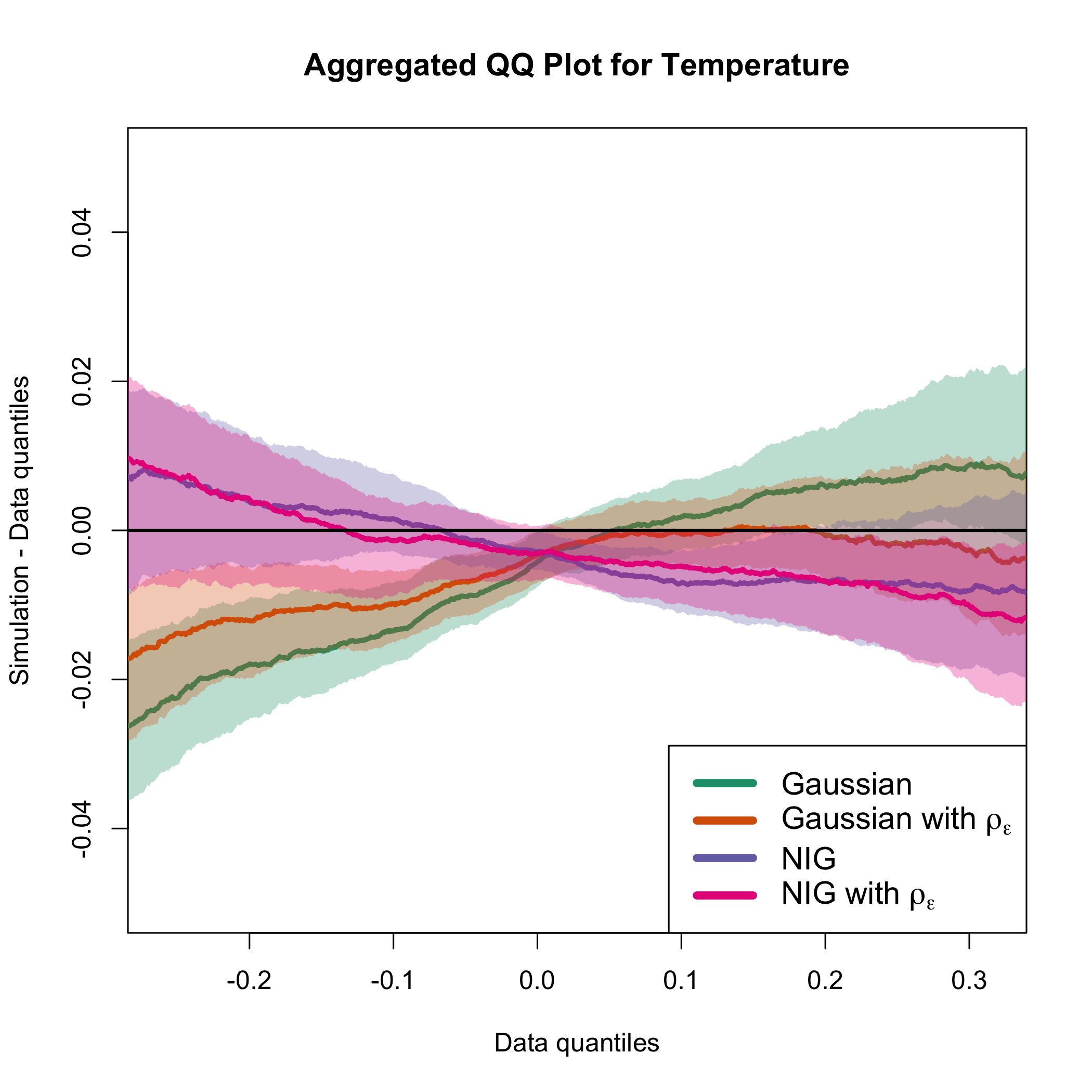}
        \caption{1000 dbar Temperature}
        \label{fig:1000_temp_qq}
    \end{subfigure}
    \hfill
    \begin{subfigure}[b]{0.4\textwidth}
        \centering
        \includegraphics[width=\textwidth]{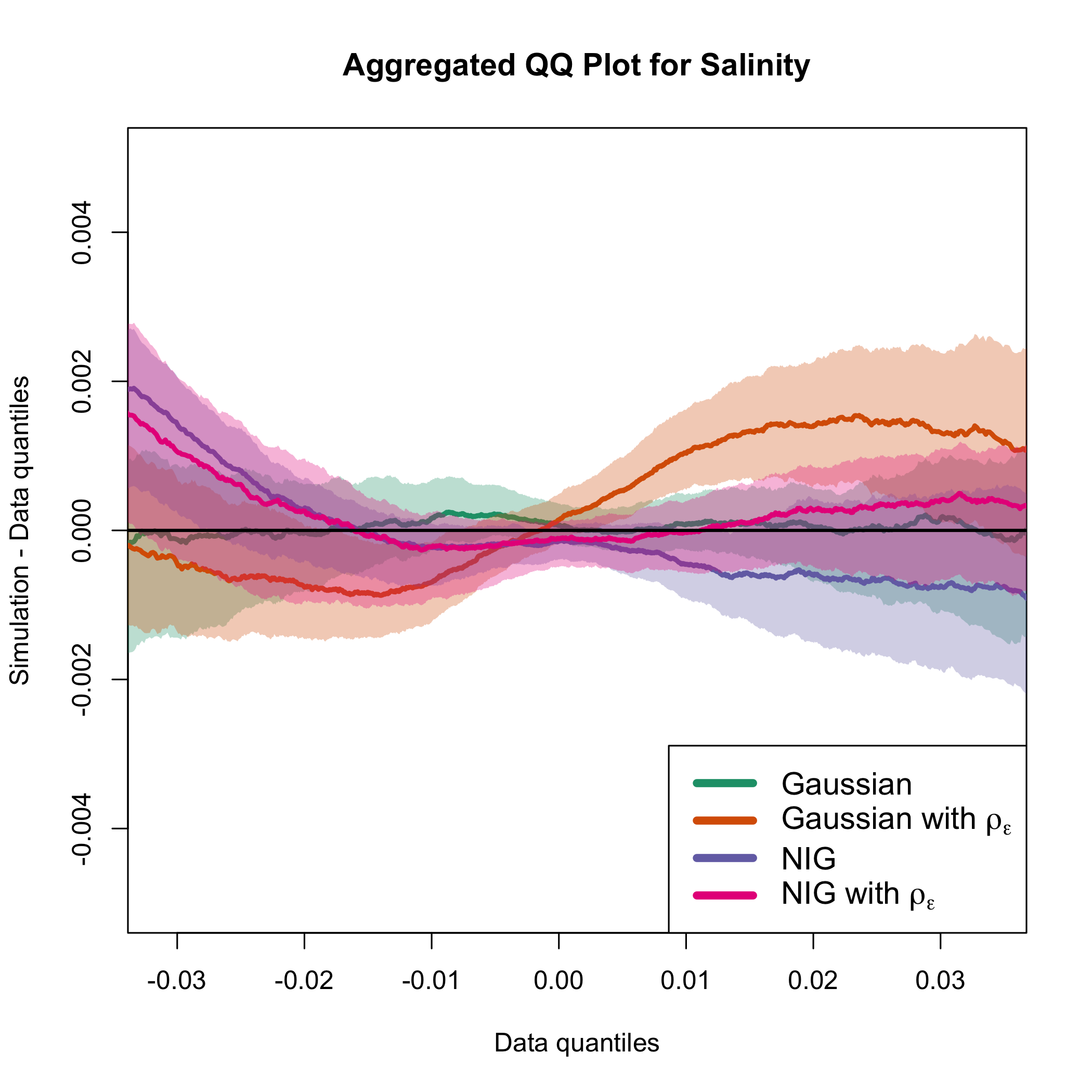}
        \caption{1000 dbar Salinity}
        \label{fig:1000_psal_qq}
    \end{subfigure}
    
    \caption{Aggregated QQ plot results for all models: qsample-qtheory vs.\ qtheory for LOOO cross-validation; Figures (a), (c), and (e) are for Temperature, while (b), (d), and (f) are for Salinity. The solid colored line is the average of the 20 simulated QQ curves for each model, and the semi-transparent band around it is the point-wise 95 \% simulation envelope.}
    \label{fig:qq-figures-horiz}
\end{figure}

%%% STRATIFIED QQ %%%%
% For 10 dbar Temperature
\begin{figure}[H]
    \centering
    \includegraphics[width=0.67\linewidth]{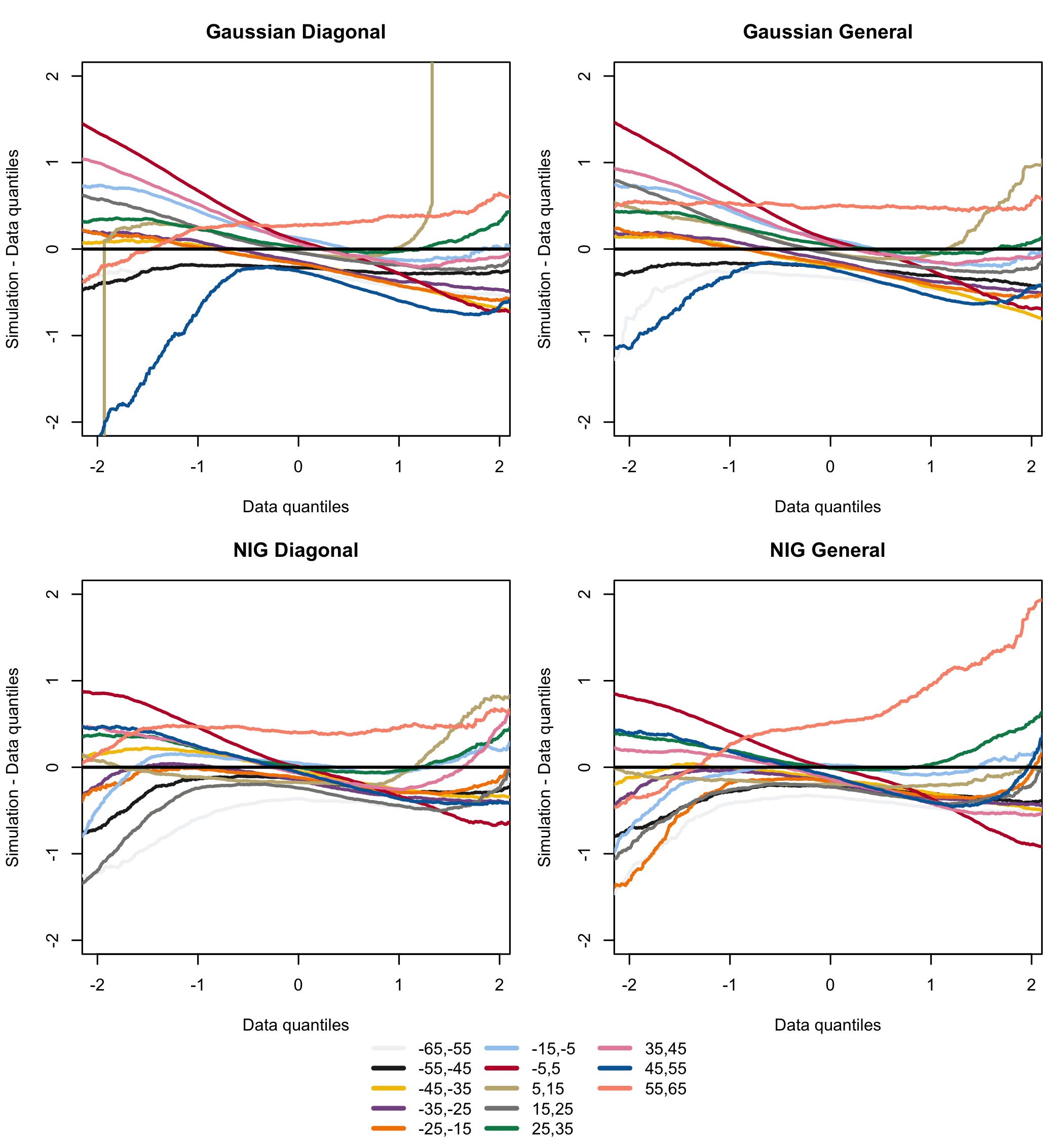}
    \caption{Stratified QQ plot by latitude corresponding to Figure~\ref{fig:10_temp_qq}, showing 10 dbar temperature.}
    \label{fig:10_temp_qq_strt}
\end{figure}
%\clearpage
\vspace{-7mm}
% For 10 dbar Salinity
\begin{figure}[H]
    \centering
    \includegraphics[width=0.67\linewidth]{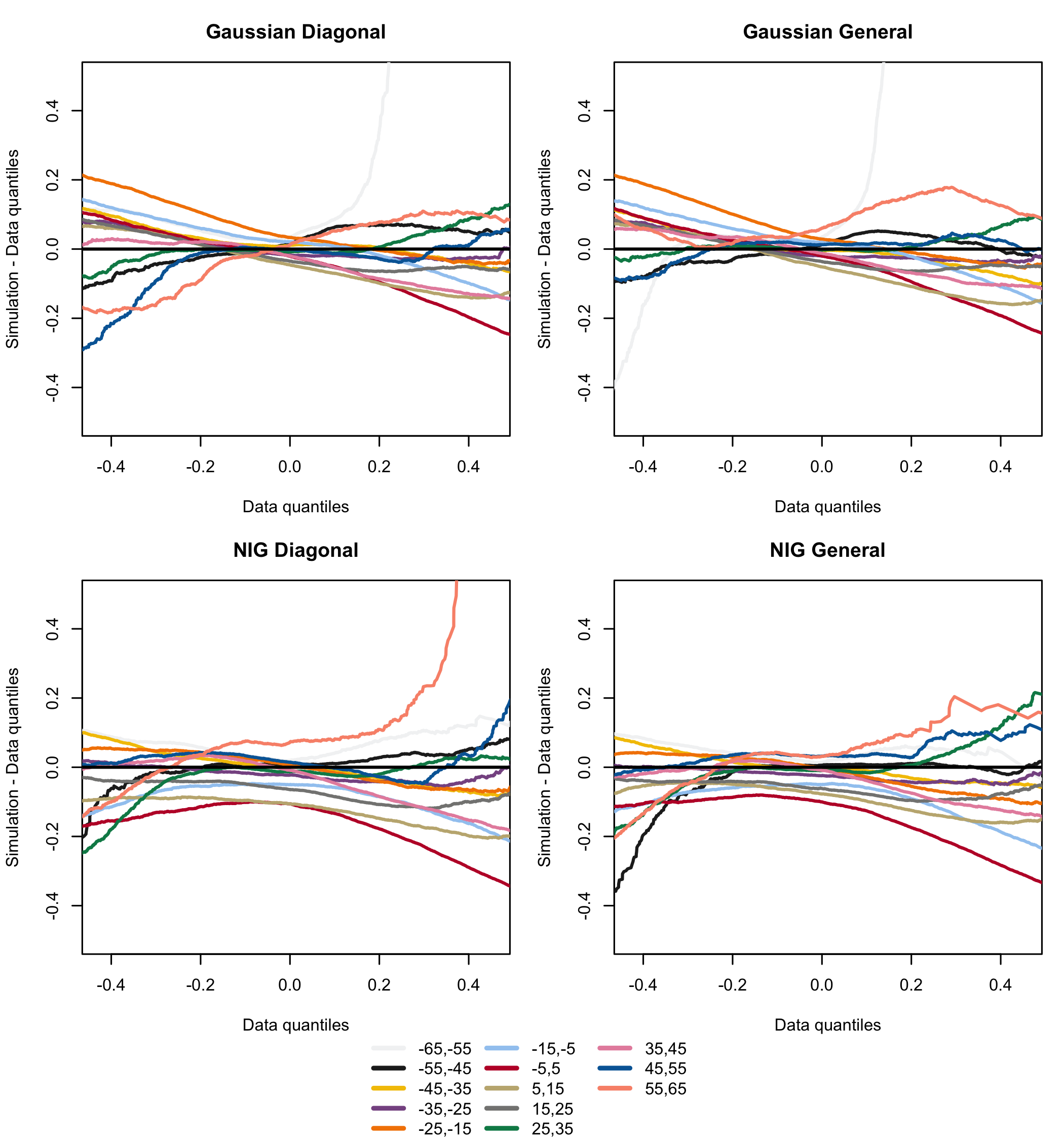}
    \caption{Stratified QQ plot by latitude corresponding to Figure~\ref{fig:10_psal_qq}, showing 10 dbar salinity.}
    \label{fig:10_psal_qq_strt}
\end{figure}

% For 300 dbar Temperature
\begin{figure}[H]
    \centering
    \includegraphics[width=0.67\linewidth]{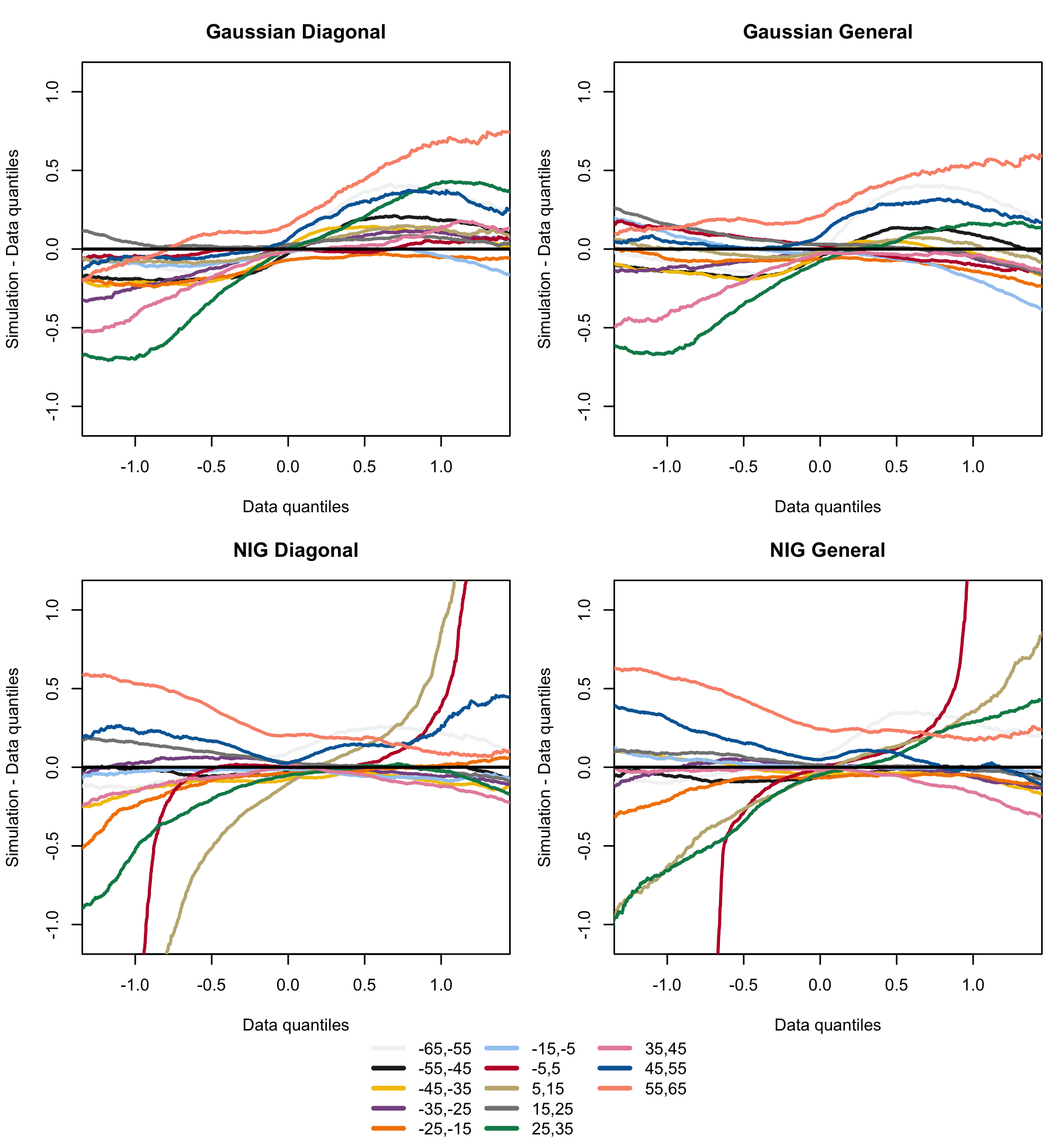}
    \caption{Stratified QQ plot by latitude corresponding to Figure~\ref{fig:300_temp_qq}, showing 300 dbar temperature.}
    \label{fig:300_temp_qq_strt}
\end{figure}
\vspace{-7mm}
% For 300 dbar Salinity
\begin{figure}[H]
    \centering
    \includegraphics[width=0.67\linewidth]{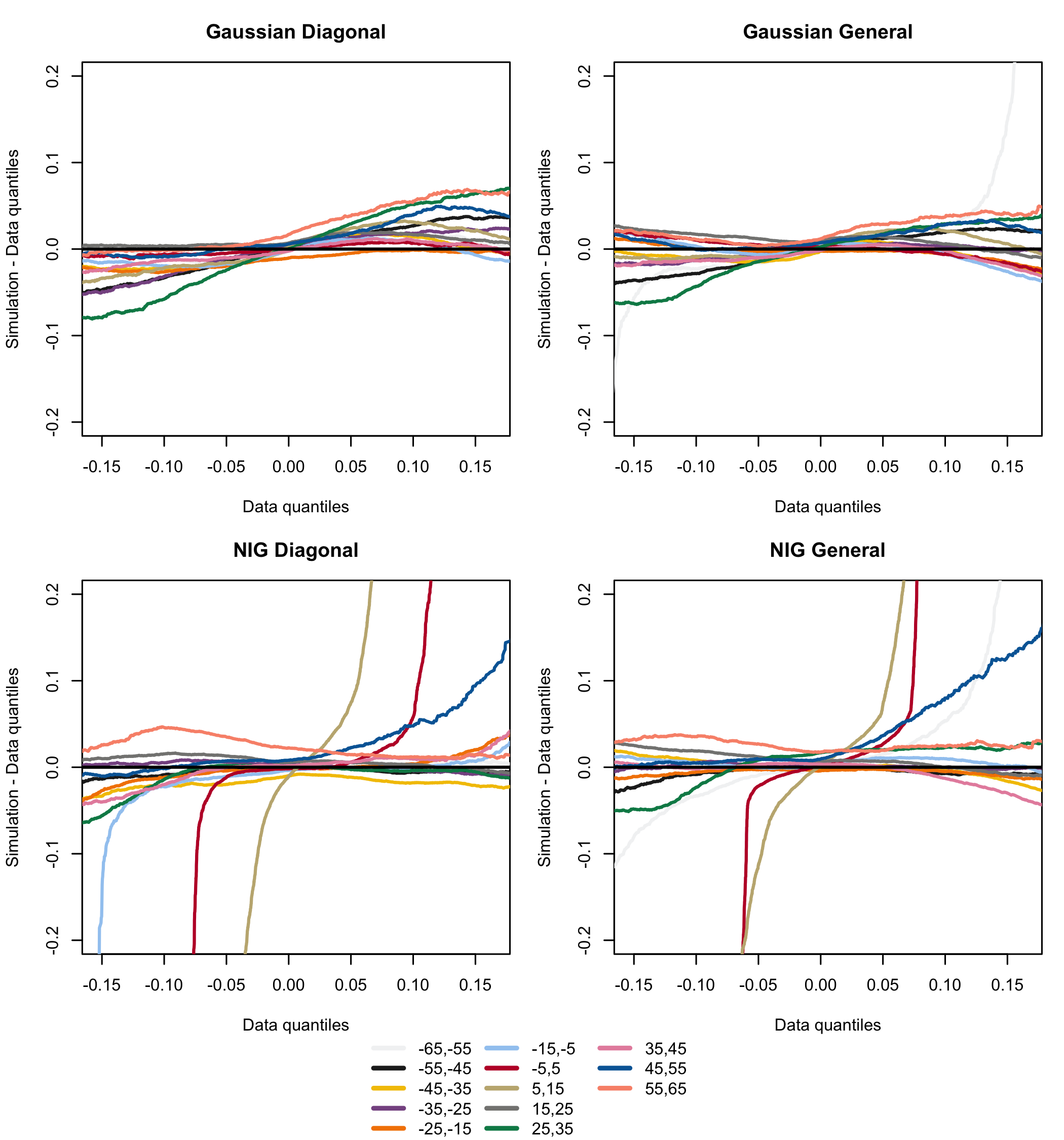}
    \caption{Stratified QQ plot by latitude corresponding to Figure~\ref{fig:300_psal_qq}, showing 300 dbar salinity.}
    \label{fig:300_psal_qq_strt}
\end{figure}

% For 1000 dbar Temperature
\begin{figure}[H]
    \centering
    \includegraphics[width=0.67\linewidth]{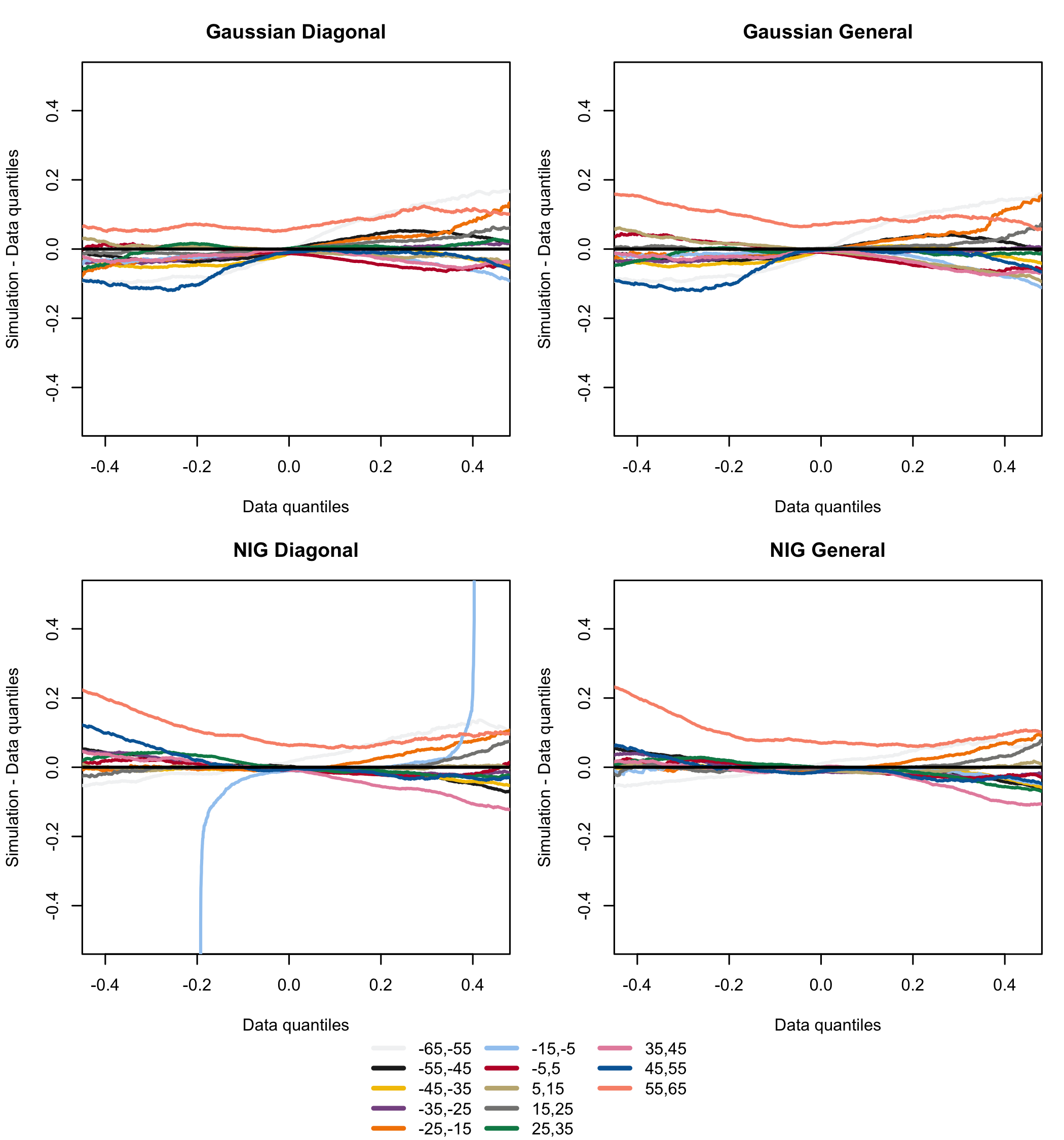}
    \caption{Stratified QQ plot by latitude corresponding to Figure~\ref{fig:1000_temp_qq}, showing 1000 dbar temperature.}
    \label{fig:1000_temp_qq_strt}
\end{figure}
\vspace{-7mm}
% For 1000 dbar Salinity
\begin{figure}[H]
    \centering
    \includegraphics[width=0.67\linewidth]{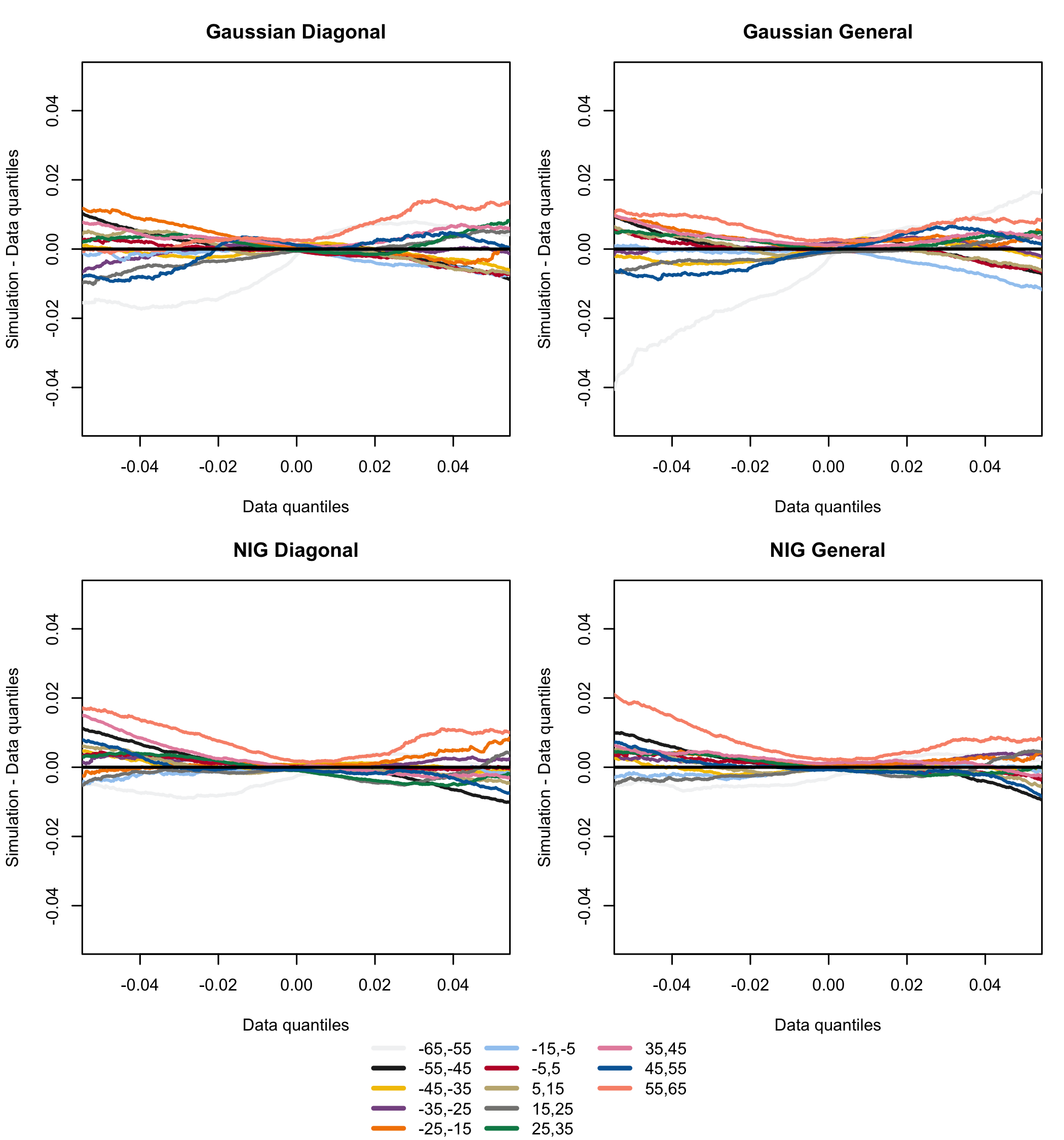}
    \caption{Stratified QQ plot by latitude corresponding to Figure~\ref{fig:1000_psal_qq}, showing 1000 dbar salinity.}
    \label{fig:1000_psal_qq_strt}
\end{figure}

\newpage
\subsection{Simulation Study}

This section presents the full results of the simulation study described in the main work.

% Sigma Plots
\begin{figure}[H]
    \centering
    \begin{subfigure}[t]{0.48\textwidth}
        \centering
        \includegraphics[width=\linewidth]{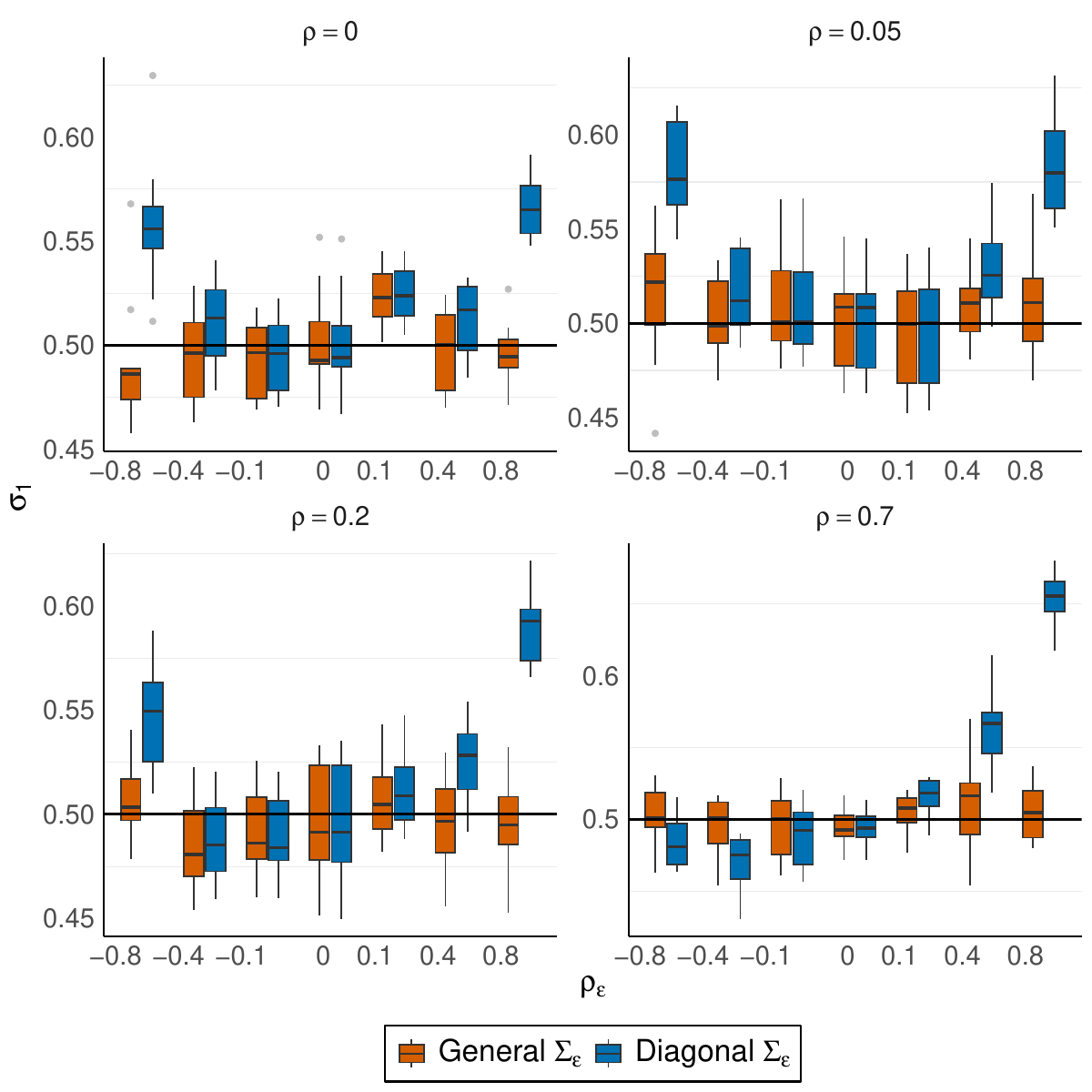}
        \caption{Estimated $\sigma_1$ parameter across simulation settings}
    \end{subfigure}%
    \hfill
    \begin{subfigure}[t]{0.48\textwidth}
        \centering
        \includegraphics[width=\linewidth]{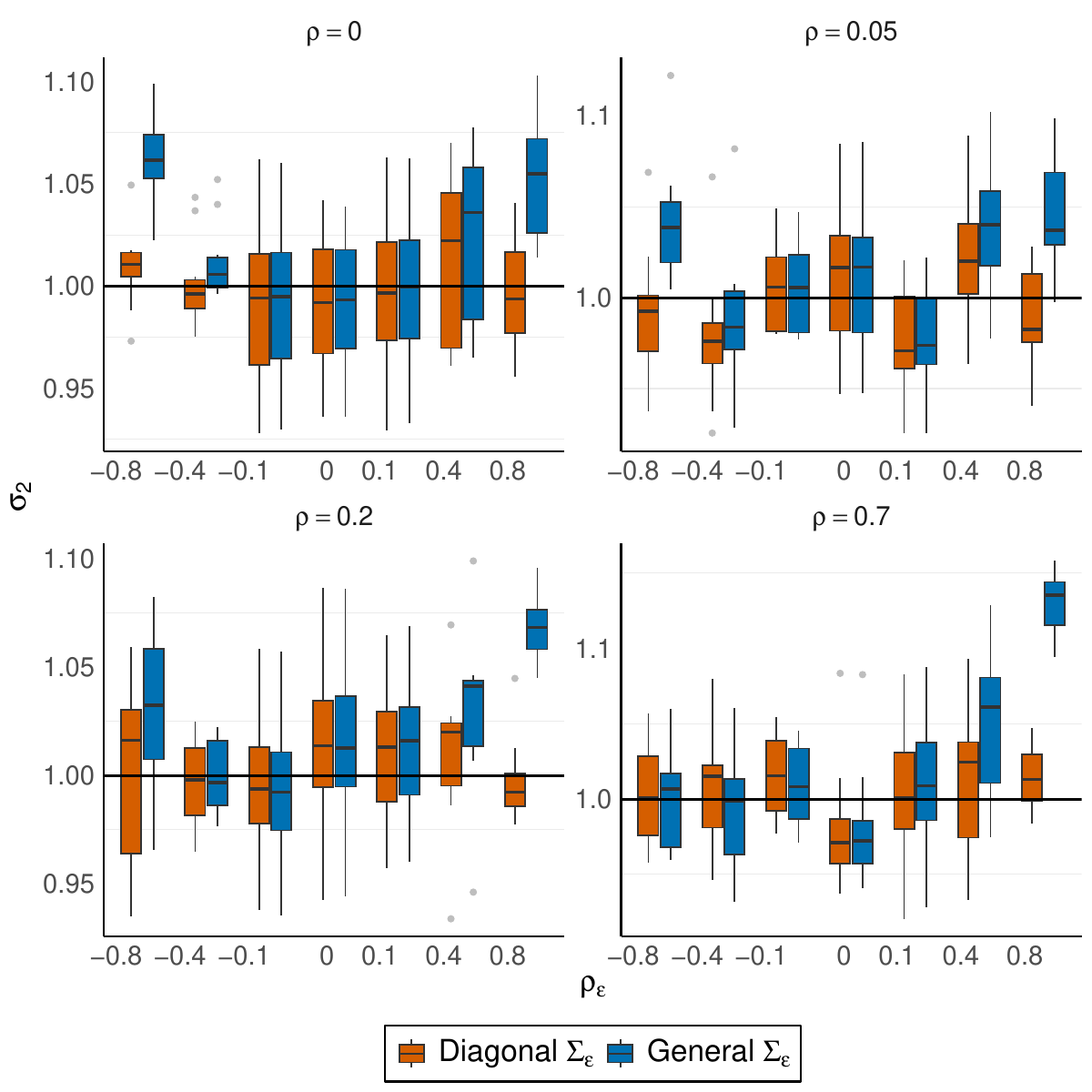}
        \caption{Estimated $\sigma_2$ parameter across simulation settings}
    \end{subfigure}
    \caption{Boxplots of $\sigma_1$ and $\sigma_2$ parameters across different values of $\rho$ and $\rho_{\epsilon}$. Each boxplot represents the variation in parameter estimation.}
    \label{fig:sigma_plots}
\end{figure}

% Signal-to-Noise Ratio Plots
\begin{figure}[H]
    \centering
    \begin{subfigure}[t]{0.48\textwidth}
        \centering
        \includegraphics[width=\linewidth]{Figures/snr1_boxplots_rho.pdf}
        \caption{Signal-to-Noise Ratio for $\sigma_1$}
        \label{fig:snr1}
    \end{subfigure}%
    \hfill
    \begin{subfigure}[t]{0.48\textwidth}
        \centering
        \includegraphics[width=\linewidth]{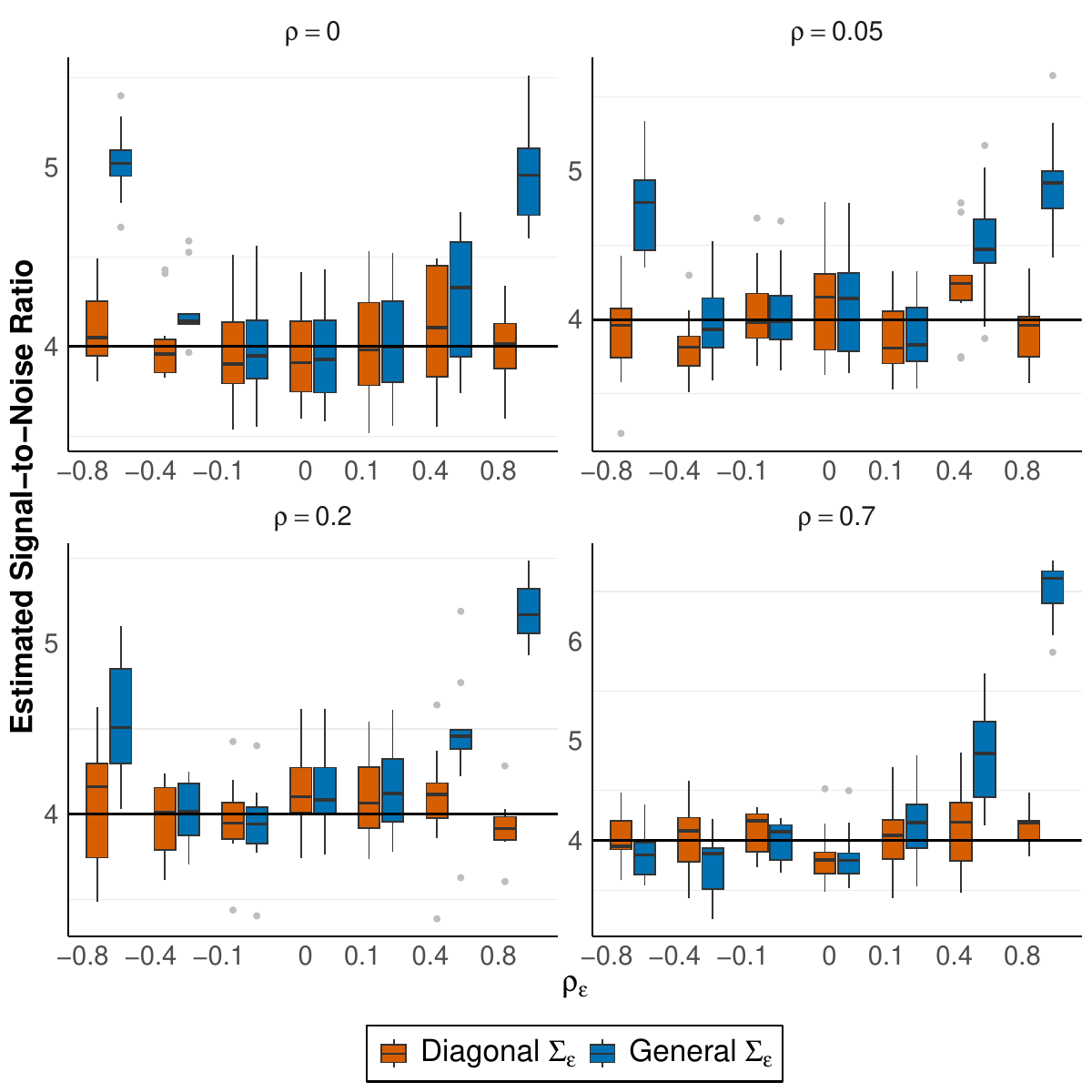}
        \caption{Signal-to-Noise Ratio for $\sigma_2$}
        \label{fig:snr2}
    \end{subfigure}
    \caption{Signal-to-Noise Ratios for $\sigma_1$ and $\sigma_2$. Each subplot shows the estimated ratio under different simulation settings.}
    \label{fig:snr_plots}
\end{figure}

% Kappa Plots
\begin{figure}[H]
    \centering
    \begin{subfigure}[t]{0.48\textwidth}
        \centering
        \includegraphics[width=\linewidth]{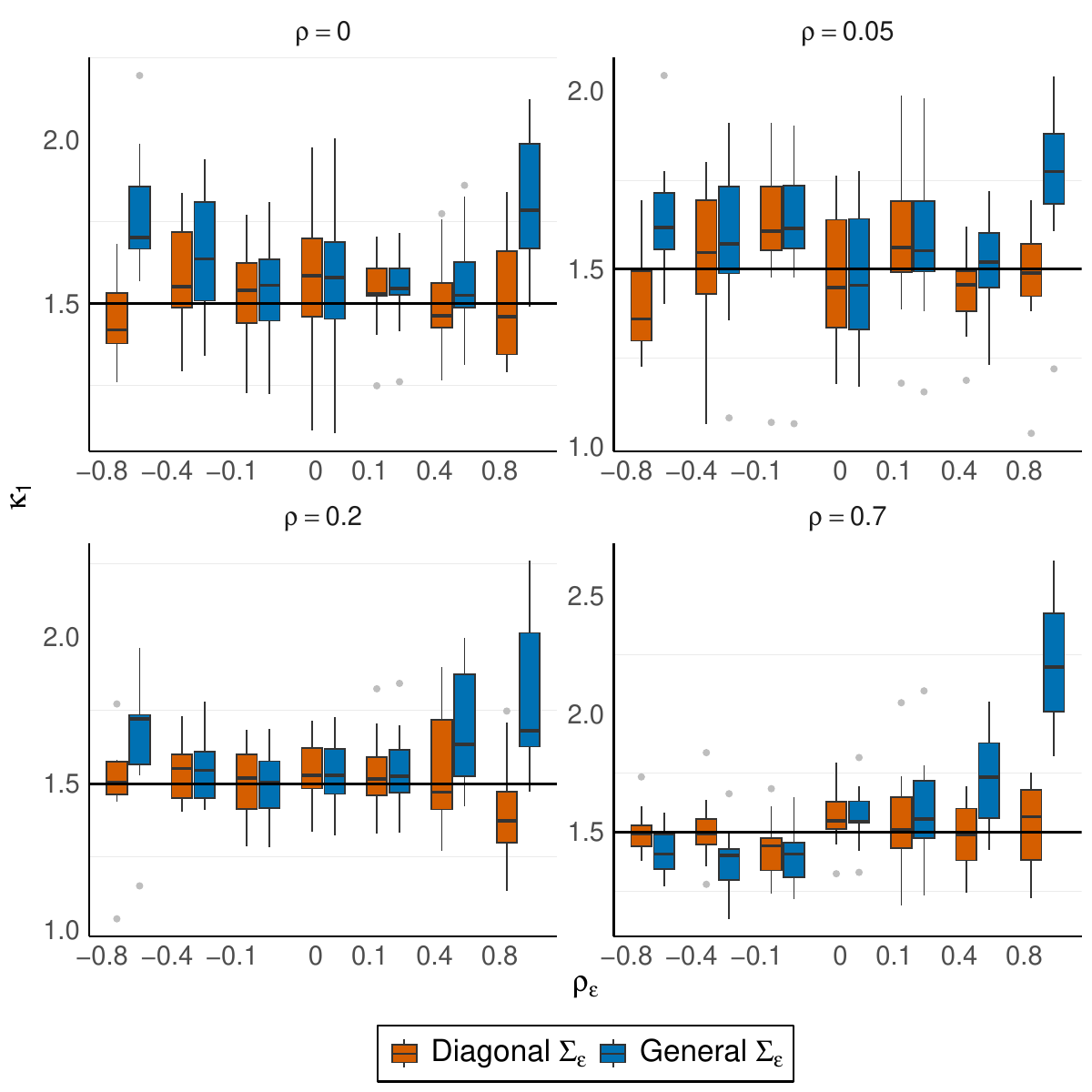}
        \caption{Estimated $\kappa_1$ parameter}
        \label{fig:kappa1}
    \end{subfigure}%
    \hfill
    \begin{subfigure}[t]{0.48\textwidth}
        \centering
        \includegraphics[width=\linewidth]{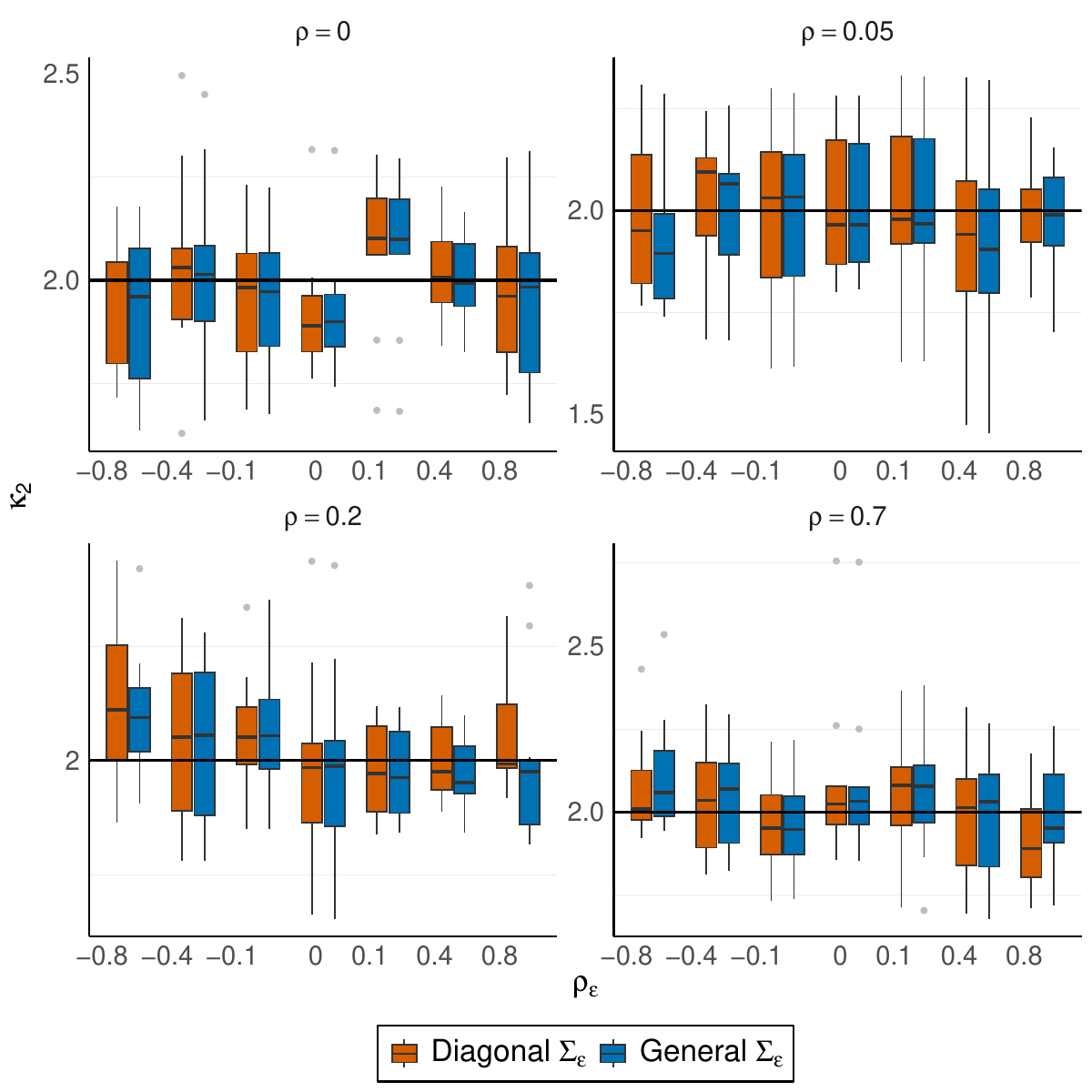}
        \caption{Estimated $\kappa_2$ parameter}
        \label{fig:kappa2}
    \end{subfigure}
    \caption{Boxplots of $\kappa_1$ and $\kappa_2$ parameters across different values of $\rho$ and $\rho_{\epsilon}$.}
    \label{fig:kappa_plots}
\end{figure}

% rho Plots
\begin{figure}[H]
    \centering
    \begin{subfigure}[t]{0.48\textwidth}
        \centering
        \includegraphics[width=\linewidth]{Figures/rho_boxplots.pdf}
        \caption{Estimated $\rho$ parameter}
        \label{fig:rho}
    \end{subfigure}%
    \hfill
    \begin{subfigure}[t]{0.48\textwidth}
        \centering
        \includegraphics[width=\linewidth]{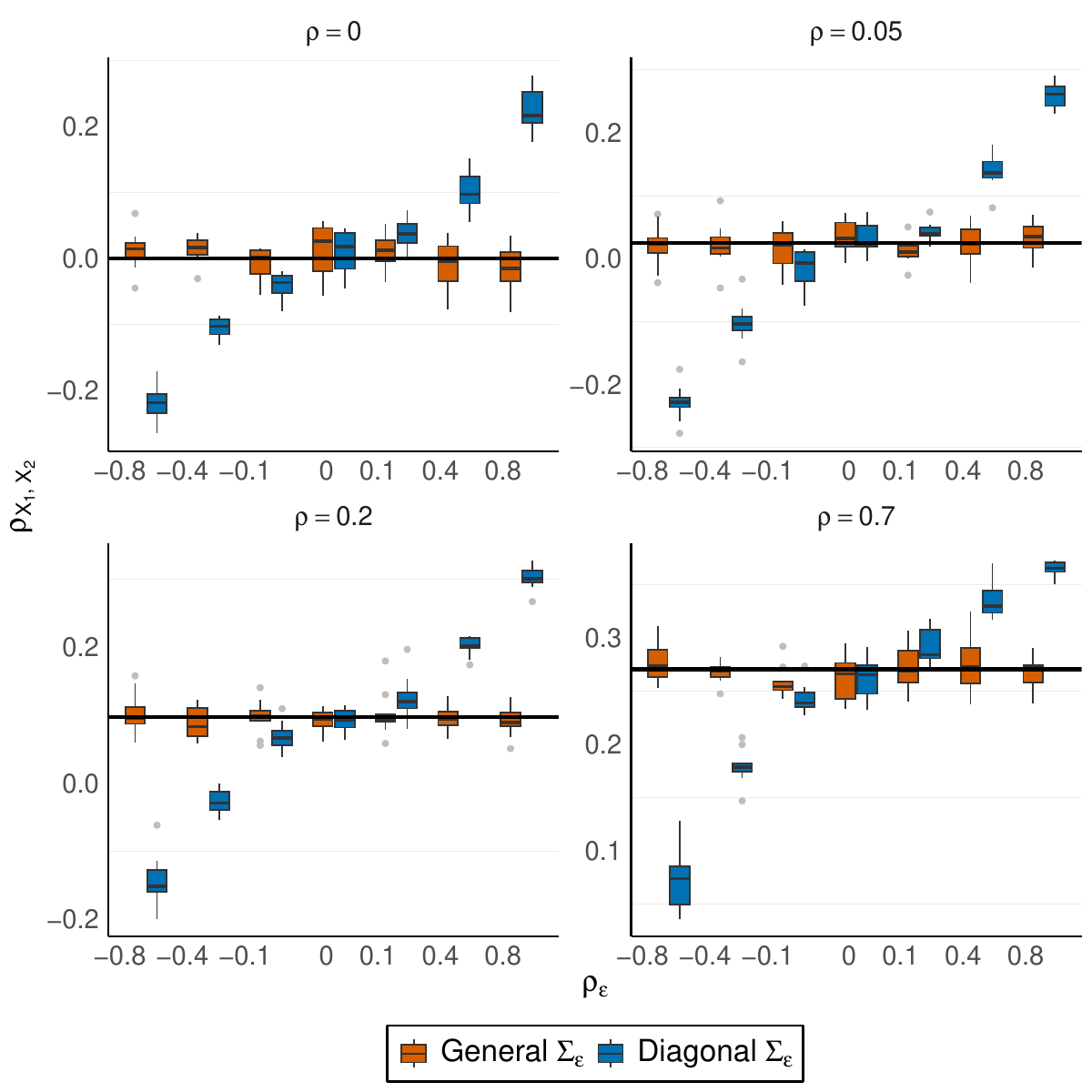}
        \caption{Estimated Pearson correlation}
        \label{fig:correlation}
    \end{subfigure}
    \caption{Boxplots for estimated values of  $\rho$ and Pearson correlation parameter across different values of $\rho$ and $\rho_{\epsilon}$.}
    \label{fig:rho_plots}
\end{figure}

\newpage

\newpage
\small 
\printbibliography

\end{document}